\documentclass[aps,pra,amsmath,notitlepage,nofootinbib,twocolumn,superscriptaddress,noeprint]{revtex4-2}
\pdfoutput=1
\usepackage[utf8]{inputenc}
\usepackage[english]{babel}
\usepackage[T1]{fontenc}

\usepackage{amsmath,amsfonts,amssymb,amsthm,bm,bbm,bbold}
\usepackage{graphicx}
\usepackage{amssymb}
\usepackage{mathtools}
\usepackage{physics}
\usepackage{makecell}
\usepackage[normalem]{ulem}
\usepackage{cancel}
\usepackage{tikz}
\usepackage{textcomp}
\usepackage{microtype}
\usepackage{xcolor}
\usepackage[breaklinks=true,colorlinks=true,linkcolor=teal,urlcolor=teal,citecolor=teal]{hyperref}
\usepackage{cleveref}
\usepackage{dsfont}

\renewcommand{\vec}{\boldsymbol}
\let\originalleft\left
\let\originalright\right
\renewcommand{\left}{\mathopen{}\mathclose\bgroup\originalleft}
\renewcommand{\right}{\aftergroup\egroup\originalright}
\renewcommand{\right}{\aftergroup\egroup\originalright}

\newtheorem{theorem}{Theorem}

\begin{document}

\preprint{APS/123-QED}

\title{Does entanglement enhance single-molecule pulsed biphoton spectroscopy?}

\author{Aiman Khan}
 \email{aiman.khan.1@warwick.ac.uk}
\affiliation{Department of Physics, University of Warwick, Coventry, CV4 7AL, United Kingdom}

\author{Francesco Albarelli}
\affiliation{Dipartimento di Fisica ``Aldo Pontremoli'', Università degli Studi di Milano, via Celoria 16, 20133 Milan, Italy}
\affiliation{Istituto Nazionale di Fisica Nucleare, Sezione di Milano, via Celoria 16, 20133 Milan, Italy}

\author{Animesh Datta}
\email{animesh.datta@warwick.ac.uk}
\affiliation{Department of Physics, University of Warwick, Coventry, CV4 7AL, United Kingdom}

\date{\today}
\begin{abstract}

It depends. For a single molecule interacting with one mode of a biphoton probe, 
we show that the spectroscopic information has three contributions, 
only one of which is a genuine two-photon contribution.
When all the scattered light can be measured, solely this contribution exists and can be fully extracted using unentangled measurements.
Furthermore, this two-photon contribution can, in principle, be matched by an optimised but unentangled single-photon probe.
When the matter system spontaneously emits into inaccessible modes, an advantage due to entanglement can not be ruled out.
In practice, time-frequency entanglement does enhance spectroscopic performance of the oft-studied weakly-pumped spontaneous parametric down conversion (PDC) probes.
For two-level systems and coupled dimers, more entangled PDC probes yield more spectroscopic information, even in the presence of emission into inaccessible modes.
Moreover, simple, unentangled measurements can capture between 60\% - 90\% of the spectroscopic information.
We thus establish that biphoton spectroscopy using source-engineered PDC probes and unentangled measurements can provide tangible quantum enhancement.
Our work underscores the intricate role of entanglement in single-molecule spectroscopy using quantum light.

\end{abstract}

\maketitle


\section{Introduction}

Over the last couple of decades, entangled states of quantum light have been explored for their potential use in linear and nonlinear optical spectroscopy. 
Linear absorption spectroscopy in a `biphoton' setup --- employing a single one-photon interaction of the sample with the signal mode~\citep{stefanov2017role} of an entangled state
--- has been experimentally performed on glass~\citep{yabushita2004spectroscopy}, crystalline~\citep{kalachev2007biphoton}, and nanoparticle~\citep{kalashnikov2014quantum} samples. 
These offer a larger coincidence signal-to-noise ratio~(SNR)~compared to photon-counting measurements of classical light.

More recently, coherent non-linear optical spectroscopies using quantum light have been proposed and theoretically studied using a fully quantum mechanical approach~\citep{dorfman2016nonlinear}.
Amongst these, two-photon absorption of entangled biphoton states has garnered much attention~\citep{dorfman2016nonlinear,schlawin2017entangled,schlawin2018entangled,mukamel2020roadmap,raymer2021entangled, Parzuchowski2021, CoronaAquino2022}. The supposed improved performance of these quantum spectroscopic methods over classical ones is often attributed to the availability of
 control variables such as the entanglement time that do not have classical counterparts~\cite{saleh1998entangled},
  or more generally to the absence of the usual Fourier limit on 
 joint temporal and spectral resolutions of entangled photons~\cite{dorfman2016nonlinear}.

\begin{figure}[t!]
        \fbox{
        \includegraphics[width=0.450\textwidth]{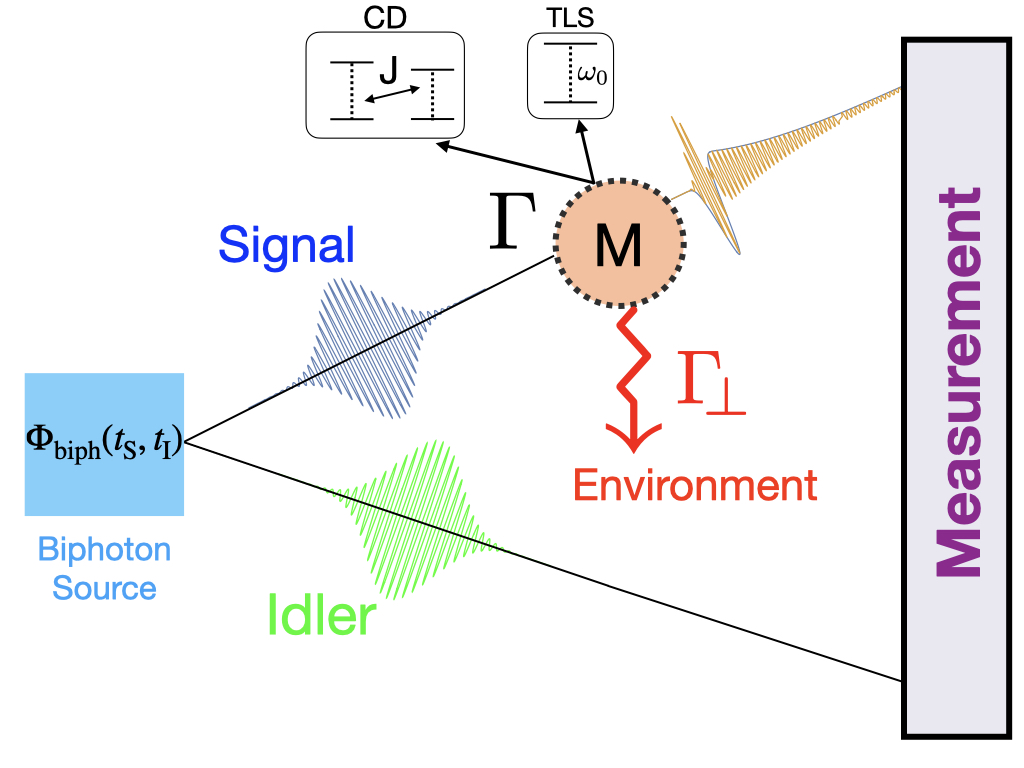}}
        \caption{Schematic of pulsed biphoton spectroscopy. Only the signal photon interacts with the molecular system M, and both photons are 
        measured after the light-matter interaction. $\Gamma$ captures the interaction of M with the incoming signal mode while $\Gamma_{\perp}$ does so for all other photonic modes, collectively dubbed the environment E.
        The specialisation of M to the two-level system (TLS) and coupled dimer (CD) is only made in Sections \eqref{sec6b} and \eqref{sec6c} respectively.}
         \label{fig:biphotonspec_diagram}
\end{figure}

However, the improved SNR in the biphoton experiments may be attributed to the use of coincidence counters with sufficiently small time windows, and not to the inherent quantum correlations of the entangled, non-classical light.
In fact, Stefanov has mathematically shown that probability distributions of outcomes of uncorrelated 
measurements on the two spatially distinguishable modes of the outgoing biphoton state can always be mimicked with a `classical' incoming 
ensemble with identical frequency correlations between the two photons, and \textit{no} entanglement~\citep{stefanov2017role}.
Entanglement with an additional `idler' mode then seemingly provides no advantage, --- or disadvantage, when the subsequent measurements 
on signal and idler modes are independent.

The general question of the role of entanglement in quantum light spectroscopy is more subtle and intricate.
It depends on the location---at the input or at the measurement stage,
on the model of the matter sample and the light-matter interaction, and on the nature and strength of the coupling to both photonic and phononic environments. 
It also depends on the type of light used as probe~(and consequently the type of entanglement, such as between time-frequency modes, photon numbers, or some other degrees of freedom), 
as well as on the time of measurement compared to the typical matter system timescale(s),
its type, and the detection system used.
Finally, it depends on what spectroscopy is defined to be.
Mathematically, a larger probability --- of absorption, emission, or general quantum measurement outcomes --- is not identical to a higher precision of estimating unknown parameter(s) of the model matter system~\cite{Albarelli2022}.
The latter is typically the operational goal of spectroscopy.

In this paper, we elucidate the role of time-frequency quantum entanglement in biphoton spectroscopy in the precise estimation of unknown parameters of individual molecules.
As illustrated in Figure \ref{fig:biphotonspec_diagram}, it captures spectroscopic setups in which only one of the spatially distinct travelling modes interacts with the sample, the state of light itself carrying two excitations.
Our interaction model is motivated by the coherent coupling of single molecules~(with strength $\Gamma$) with optical fields. 
We also include a photonic environment characterised by the coupling strength $\Gamma_{\perp}$ modelling emission into inaccessible modes. 
Our conclusions are based on the quantum information-theoretic methodology for quantum light spectroscopy developed in our recent paper~\cite{Albarelli2022}.

Of proximate experimental relevance, we establish the functional utility of time-frequency entanglement 
for the spectroscopy of a two-level system~(TLS) and a coupled dimer~(CD) using states generated from weakly-pumped spontaneous parametric down conversion (PDC).
Such states are typical of spectroscopic techniques employing entangled light~\cite{yabushita2004spectroscopy,kalachev2007biphoton,kalashnikov2014quantum,schlawin2017entangled,schlawin2018entangled,raymer2021entangled,carnio2021optimize,fujihashi2022probing}. 
Our choice of matter systems also has immediate practical relevance. 
Under certain symmetries of the dipole operator or isotropic pumping, alkali-metal atomic vapours are approximately TLS~\citep{Steck2007}. 
The CD is often employed as a minimal model~\citep{ishizaki2009adequacy,marcus2020towards,khan2021model} for exciton-hopping quantum dynamics
in a wide class of light-harvesting complexes such as the 8-site Fenna-Matthews-Olsen~(FMO) complex and the 27-site B800-B850 light-harvesting-2 complexes.
It is more substantively relevant for the cyanobacterial allophycocyanin~(APC) complex~\citep{womick2009exciton}, as well as conjugated polymers~\citep{barford2014theory,barford2022exciton}.

Our main results for single-molecule biphoton spectroscopy are as follows:
\begin{enumerate}
    \item At long times, the spectroscopic information is bounded by a sum of three --- a two-photon, a one-photon, and a classical, contributions~[Eq.~(\ref{eq:rhoPouttwinQFI})].  
    
    \item Entangled measurements across the signal and idler modes are not necessary to attain the bound in Eq.~(\ref{eq:rhoPouttwinQFI}) [see Section \eqref{sec:idlertosignal})]. 

    \item Time-frequency entanglement of the input probe provides no in-principle advantage when $\Gamma_{\perp} = 0$ [see Section \eqref{sec:noadvantage}].
    
    \item A class of optimised unentangled measurements always yield more information than any measurements on the signal photon state only~[Eq.~(\ref{eq:lowerboundi2s})], i.e., by tracing out the idler mode.
   
    \item Time-frequency entanglement of PDC probes enhances the spectroscopy of TLS and CD systems. This can be used to engineer PDC sources for practical quantum-enhanced spectroscopy. [See Section \eqref{sec6}].
    
\end{enumerate}

Our results advance the understanding of single-molecule spectroscopy using entangled light.
They pave the path towards capturing experimental scenarios that could be the successors of absorption-based techniques~\citep{chong2010ground,celebrano2011single}
or fluorescence-based ones such as single-molecule pump-probe~(SM2P) spectroscopy~\citep{van2005single,van2005single1,moya2022observation} using quantum light.
To the best of our knowledge, all these single-molecule techniques have employed pulses of classical light, with or without a fixed phase relation between them.
Our work also stand apart from other recent ones on ``quantum-enhanced'' spectroscopy that model the ensemble matter system as an infinite chain of beamsplitters~\cite{nair2018quantum,dinani2016quantum,shi2020entanglement}.

In addition to treating both the light and matter quantum mechanically, our work advances the 
quantum-information theoretic understanding of spectroscopy using entangled light. 
It does so by clarifying the spectroscopic potential of entangled measurements across signal and idler modes,
and a simpler class using local operations and classical communication~(LOCC).
These place our work beyond that of Stefanov~\citep{stefanov2017role}.

The paper is structured as follows: 
in Section \ref{sec2}, we recast quantum spectroscopy as a local estimation problem, which will be the basis of our evaluations of fundamental error bounds. 
In Section \ref{sec3}, we describe the fully quantum model of pulsed light-matter interaction in biphoton spectroscopy, as well as define the form of the most general incoming biphoton probes that can be employed in the spectroscopic setup in Figure \ref{fig:biphotonspec_diagram}. 
In Section \ref{sec4}, we calculate explicit expressions for the quantum Fisher information (QFI) of the outgoing state for arbitrary incoming biphoton states and in the light-matter interaction Hamiltonian for the biphoton setup. 
In Section \ref{sec5}, we will establish that unentangled measurements can attain the QFI, and also identify near-optimal measurements that should be more practical to implement. 
In Section \ref{sec6}, we apply the theoretical machinery of the previous sections to the experimentally viable PDC states, for which we
also show that entanglement has a functional usefulness in TLS and CD spectroscopy.
Finally, we conclude in Section \ref{sec7}.

\section{Quantum Light Spectroscopy As An Estimation Problem}\label{sec2}

Spectroscopy uses light --- quantum or classical, to probe matter systems via travelling field states.
Following the light-matter interaction, the probe(s)~(P) carry away information about parameters of the matter~(M) system.
When both the light and the matter are described quantum mechanically, the quantum state of the probe light
just before detection, can be generally represented as 
\begin{equation}\label{eq:rhooutdef}
    \rho^{\mathrm{P}}_{\mathrm{out}}(\theta) = \mathrm{Tr}_{\mathrm{ME}}\left(\,\mathcal{V}^{\theta}_{\mathrm{lm}}\left[\rho^{\mathrm{M}}\otimes\rho^{\mathrm{P}}\otimes\ket{0^{\mathrm{E}}}\bra{0^{\mathrm{E}}}\right]\, \right),
\end{equation}
where $\rho^{\mathrm{P}}$ is the incoming probe state of light, $\rho^{\mathrm{M}}$
is the initial state of the matter system, $\mathcal{V}^{\theta}_{\mathrm{lm}}$ 
captures the quantum interaction between light and matter and labelled by the single, real physical parameter $\theta$ that is to be estimated.
$\ket{0^{\mathrm{E}}} \equiv \bigotimes_l\ket{0^l}$ captures all environmental modes~(E) of the electromagnetic field that may couple to the matter system M, unoccupied at the start of the experiment.
The tracing out of the matter and the environmental modes captures the fact that these parts of the global state are inaccessible to the measurement apparatus. 

The parametric model corresponding to POVM measurement~$\{M_i:~ M_i>0,\, \sum_i M_i = \mathds{I}^{\mathrm{P}}\}$ on the output state $\rho^{\mathrm{P}}_{\mathrm{out}}(\theta)$ is given by the Born rule $\{p(i|\theta) = \mathrm{Tr}\left[\,\rho^{\mathrm{P}}_{\mathrm{out}}(\theta)M_i\,\right]\,|\,\theta\in\mathds{R}\}$. Statistical inference then involves constructing estimators $\hat{\theta} = \theta(X_1,X_2,\dots,X_n)$, where $\{X_i\}$ are random variables corresponding to each of $n$ measured values, independent and identically distributed.
The variance of the estimator $V(\theta|\{M_i\}) = \mathbb{E}_{\theta}[(\hat{\theta}-\theta)]^2$ is the mean square error of the estimator statistic~($\mathbb{E}_{\theta}$ denotes expectation with respect to $X_1,X_2,\dots,X_n~\thicksim~p(i,\theta))$.
It is upper-bounded by the Cr\'{a}mer-Rao bound~(CRB)~\citep{rao1992information,cramer2016mathematical}
\begin{equation}
    V(\theta|\{M_i\}) \geq \frac{1}{n\,\mathcal{C}(\theta|\{M_i\})},
\end{equation}
and $\mathcal{C}(\theta|\{M_i\})$ is the (classical) Fisher information, defined as
\begin{equation}\label{eq:cfidefinition}
    \mathcal{C}(\theta|\{M_i\}) = \mathrm{Var}_{\theta}\left[ \frac{\partial}{\partial \theta} \mathrm{log}~ p(i|\theta) \right] = -\mathbb{E}_{\theta}\left[ \frac{\partial}{\partial \theta}\mathrm{log}~ p(i|\theta)  \right]^2.
\end{equation} 
The model, and therefore the optimal estimators themselves, depend on the
POVM $\{M_i\}$.

A stronger and more fundamental bound on the precision of estimating $\theta$ can be obtained by minimising $V(\theta|\{M_i\})$ over all possible measurements $\{M_i\}$
allowed by the laws of quantum mechanics. This is
referred to as the quantum Cr\'{a}mer-Rao bound~(QCRB)~\citep{braunstein1994statistical,paris2009quantum}
\begin{equation}
    V(\theta|\{M_i\}) \geq \frac{1}{n\,\mathcal{C}(\theta|\{M_i\})} \geq \frac{1}{n\,\mathcal{Q}(\theta;\rho^{\mathrm{P}}_{\mathrm{out}}(\theta))},  
\end{equation}
where $\mathcal{Q}(\theta;\rho^{\mathrm{P}}_{\mathrm{out}}(\theta))$ is the quantum Fisher information~(QFI) corresponding to the parameter $\theta$ in the outgoing state $\rho^{\mathrm{P}}_{\mathrm{out}}(\theta)$,
\begin{equation}
\label{eq:qfi}
    \mathcal{Q}(\theta;\rho^{\mathrm{P}}_{\mathrm{out}}(\theta))  = \mathrm{Tr}\,(\,\rho^{\mathrm{P}}_{\mathrm{out}}(\theta)\,L_{\theta}^2\,)  \geq \mathcal{C} (\theta|\{M_i\}),
\end{equation}
with the  self-adjoint symmetric logarithmic derivative~(SLD) operators defined via
\begin{equation}\label{eq:lyapunovsld}
   L_{\theta}\,\rho^{\mathrm{P}}_{\mathrm{out}}(\theta) + \rho^{\mathrm{P}}_{\mathrm{out}}(\theta)\,L_{\theta} = 2\,\frac{\partial \rho^{\mathrm{P}}_{\mathrm{out}}(\theta)}{\partial \theta} .
\end{equation}

For the estimation of a single parameter $\theta$, the corresponding QCRB can be saturated by the projective measurement corresponding to eigenvectors of the SLD operator $L_{\theta}$~\citep{paris2009quantum}.
For rank-deficient $\rho^{\mathrm{P}}_{\mathrm{out}}(\theta)$,
however, this is only a necessary condition and eigenvectors of the SLD operator are only one of many QCRB-saturating POVMs~\cite{kurdzialek2023measurement}. 
Indeed, for pure states $\rho^{\mathrm{P}}_{\mathrm{out}}(\theta) = \ket{\psi_{\theta}} \bra{\psi_{\theta}}$, the SLD operator has the simpler form
\begin{equation}\label{eq:sldpure}
    L_{\theta} = \ket{\partial_{\theta}\psi_{\theta}}\bra{\psi_{\theta}} + \ket{\psi_{\theta}}\bra{\partial_{\theta}\psi_{\theta}}
\end{equation}
and the QFI is
\begin{equation}\label{eq:qfipure}
    \mathcal{Q}({\theta};\rho^{\mathrm{P}}_{\mathrm{out}}(\theta)) = 4\,\left(\,\langle\partial_{\theta}\psi_{\theta}|\partial_{\theta}\psi_{\theta}\rangle - |\langle\psi_{\theta}|\partial_{\theta}\psi_{\theta}\rangle|^2\,\right).
\end{equation}
For a general mixed state expressed as its spectral decomposition, the QFI is~\citep{liu2014quantum,liu2019quantum}
\begin{align}\label{eq:qfiarbitraryrank}
&\mathcal{Q}\left(\theta;\sum_{n}\,p_n\,\ket{\psi_{n}}\bra{\psi_{n}}\right)  = \sum_{n}\,\frac{(\partial_{\theta} p_n)^2}{p_n}\noindent\\
    &+ \sum_n\,4p_n\,\langle\partial_{\theta}\psi_{n}|\partial_{\theta}\psi_{n}\rangle
    -\sum_{m,n}\frac{8p_m p_n}{p_m + p_n}\,|\langle\partial_{\theta}\psi_{m}|\psi_{n}\rangle|^2 . \nonumber
\end{align}

\section{Light-Matter Interaction for a Biphoton Probe}
\label{sec3}
The dynamics of a quantum matter system interacting with quantised light can be described via the Hamiltonian
\begin{equation}
    H = H^{\mathrm{M}} + H^{\mathrm{F}} + H^{\mathrm{MSE}},
\end{equation}
where $H^{\mathrm{M}}$ corresponds to matter $\mathrm{M}$ dynamics only, $H^{\mathrm{F}}$ is the free field Hamiltonian corresponding to the incoming signal~(S) and idler~(I) modes, as well as the electromagnetic environmental E modes.
Each term in $H^{\mathrm{MSE}}$ is of the dipole-field coupling form $-\vec{d}.\vec{E}$, where $\vec{d}$ is {a} transition dipole moment operator for the matter system, and $\vec{E}$ is the (total)~quantised electric field operator~(see Figure \ref{fig:biphotonspec_diagram}). 
The dipole coupling term is appropriate for single molecules which are small compared to typical optical wavelengths, 
thus allowing the dipole approximation~\citep{mandel1995optical}.

While our arguments can be extended to general molecular systems, we restrict our discussion in this paper to vibrationless $P$-site Hamiltonians~($P=1,2$) of the form
\begin{equation}\label{eq:matterhamiltonian}
    H^{\mathrm{M}} = \sum_{j=1}^P\hbar\omega_j\ket{j}\bra{j} + \sum_{i\neq j }J_{ij}\ket{i}\bra{j},
\end{equation}
where $\ket{j}$ is the excited level corresponding to the $j$-th site~(with frequency $\omega_j$), and $J_{ij}$ is the Coulomb dipole-dipole coupling between sites $i$ and $j$. $P=1$ corresponds to a two-level system~(TLS) Hamiltonian, whereas $P=2$ corresponds to the coupled dimer~(CD) system, composed of two sites that are coupled to each other via the single coupling constant $J$.
The transition dipole operator connecting the ground state of the matter system with the singly-excited manifold~(SEM) is of the form $\vec{d} = \sum_j (\vec{\mu}_{jg}\ket{g}\bra{j} + \mathrm{h.c.})$ where the matrix elements are  $\vec{\mu}_{jg} = \langle g|\vec{\mu}|j \rangle$.

The free field Hamiltonian can be decomposed into a countably infinite number of one-dimensional~(1-D) electromagnetic fields~\citep{blow1990continuum,kolobov1999spatial,loudon2000quantum,ko2022dynamics},
\begin{align}\label{eq:fieldhamiltonian}
    H^{\mathrm{F}} &= \sum_{\vec{\epsilon}}~\int\,d^3\vec{k}\,\hbar c|\vec{k}|\,a_{\vec{\epsilon}}^{\dag}(\bm{k})a_{\vec{\epsilon}}(\bm{k}) \noindent\nonumber \\
    &= \sum_l ~\int_0^{\infty}~d\omega\,\hbar\omega\,a_l^{\dag}(\omega)a_l(\omega),
\end{align}
where $\vec{k}$ and $\vec{\epsilon}$ are respectively the wavevector and polarisation indices for the electromagnetic mode, 
and the subsequent index $l$ labels the resulting 1-D modes.
Although the sum over the index $l$ necessarily runs to infinity and subsumes both the incoming signal/idler modes, as well as the environmental E modes, 
we only need consider, in the description of the light-matter interaction, modes that couple to the matter system M due to $H^{\mathrm{MSE}}$ itself.

These modes can be identified using the slowly-varying envelope approximation~(SVEA) 
in the optical domain, where the frequency bandwidth of the incoming field is assumed to be much smaller than the carrier wave frequency $B\ll\bar{\omega}_{\mathrm{S}}$.
Furthermore, the propagating, incoming beam of the signal S arm can be approximated to be paraxial.
$1$-D quantisation of the solutions of the classical paraxial equation for the signal arm along the direction of propagation yields~\citep{deutsch1991paraxial}
\begin{equation}
    \vec{E}_{\mathrm{S}}(t) = i\vec{\epsilon}_{\mathrm{S}}\,\mathcal{A}_{\mathrm{S}}(\bar{\omega}_{\mathrm{S}})\,\int_{-\infty}^{\infty}d\omega_{\mathrm{S}}~\,a_{\mathrm{S}}(\omega_{\mathrm{S}})e^{-i\omega_{\mathrm{S}} t},  
\end{equation}
where $\mathcal{A}_{\mathrm{S}}(\bar{\omega}_{\mathrm{S}}) = \sqrt{\frac{\bar{\omega}_{\mathrm{S}}}{2\epsilon_0 c A \hbar}}$ is the collective pulse factor~($A$ is the transverse quantisation area of the signal beam), and $\vec{\epsilon_{\mathrm{S}}}$ denotes the unit polarisation vector of the signal beam. 
Note that the emergent Fourier transform of the field operators can be notated as
\begin{equation}\label{eq:whitenoisedef}
    a_{\mathrm{S}}(t) = \frac{1}{\sqrt{2\pi}}\,\int_{-\infty}^{\infty}d\omega_{\mathrm{S}}~\,a_{\mathrm{S}}(\omega_{\mathrm{S}})e^{-i\omega_{\mathrm{S}} t}.
\end{equation}
These are known as white-noise operators and are $\delta$-correlated in time as $[a_{\mathrm{S}}(t),a_{\mathrm{S}}^{\dag}(t')] = \delta(t-t')$, where $\delta(t)$ is the Dirac delta function.

Description of the electromagnetic environment E is simplified by the practical fact that they are inaccessible to experiments, and must be considered in terms of their effects on reduced dynamics only. This effect can then be recovered by using a single bosonic degree of freedom~(as opposed to the infinitude of environmental spatial modes that the matter system M may decay into), labelled by the `$b$' operators:
\begin{equation}
    \vec{E}_{\mathrm{E}}(t) = i\vec{\epsilon}_{\mathrm{E}}\,\mathcal{A}_{\mathrm{E}}(\bar{\omega}_{\mathrm{S}})\,\int_{-\infty}^{\infty}d\omega_{\mathrm{E}}~\,b(\omega_{\mathrm{E}})e^{-i\omega_{\mathrm{E}} t} ,
\end{equation}
where $\mathcal{A}_{\mathrm{E}}(\bar{\omega}_{\mathrm{S}})$ is a collective factor characterising the effect of the continuum electromagnetic environment.
For a more detailed description, see Ref.~\cite[Appendix A]{Albarelli2022}.

In the interaction frame generated by $H_{0} = \sum_j\,\hbar\bar{\omega}_{\mathrm{S}}\ket{j}\bra{j} + H^{\mathrm{F}}$, where the field Hamiltonian can be taken to include only modes that participate in the interaction
\begin{equation}\label{eq:fieldHamiltonian}
    H^{\mathrm{F}} = \int d\omega\, \hbar\omega\,a_{\mathrm{S}}^{\dag}(\omega)a_{\mathrm{S}}(\omega) + \int d\omega\,\hbar\omega\,b^{\dag}(\omega)b(\omega),
\end{equation}
the total Hamiltonian of the biphoton setup in Figure \ref{fig:biphotonspec_diagram} is
\begin{align}\label{eq:TLSbiphotonHamiltonian_shifted}
     H(t) = H_I^{\mathrm{M}} 
     - i\hbar\, (&\sqrt{\Gamma}\,L^{\dag}\otimes a_{\mathrm{S}}(t)\otimes\mathds{1}^{\mathrm{I}}\otimes\mathds{1}^{\mathrm{E}}~~\nonumber\noindent\\
    + &\sqrt{\Gamma_{\perp}}\,L^{\dag}\otimes \mathds{1}^{\mathrm{S}}\otimes\mathds{1}^{\mathrm{I}}\otimes b(t) - \mathrm{h.c.})\,,
\end{align}
where $H_{I}^{\mathrm{M}} = \sum_j \hbar(\omega_j-\bar{\omega}_{\mathrm{S}}) + \sum_{j\neq k} J_{jk}\ket{j}\bra{k}$, and $\mathds{1}^{\mathrm{S}}$~($\mathds{1}^{\mathrm{I}}$) is identity operator on the signal~(idler) Hilbert space. Further, the collective dipole operators, weighted by the strength of interaction, are 
\begin{align}\label{eq:Gammadef}  
    \sqrt{\Gamma} L &= \sqrt{2\pi}\,\mathcal{A}_{\mathrm{S}}(\bar{\omega}_{\mathrm{S}})\sum_j\,\frac{(\vec{\epsilon}_{\mathrm{S}}.\vec{\mu}_{jg})}{\hbar}\,\ket{g}\bra{j}, \nonumber\noindent\\
    \sqrt{\Gamma_{\perp}} L &= \sqrt{2\pi}\,\mathcal{A}_{\mathrm{E}}(\bar{\omega}_{\mathrm{S}})\,\sum_j\,\frac{(\vec{\epsilon}_{\mathrm{E}}.\vec{\mu}_{jg})}{\hbar}\,\ket{g}\bra{j}.
\end{align}

\subsubsection{Incoming Probe State of Entangled Light}
An arbitrary incoming biphoton state in terms of continuous frequency variables $\omega_{\mathrm{S}}$ and $\omega_{\mathrm{I}}$, corresponding to signal and idler modes respectively, can be expressed as
\begin{equation}\label{eq:twophotonfreqstate}
    \ket{\Phi_{\mathrm{biph}}} = \int d\omega_{\mathrm{S}} \int d\omega_{\mathrm{I}}\,\tilde{\Phi}(\omega_{\mathrm{S}},\omega_{\mathrm{I}})\,a^{\dag}_{\mathrm{S}}(\omega_{\mathrm{S}})a_{\mathrm{I}}^{\dag}(\omega_{\mathrm{I}})\ket{0}
\end{equation}
where $\tilde{\Phi}(\omega_{\mathrm{S}},\omega_{\mathrm{I}})$ is the joint spectral amplitude~(JSA) of the entangled state in frequency space, and contains all two-photon correlations. The bivariate JSA function admits a Schmidt decomposition~\citep{lamata2005dealing}
\begin{equation}\label{eq:JSAschmidtdecomp}
    \tilde{\Phi}(\omega_{\mathrm{S}},\omega_{\mathrm{I}}) = \sum_{n}\,r_n\,\tilde{\xi}_n^{\mathrm{S}}(\omega_{\mathrm{S}})\tilde{\xi}_n^{\mathrm{I}}(\omega_{\mathrm{I}}),
\end{equation}
which in turn can be used to express the biphoton state in Eq.~(\ref{eq:twophotonfreqstate}) in terms of discrete orthonormal Schmidt modes
\begin{equation}\label{eq:Schmidttwinstate}
     \ket{\Phi_{\mathrm{biph}}}  = \sum_{n}\,r_n\,a^{\dag}_{n,\mathrm{S}}a^{\dag}_{n,\mathrm{I}}\ket{0} \equiv \sum_{n}\,r_n\,\ket{\xi_n^{\mathrm{S}}}\ket{\xi_n^{\mathrm{I}}},
\end{equation}
where
\begin{equation}
    a^{\dag}_{n,\mathrm{S}} = \int\, d\omega_{\mathrm{S}}\,\tilde{\xi}_n^{\mathrm{S}}(\omega_{\mathrm{S}})a_{\mathrm{S}}^{\dag}(\omega_{\mathrm{S}}),\,a^{\dag}_{n,\mathrm{I}} = \int\, d\omega_{\mathrm{I}}\,\tilde{\xi}_n^{\mathrm{I}}(\omega_{\mathrm{I}})a_{\mathrm{I}}^{\dag}(\omega_{\mathrm{I}}),
\end{equation}
are Schmidt mode creation operators, and $\ket{\xi_n^\mathrm{X}} = a^{\dag}_{n,\mathrm{X}}\ket{0},\,\mathrm{X} = \mathrm{S},\mathrm{I}$ are the corresponding Schmidt basis kets.

We make the additional assumption that the idler spectral amplitude is peaked around a central frequency $\bar{\omega}_{\mathrm{I}}$, so that $a_{\mathrm{I}}^{\dag}(\omega_{\mathrm{I}})\rightarrow a_{\mathrm{I}}^{\dag}(\omega_{\mathrm{I}}-\bar{\omega}_{\mathrm{I}})$ yields idler operators peaked around $\omega_{\mathrm{I}} = 0$. This is justified if the biphoton entangled states are produced in physical processes in which (one or more) pump pulse photons, derived from a spectral amplitude distribution centred around some central frequency $\bar{\omega}_{\mathrm{P}}$, are converted into daughter signal~(centred around $\bar{\omega}_{\mathrm{S}}$) and idler~(centred around $\bar{\omega}_{\mathrm{I}}$) photons. For biphoton states produced in the weak downconversion limit of type-II spontaneous PDC using $\chi^{(2)}$-nonlinear crystals, conservation of energy dictates that $\bar{\omega}_{\mathrm{P}} = \bar{\omega}_{\mathrm{S}}+\bar{\omega}_{\mathrm{I}}$. On the other hand, for biphoton states produced in four-wave mixing~(FWM) schemes~\cite{takesue2004generation,chen2005two,glorieux2012generation} that exploit $\chi^{(3)}$-nonlinearities, conservation of energy dictates~(for degenerate schemes) that $2\bar{\omega}_{\mathrm{P}} = \bar{\omega}_{\mathrm{S}}+\bar{\omega}_{\mathrm{I}}$.
This assumption is distinct from the SVEA, and is motivated by  the details of the physical process used to produce the biphoton entangled state. 
In terms of the re-centred signal and idler field operators, the JSA of the biphoton state in Eq.~(\ref{eq:twophotonfreqstate}) then correspondingly transforms as $\tilde{\Phi}_{\mathrm{biph}}(\omega_{\mathrm{S}},\omega_{\mathrm{I}})\rightarrow \tilde{\Phi}_{\mathrm{biph}}(\omega_{\mathrm{S}}-\bar{\omega}_{\mathrm{S}},\omega_{\mathrm{I}}-\bar{\omega}_{\mathrm{I}})$.

In the rest of this paper, we assume that the JSA fucntion $\tilde{\Phi}_{\mathrm{biph}}(\omega_{\mathrm{S}},\omega_{\mathrm{I}})$, as well as photon operators $a_{\mathrm{S}}^{\dag}(\omega_{\mathrm{S}})$ and  $a_{\mathrm{I}}^{\dag}(\omega_{\mathrm{I}})$ to have been appropriately re-centred so that $\bar{\omega}_{\mathrm{S}}=\bar{\omega}_{\mathrm{I}}=0$.
Positing corresponding white noise operators obtained as Fourier transforms of centred idler operators, $a^{\dag}_{\mathrm{I}}(t_{\mathrm{I}}) = \frac{1}{\sqrt{2\pi}}\int d\omega_{\mathrm{I}}\, e^{-\omega_{\mathrm{I}} t_{\mathrm{I}}}a_{\mathrm{I}}^{\dag}(\omega_{\mathrm{I}})$, we can obtain an equivalent representation of the biphoton entangled state in terms of operators $a_{\mathrm{S}}^{\dag}(t)$ and $a_{\mathrm{I}}^{\dag}(t)$,
\begin{equation}
    \ket{\Phi_{\mathrm{biph}}} = \int dt_{\mathrm{S}}\int dt_{\mathrm{I}}\,\Phi_{\mathrm{biph}}(t_{\mathrm{S}},t_{\mathrm{I}})\,a_{\mathrm{S}}^{\dag}(t_{\mathrm{S}})\,a_{\mathrm{I}}^{\dag}(t_{\mathrm{I}})\ket{0},
\end{equation}
where the time-axis joint temporal amplitude~(JTA) $\Phi_{\mathrm{biph}}(t_{\mathrm{S}},t_{\mathrm{I}})$ is the two-dimensional Fourier transform of the centred frequency-axis JSA $\tilde{\Phi}(\omega_{\mathrm{S}},\omega_{\mathrm{I}})$,
\begin{equation}
    \Phi_{\mathrm{biph}}(t_{\mathrm{S}},t_{\mathrm{I}}) = \frac{1}{2\pi}\,\int d\omega_{\mathrm{S}} \int d\omega_{\mathrm{I}}\,e^{i(\omega_{\mathrm{S}} t_{\mathrm{S}} + \omega_{\mathrm{I}} t_{\mathrm{I}})}\,\tilde{\Phi}_{\mathrm{biph}}(\omega_{\mathrm{S}},\omega_{\mathrm{I}}).
\end{equation}
The JTA $\Phi_{\mathrm{biph}}(t_{\mathrm{S}},t_{\mathrm{I}})$ then admits an analogous Schmidt decomposition
\begin{equation}\label{eq:timeaxisschmidt}
    \Phi_{\mathrm{biph}}(t_{\mathrm{S}},t_{\mathrm{I}}) = \sum_{n}\,r_n\,\xi^{\mathrm{S}}_n(t_{\mathrm{S}})\xi^{\mathrm{I}}_n(t_{\mathrm{I}}),
\end{equation}
where 
\begin{equation}
\xi_n^{\mathrm{X}}(t_{\mathrm{X}}) = \frac{1}{\sqrt{2\pi}}\,\int d\omega_{\mathrm{X}}\,e^{i\omega_{\mathrm{X}} t_{\mathrm{X}}}\,\tilde{\xi}^{\mathrm{X}}_n(\omega_{\mathrm{X}}),\,\mathrm{X=S,I},
\end{equation}
are the time-domain Schmidt basis functions.

\section{QFI for a Biphoton Probe}
\label{sec4}
We start with the molecule in its ground state  $\ket{g}\bra{g}$ at $t=0$. Then,
the joint SI-E state at asymptotically long times $t$, where $\max[\Gamma,\Gamma_{\perp}]t\gg1$, is effected by the 
completely-positive, trace-preserving (CPTP) map
\begin{align}
\label{eq:LMsuperoperator} 
 &\mathcal{W}_g[\rho^{\mathrm{SI}}\otimes\ket{0^{\mathrm{E}}}\bra{0^{\mathrm{E}}}] = ~~~~~~~~~~~~~~~~~~~~~~~~~~~~~~~~~~~~~~~~~~~~~~~~~~~~~~~~~~~  \nonumber\noindent\\ & 
\mathrm{Tr}_{\mathrm{M}}\big[\lim_{\,t\rightarrow\infty} U(t)\,\ket{g}\bra{g}\otimes \rho^{\mathrm{SI}}\otimes\ket{0^{\mathrm{E}}}\bra{0^{\mathrm{E}}}\, U^{\dag}(t)\big],
\end{align} 
where $U(t) = \mathcal{T}\left[\exp(-\frac{i}{\hbar}\int_{-\infty}^{t}dt'H(t')) \right]$ is the unitary propagator corresponding to Eq.~\eqref{eq:TLSbiphotonHamiltonian_shifted},
and $\rho^{\mathrm{SI}}$ is the signal-idler probe state.
At $t \rightarrow \infty$, the molecule decays back to the ground state $\ket{g}.$ 
Thus, for a pure biphoton input
$\rho^{\mathrm{SI}} = \ket{\Phi_{\mathrm{biph}}} \bra{\Phi_{\mathrm{biph}}} $, the transformed state $\mathcal{W}_g~[\ket{\Phi_{\mathrm{biph}}}\bra{\Phi_{\mathrm{biph}}}\otimes\ket{0^{\mathrm{E}}}\bra{0^{\mathrm{E}}}]$ is also pure.
The linearity of the CPTP map $\mathcal{W}_g$ can be employed to obtain the outgoing SI-E state as the piecewise transformation of the Schmidt component wavefunctions on the signal-environment SE subspace, while the idler I components remain unchanged, giving
\begin{align}\label{eq:outgoingtwinphotonstate}
    &\mathcal{W}_g\left[\sum_n r_n \ket{\xi_n^{\mathrm{S}}}\ket{\xi_n^{\mathrm{I}}}\otimes\ket{0^{\mathrm{E}}}\right]  = \sum_n r_n\,\mathcal{W}_g\left[\ket{\xi_n^{\mathrm{S}}}\ket{\xi_n^{\mathrm{I}}}\otimes\ket{0^{\mathrm{E}}}\right] \noindent\nonumber\\
    &=\sum_{n}r_n\left( \ket{\phi_n^{\mathrm{S}}}\ket{\xi_n^{\mathrm{I}}}\otimes\ket{0^{\mathrm{E}}} +  \ket{0^{\mathrm{S}}}\ket{\xi_n^{\mathrm{I}}}\otimes\ket{\pi_n^{\mathrm{E}}}\right),
\end{align}
where the first term captures the signal photon being emitted into its original mode after absorption,
while the second captures the absorbed signal photon being emitted into the environment. 
Here, the $n$-th components
\begin{align}
    \ket{\phi_n^{\mathrm{S}}}&= \ket{\xi_n^{\mathrm{S}}} - \Gamma\,\ket{\varepsilon_n^{\mathrm{S}}},~
    \ket{\pi_n^{\mathrm{E}}} = -\sqrt{\Gamma \Gamma_{\perp}}\,\ket{\varepsilon_n^{\mathrm{E}}}
\end{align}
are obtained using the single-mode solutions~\cite{konyk2016quantum} as used in Ref.~\cite{Albarelli2022}.
We have abbreviated the distortion in the $n$-th Schmidt mode of the signal space as
\begin{equation}
    \ket{\varepsilon_n^{\mathrm{S}}} = \int_{-\infty}^{\infty} dt_1 \left[ \int_{-\infty}^{t_1}d\tau\,f_{\mathrm{M}}(t_1-\tau)\,\,\xi_n^{\mathrm{S}}(\tau)\right]a_{\mathrm{S}}^{\dag}(t_1)\ket{0},
\end{equation}
where 
\begin{equation}\label{eq:characteristicfn_main}
    f_{\mathrm{M}}(t) = \langle g|\,L\,\mathrm{exp}\left[ \left( -iH^{\mathrm{M}}_I  - \frac{\Gamma+\Gamma_{\perp}}{2}\,L^{\dag}L  \right)t\right]\,L^{\dag}\,|g\rangle
\end{equation}
is the characteristic response function of the molecule M. Analogous definitions can be made for $\ket{\varepsilon_n^{\mathrm{E}}}$.

Partial trace over the E subspace yields the outgoing state as the mixture of single photon (in the idler mode) and biphoton states,
\begin{equation}\label{eq:outgoingrhoP_sum}
    \rho_{\mathrm{biph,out}}^{\mathrm{SI}} = (1-N) \ket{\Phi_{\mathrm{biph,out}}} \bra{\Phi_{\mathrm{biph,out}}} + N\ket{0^{\mathrm{S}}}\bra{0^{\mathrm{S}}}\otimes\sigma^{\mathrm{I}}
\end{equation}
where 
\begin{equation}\label{eq:biphoton_outgoing}
     \ket{\Phi_{\mathrm{biph,out}}} = \frac{1}{\sqrt{1-N}}~\sum_n\,r_n\ket{\phi_n^{\mathrm{S}}}\ket{\xi_n^{\mathrm{I}}},
\end{equation}
the normalisation factor being $N  = \Gamma\Gamma_{\perp}\sum_n\,r_n^2\,\langle\varepsilon_n^{\mathrm{S}}|\varepsilon_n^{\mathrm{S}}\rangle$, and 
\begin{equation}\label{eq:sigmaIdef}
    \sigma^{\mathrm{I}} = \frac{\Gamma \Gamma_{\perp}}{N}\,\sum_{mn}\,r_m r_n\,\langle \varepsilon_n^{\mathrm{E}} | \varepsilon_m^{\mathrm{E}} \rangle\,\ket{\xi_m^{\mathrm{I}}}\bra{\xi_n^{\mathrm{I}}}
\end{equation}
is the conditional idler state when the excitation due to the signal is lost to E. 
Note that the transformed signal states are no longer orthonormal
\begin{equation}
    \langle\phi_m^{\mathrm{S}}|\phi^{\mathrm{S}}_n\rangle = \delta_{mn} - \Gamma\Gamma_{\perp}\,\langle\varepsilon_m^{\mathrm{S}}|\varepsilon_n^{\mathrm{S}}\rangle.
\end{equation}
This is, however, recovered in the limit of perfect coupling so that $\lim_{\,\Gamma_{\perp\,}\rightarrow 0}\,\langle\phi_m^{\mathrm{S}}|\phi_n^{\mathrm{S}}\rangle = \delta_{mn}$.
As 
$\sigma^{\mathrm{SI}} = \ket{0^{\mathrm{S}}}\bra{0^{\mathrm{S}}}\otimes \sigma^{\mathrm{I}}$ has no excitation in the S space, 
whereas $ \ket{\Phi_{\mathrm{biph,out}}}$ does, 
the two contributions Eq.~\eqref{eq:biphoton_outgoing} to the mixture live in mutually orthogonal subspaces. Thus,
\begin{equation}
    \langle \Phi_{\mathrm{biph,out}}\,|\, \sigma^{\mathrm{SI}} \,|\, \Phi_{\mathrm{biph,out}} \rangle = 0
\end{equation}
yielding the form of the QFI~\cite{liu2014quantum} of the outgoing state with respect to the Hamiltonian parameter $\theta$ as
\begin{align}\label{eq:rhoPouttwinQFI}
    &\mathcal{Q}(\theta; \rho_{\mathrm{biph,out}}^{\mathrm{SI}}) = \mathcal{C}(N,1-N)\, +~~~~~~~~~~~~~~~ \nonumber\noindent\\
    &~~~~~~~~~~~~  N\mathcal{Q}(\theta;\sigma^{\mathrm{I}}) + (1-N)\,\mathcal{Q}(\theta; \ket{\Phi_{\mathrm{biph,out}}} ),
\end{align}
where $\mathcal{C}(N,1-N) = N_{\theta}/N(1-N)$~(with $N_{\theta} \equiv \partial_{\theta}N$)
is the Fisher information associated with classical mixing of the $\ket{\Phi_{\mathrm{biph,out}}}$ and $\ket{0^{\mathrm{S}}}\bra{0^{\mathrm{S}}}\otimes\sigma^{\mathrm{I}}$ quantum states. 
The conditional idler QFI $\mathcal{Q}(\theta;\sigma^{\mathrm{I}}) $ can be obtained by solving Eq.~\eqref{eq:lyapunovsld} for $\sigma^{\mathrm{I}}$, and using Eq.~\eqref{eq:qfi}.
The biphoton QFI term can be shown to be~(for details, see Appendix \ref{appendix:twinQFIdetails}) 
\begin{align}\label{eq:biphotonQFIexpr}
    \mathcal{Q}(\theta;\ket{\Phi_{\mathrm{biph,out}}}) &= \frac{1}{1-N}\sum_n\,|r_n|^2\langle\partial_{\theta}\phi_n^{\mathrm{S}}|\partial_{\theta}\phi_n^{\mathrm{S}}\rangle \nonumber\noindent\\
    &- \frac{1}{(1-N)^2}\left| \sum_n |r_n|^2\langle\phi_n^{\mathrm{S}}|\partial_{\theta}\phi_n^{\mathrm{S}}\rangle \right|^2.
\end{align}

Eq.~\eqref{eq:rhoPouttwinQFI} is one of our main results, that the spectroscopic information about the molecule M has three distinct contributions - 
from the biphoton state whose signal mode is modified by its interaction with M, 
the one-photon idler state when the absorbed photon is lost to E, and finally the classical mixture of the two.

In the absence of entanglement, that is a product JSA where  $r_n=0\,\forall\,n>1$,  $\sigma^{\mathrm{I}} = \mathds{1}^{\mathrm{I}}$, and $\mathcal{Q}(\theta;\sigma^{\mathrm{I}}) = 0$.  Eq.~\eqref{eq:rhoPouttwinQFI} then reduces to single-photon spectroscopy~\citep{Albarelli2022}.

In the presence of entanglement, the contributions of the three terms in Eq.~(\ref{eq:rhoPouttwinQFI}) 
depend on the relative magnitudes of M-S and M-E coupling.
These corresponds to different flavours of experimental setups --- in free space scenarios where $\Gamma_{\perp}\gg\Gamma$, the first two terms dominate in Eq.~(\ref{eq:rhoPouttwinQFI}) as most of the signal excitation are lost to the E space. In contrast, for geometries engineered such that $\Gamma_{\perp}\ll\Gamma$, the biphoton QFI will be the major contributor as few excitations are lost to E. This is summarised in Table \ref{table1}.

\begin{table}
    \centering
 \begin{tabular}{|c  |c |c |c |}
    \hline
     $\Gamma_{\perp}/\Gamma$ & $\mathcal{C}(N,1-N)$ & $ N\mathcal{Q}(\sigma^{\mathrm{I}})$ & $ (1-N)\mathcal{Q}(\ket{\Phi_{\mathrm{biph,out}}})$ \\
         \hline
         $\ll 1$ & $\cross$ & $\cross$ & \checkmark \\ \hline
         $O(1)$ & \checkmark &\checkmark &\checkmark \\ \hline 
         $\gg1$ & \checkmark & \checkmark & $\cross$ \\
         \hline
    \end{tabular}
    \caption{Dominant contributions to spectroscopic information QFI $\mathcal{Q}(\theta;\rho_{\mathrm{biph,out}}^{\mathrm{SI}})$ in the asymptotic time limit $t\rightarrow \infty$ for different regimes of relative M-S and M-E couplings. A fuller dependance on the relative magnitudes of M-S coupling $\Gamma$ and M-E coupling $\Gamma_{\perp}$ is provided in Appendix \ref{app:relativemag_orders}.}
    \label{table1}
\end{table}

\subsection{Attaining the QFI in Eq.~\eqref{eq:rhoPouttwinQFI}}

\begin{figure}[h!]
        \includegraphics[width=0.50\textwidth]{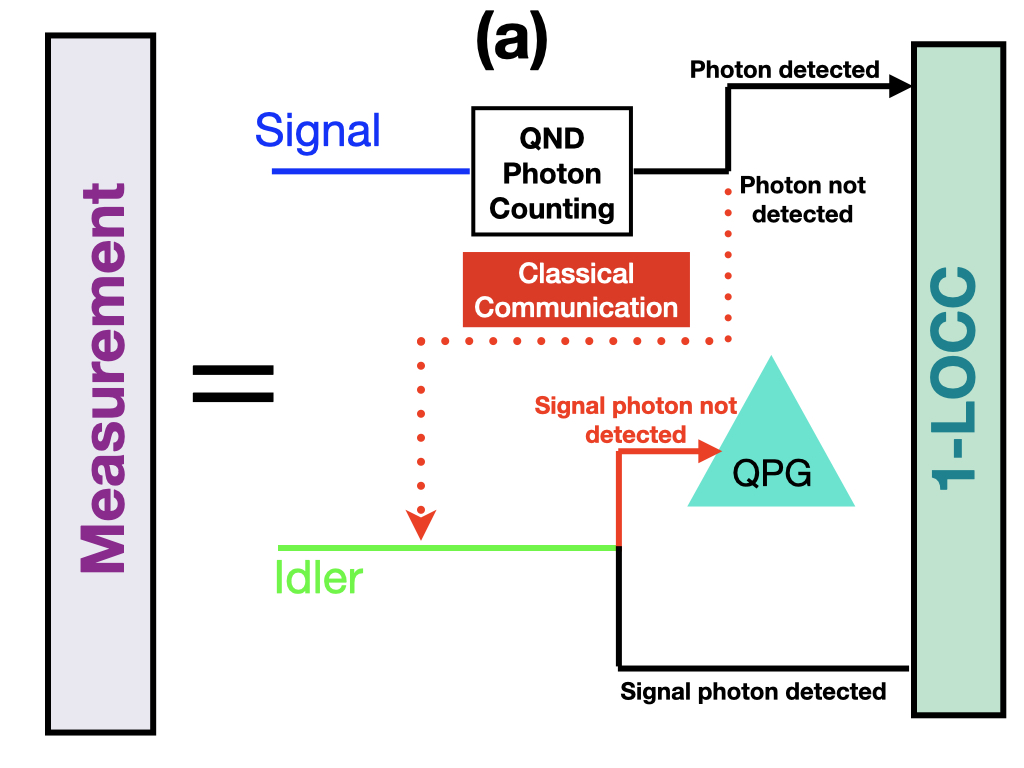}
        \includegraphics[width=0.50\textwidth]{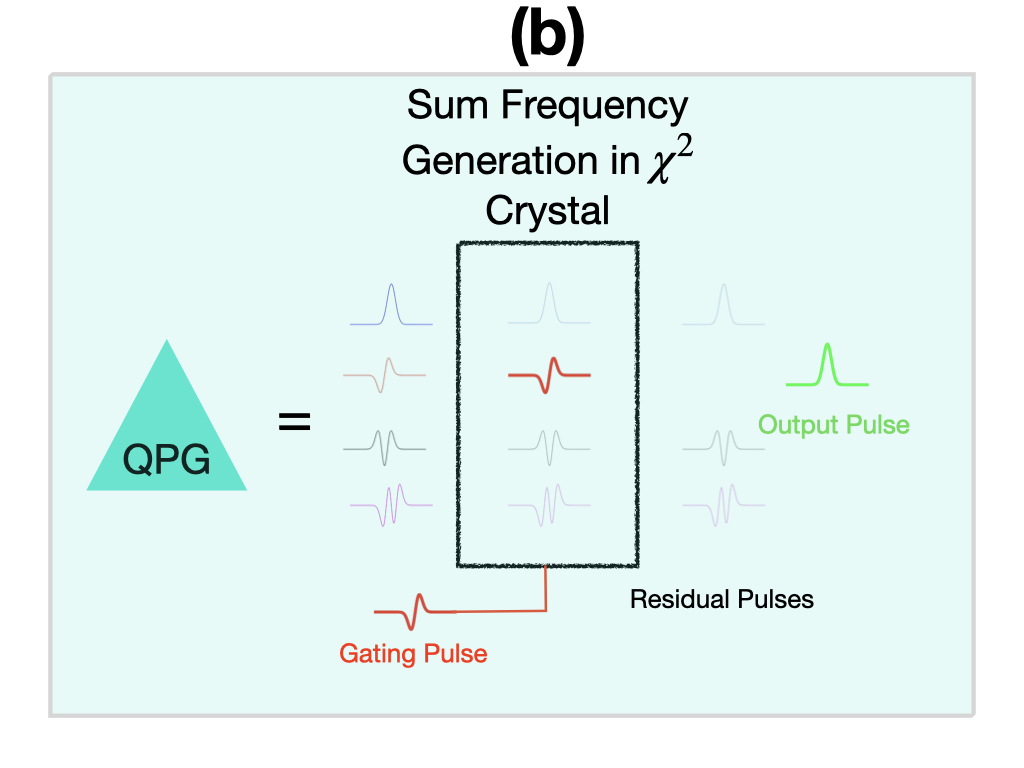}
        \caption{Detection scheme that fetches estimator precision corresponding to full QFI in Eq.~(\ref{eq:rhoPouttwinQFI}). (a) Schematic of cascade of mutually admissibe measurements. As a first step, QND photon counting is performed on the signal substation. Based on the outcome~(classically communicated to the idler substation), the idler photon is either detected in mode-resolved projective measurement, or sent to the 1-LOCC box, which is discussed in detail in Section \ref{sec5}~(schematic illustration in Figure \ref{fig:idlertosignalscheme}). (b) QPG implementation of mode-resolved photon counting.}
         \label{fig:biphotonspec_measurement}
\end{figure}

The three terms in Eq.~(\ref{eq:rhoPouttwinQFI}) may be successively saturated in a cascade of mutually commuting 
 measurements on orthogonal subspaces in the SI space as illustrated schematically in Figure \ref{fig:biphotonspec_measurement} (a). 

The first term $\mathcal{C}(N,1-N)$ can be attained using quantum non-demolition~(QND) photon counting measurement effected by the set of signal projectors $\{\Pi^{\mathrm{S}}_{0},\Pi^{\mathrm{S}}_{1}\}$, where $\Pi_0^{\mathrm{S}} = \ket{0^{\mathrm{S}}}\bra{0^{\mathrm{S}}}$, and $\Pi_1^{\mathrm{S}} = \int d\omega\,a^{\dag}_{\mathrm{S}}(\omega)\ket{0^{\mathrm{S}}}\bra{0^{\mathrm{S}}}a_{\mathrm{S}}(\omega)$. 
A QND measurement is advisable as destructive photon counting at this stage can only fetch as much as the information as the classical $\mathcal{C}(N,1-N)$ term, the collapsed photon states carrying no more quantum information.
Practically, such QND photon counting has been achieved using either cross-Kerr mapping of photons numbers onto phase shifts of a secondary optical probe~\citep{imoto1985quantum,munro2005high,gerry2008quantum}, or by strongly coupling the photonic state to atoms in cavity electrodynamics that maps photon numbers to atomic phases, which can then be detected using interferemetric techniques~\citep{nogues1999seeing,guerlin2007progressive,malz2020nondestructive}. 
The photon counting measurements, non-demolition or not, are effectively absorption measurements, and the magnitude of $N$ can be estimated from these measurement outcomes

The second term $ N\mathcal{Q}(\sigma^{\mathrm{I}})$ can, in general, be attained by measuring (idler) projectors corresponding to the eigenvectors of the SLD for $\sigma^{\mathrm{I}}$.
A practical setup that can implement approximately optimal single-photon projectors as mode-resolved photon counting may be achieved using quantum pulse gating~(QPG) techniques~\citep{eckstein2011quantum,donohue2018quantum,Reddy2018,de2021effects,ansari2021achieving,garikapati2022programmable} for ultrafast pulses. This involves an incoherent train of pulses coupling with a sufficiently shaped gating pulse in a sum-frequency interaction inside a nonlinear crystal. The shape of the gating pulse determines the mode the incoming pulse is effectively projected on to, presenting at the output as a higher frequency signal than the incoming pulse~(see Figure \ref{fig:biphotonspec_measurement} (b)).

The third term in Eq.~(\ref{eq:rhoPouttwinQFI}) may, in general require measurements entangled across the signal and idler on the 
pure state $\ket{\Phi_{\mathrm{biph,out}}}$ 
to be attained. In the next section, we show that an unentangled measurement suffices.

\section{1-LOCC Detection Schemes}\label{sec5}

Measurement protocols for multipartite quantum systems can be divided into three classes~\cite{horodecki2009quantum,zhou2020saturating} --- 
(a) uncorrelated local measurements~(LM) with no classical communication between individual substations, 
(b) correlated local operations and classical communication~(LOCC) where results of local measurement operations may be conveyed back and forth between the various substations using classical bits, and 
(c) global measurements~(GM) which are the most general class of measurements that can be performed on multipartite quantum systems.
In terms of their ability to extract quantum information and resource intensiveness of practical implementation~\cite{friis2017flexible}, $\mathrm{LM}\subseteq\mathrm{LOCC}\subseteq\mathrm{GM}$.
In our spectroscopic setup, entangled measurements across the signal and the idler would be in GM, but not LM or LOCC.

We show that such entangled measurements are \emph{not} necessary to attain the third term in  Eq.~(\ref{eq:rhoPouttwinQFI}).
In fact, we show that a one-way idler-to-signal LOCC measurement scheme --- that we will henceforth refer to as ``1-LOCC'', always attains the third term in  Eq.~(\ref{eq:rhoPouttwinQFI}).
For the biphoton setup, the most general 1-LOCC is schematically illustrated in Figure \ref{fig:idlertosignalscheme}.
In such a detection scheme, the results of local measurement on the idler substation are classically communicated on to the signal substation, where then measurement operators for local detection are chosen accordingly\footnote{Signal-to-idler 1-LOCC detection schemes can be constructed in similar fashion, but with some important differences. See Appendix \ref{appendix:S2Iprepareandmeasure}. Signal-to-idler schemes may be more challenging to implement practically as the preparation step must necessarily follow the light-matter interaction, outcomes of which must subsequently be communicated to the idler substation.}. 
This is another of our main results.

Experimentally, LOCC operations on continuous-variable~(CV) time-frequency entangled states have been successfully implemented 
as part of CV teleportation of light states~\citep{lee2011teleportation,huo2018deterministic}.
Our 1-LOCC detection scheme is thus a potentially attractive class of measurements for biphoton spectroscopy, and must be contrasted against interferometric quantum spectroscopies that propose global measurements by bringing together the two photons in linear~\citep{chen2022quantum,asban2022nonlinear} or non-linear interferometers~\citep{panahiyan2023nonlinear} at the detection stage.

Operationally,  the idler-to-signal 1-LOCC detection scheme can also be viewed as a heralding scheme where a measurement is 
 performed on the idler photon, and outcomes communicated to the signal station independently of the light-matter interaction which has support on the MSE subspace.

We construct a spectroscopically useful subclass of 1-LOCC detection schemes by optimising the CFI over POVMs 
on the signal mode only.
We call this the ``measurement-optimal'' 1-LOCC detection scheme, and show that it
(i) always includes a measurement whose CFI equals $\mathcal{Q}(\theta; \ket{\Phi_{\mathrm{biph,out}}(\theta)})$ in a specific choice of the 1-LOCC scheme, and 
(ii) the associated CFI for \emph{all} members exceeds that of any single-photon measurement on the reduced signal state only. 

For perfect coupling geometries~($\Gamma_{\perp} = 0$), (i) implies that entanglement in the incoming biphoton state is not,
in-principle, a resource.
This is because the QFI of an entangled biphoton state may be attained with a suitably heralded Fock state and uncorrelated LM measurement.

\subsection{1-LOCC Fisher Information in Biphoton Setup}\label{sec:idlertosignal}

\begin{figure}[t!]
        \includegraphics[width=0.5\textwidth]{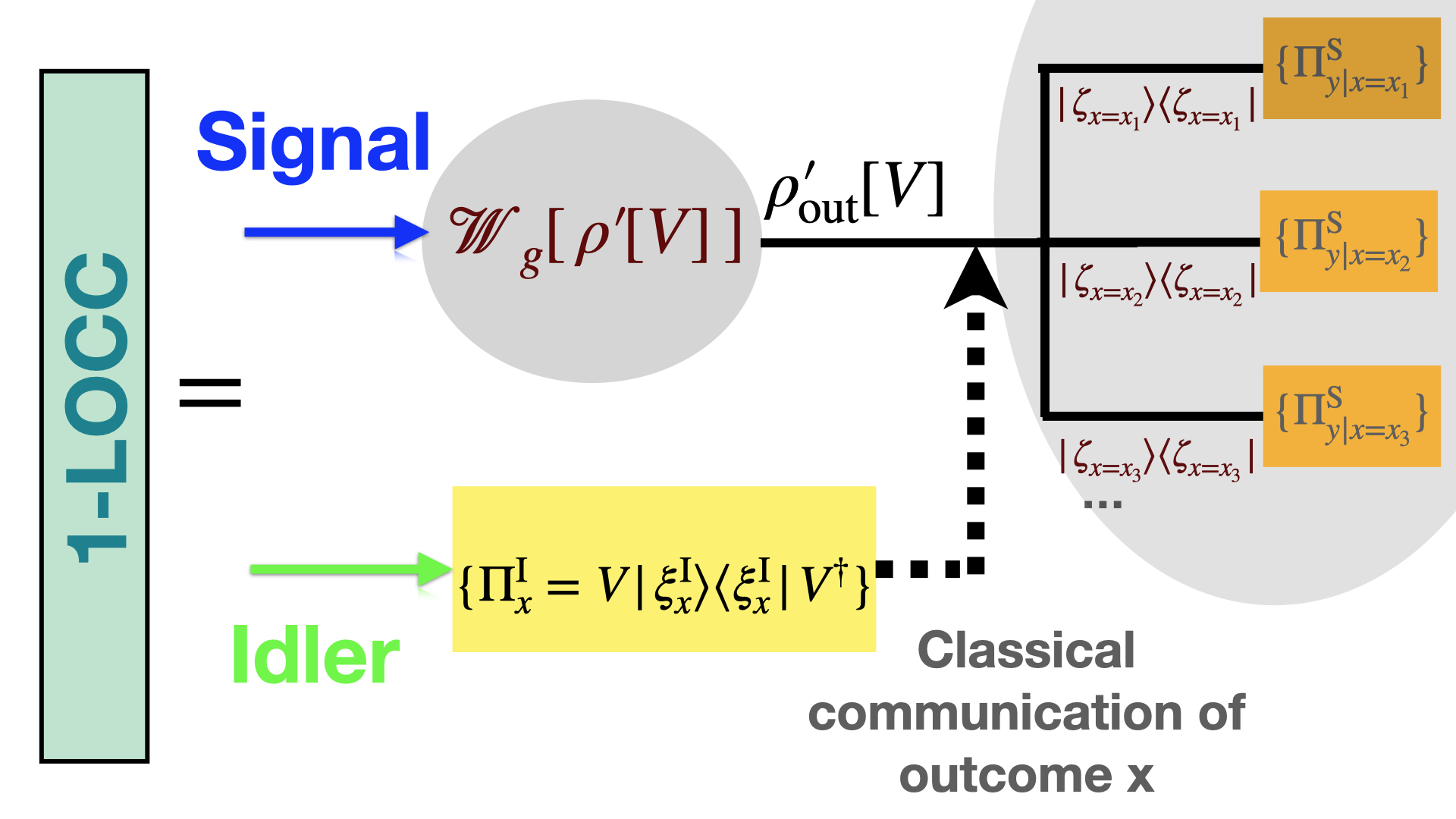}
        \caption[Idler-to-signal prepare-and-measure local operations and classical communication~(LOCC) scheme.]{Idler-to-signal 1-LOCC scheme. The biphoton Kraus operators for the overall LOCC scheme are given as $\{\Pi_{y,x} = \Pi_{y|x}^{\mathrm{S}}\otimes\Pi_x^{\mathrm{I}}$\}.
        }
        \label{fig:idlertosignalscheme}
\end{figure}

The most general idler-to-signal 1-LOCC detection scheme proceeds in the following three steps:
\begin{enumerate}
    \item Projectively measure the idler photon in the basis $\{V\ket{\xi_x^{\mathrm{I}}}\bra{\xi_x^{\mathrm{I}}}V^{\dag}\}$, where $V$ is an 
    arbitrary unitary operator on the idler Hilbert space. This transforms the incoming entangled biphoton probe state via Kraus operators 
    $\Pi_x^{\mathrm{I}} = \mathds{1}^{\mathrm{S}}\otimes V\ket{\xi_x^{\mathrm{I}}}\bra{\xi_x^{\mathrm{I}}} V^{\dag}$ to
    \begin{align}
        \rho'[V] &= \sum_x\, \left( \Pi_x^{\mathrm{I}} \right)^{\dag} \,\ket{\Phi_\mathrm{biph}}\bra{\Phi_{\mathrm{biph}}}\,\Pi_x^{\mathrm{I}}\noindent\nonumber\\
        &= \sum_x\,\ket{\psi_x}\bra{\psi_x}\otimes V\ket{\xi_x^{\mathrm{I}}}\bra{\xi_x^{\mathrm{I}}}V^{\dag},
    \end{align}
    where $\ket{\psi_x} = \sum_n\,r_n V^{*}_{nx}\ket{\xi_n^{\mathrm{S}}}$, and $V_{mn} = \langle\xi_m^{\mathrm{I}}|V|\xi_n^{\mathrm{I}}\rangle$ are elements of unitary matrix $V$ in the idler Schmidt basis.
    
    \item Communicate (classically) the outcome of projective measurement $\{\Pi_x^{\mathrm{I}}\}$  to the signal substation.
     The M-S-E interaction, given by the CPTP map $\mathcal{W}_g$ in Eq.~(\ref{eq:LMsuperoperator}), transforms the signal Schmidt basis $\{\ket{\xi_n^{\mathrm{S}}}\}$ onto the non-orthonorgonal set\footnote{The preparation step~(characterised by Kraus operators $\{\Pi_x^{\mathrm{I}}\}$) quantum operation and the M-S-E interaction~(characterised by superoperator $\mathcal{W}_g$) commute with each other, and can be applied in any order. If, in the final step, the signal subensembles $\{\ket{\zeta_x}\}$ are all projected onto a common measurement basis, the LOCC scheme reduces to an LM scheme with independent measurements performed at signal and idler substations.}  $\{\ket{\phi_n^{\mathrm{S}}}\}$, and renormalises the outgoing state as in Eq.~(\ref{eq:biphoton_outgoing}).
     The resulting SI state is given by
    \begin{equation}
        \rho'_{\mathrm{out}}[V] =  \sum_x\, \ket{\zeta_x}\bra{\zeta_x}\otimes V\ket{\xi_x^{\mathrm{I}}}\bra{\xi_x^{\mathrm{I}}}V^{\dag},
    \end{equation}
    where $\ket{\zeta_x} = \frac{1}{\sqrt{1-N}}\sum_n\,r_n V^{*}_{nx}\ket{\phi_n^{\mathrm{S}}}$, is the unnormalised conditional signal state for the outcome $x$.
    
    \item Measure the signal photon using operators $\{\Pi^{\mathrm{S}}_{y|x=x_m}\}$ depending on the communication $x=x_m$ received from the idler.
\end{enumerate}
These stages of the 1-LOCC scheme are illustrated in Figure \ref{fig:idlertosignalscheme}.

The maximum CFI attainable using a 1-LOCC scheme is formally given by
\begin{align}\label{eq:optimise1LOCCCFI}
&\max_{\Pi_{y|x}^{\mathrm{S}},\Pi_x^{\mathrm{I}}}\,\mathcal{C}(\theta | \{ \Pi_{y,x} = \Pi_{y|x}^{\mathrm{S}}\otimes\Pi_x^{\mathrm{I}}\}) \nonumber\noindent\\
~~~= &\max_{\Pi_x^{\mathrm{I}}}\,\,\left(\max_{\Pi_{y|x}^{\mathrm{S}}}\,\,\mathcal{C}(\theta | \{ \Pi_{y,x} = \Pi_{y|x}^{\mathrm{S}}\otimes\Pi_x^{\mathrm{I}}\})\right),
\end{align}
$\mathrm{s.t.}\,\,\sum_x\,\Pi_x^{\mathrm{I}} = \mathds{1}^{\mathrm{I}},\,\,\sum_{y}\Pi_{y|x}^{\mathrm{S}} = \mathds{1}^{\mathrm{S}}\,\forall\,\,x.$
The maximal 1-LOCC CFI is upper bounded as
\begin{equation}\label{eq:cfiloccqfi_inequality}
\max_{\Pi_{y|x}^{\mathrm{S}},\Pi_x^{\mathrm{I}}}\,\mathcal{C}(\theta | \{ \Pi_{y,x} = \Pi_{y|x}^{\mathrm{S}}\otimes\Pi_x^{\mathrm{I}}\}) \leq \mathcal{Q}(\theta;\ket{\Phi_{\mathrm{biph,out}}}).
\end{equation}
since 1-$\mathrm{LOCC} \subseteq \mathrm{LOCC} \subseteq \mathrm{GM}$.

The maximisation of the CFI functional may now proceed in two steps, following the RHS of Eq.~(\ref{eq:optimise1LOCCCFI}): 
first, for a given unitary $V$, the CFI is maximised over all allowed $\{\Pi^{\mathrm{S}}_{y|x=x_m}\}$, and
second, the resulting quantity is maximised over all $\Pi_{x}^{\mathrm{I}} = \mathds{1}^{\mathrm{S}}\otimes V\ket{\xi_x^{\mathrm{I}}}\bra{\xi_x^{\mathrm{I}}} V^{\dag}$, which amounts to a maximisation over all unitary operations $V$. All 1-LOCC for which maximisation over signal POVM $\{\Pi^{\mathrm{S}}_{y|x=x_m}\}$ has been performed will be termed ``measurement-optimal", and the corresponding CFI quantity, now just a function of the preparation unitary $V$, is given as 
\begin{equation}\label{eq:cmax_defmaintext}
    \mathcal{C}_{\mathrm{max}}(\theta;V) = \max_{\Pi_{y|x}^{\mathrm{S}}}\,\,\mathcal{C}(\theta | \{ \Pi_{y,x} = \Pi_{y|x}^{\mathrm{S}}\otimes\Pi_x^{\mathrm{I}}\}).
\end{equation}

Constructing the orthogonal complement of $\ket{\Phi_{\mathrm{biph,out}}}$
in the two-dimensional $\mathrm{Span}[\ket{\Phi_{\mathrm{biph,out}}},\ket{\partial_{\theta}\Phi_{\mathrm{biph,out}}}]$ as~\cite{kurdzialek2023measurement}
\begin{equation}\label{eq:perpbiph_eqn_main}
    \ket{\Phi_{\mathrm{biph,out}}^{\perp}} \equiv \left(1-\ket{\Phi_{\mathrm{biph,out}}}\bra{\Phi_{\mathrm{biph,out}}}\right) \,\,\ket{\partial_{\theta}\Phi_{\mathrm{biph,out}}},
\end{equation}
the following result holds:
\\ \\
\begin{theorem}\label{theorem1}
For a preparation step unitary $V_0$ that satisfies 
\begin{equation}\label{eq:optconditioni2s_main}
\bra{\xi_m^{\mathrm{I}}}V_0^{\dag}\,\mathrm{Tr}_{\mathrm{S}}\,\ket{\Phi_{\mathrm{biph,out}}}\bra{\Phi_{\mathrm{biph,out}}^{\perp}}\,V_0\ket{\xi_m^{\mathrm{I}}} = 0\,\,\, \forall m,
\end{equation}
$\mathcal{C}_{\mathrm{max}}(\theta;V_0) = \mathcal{Q}(\theta;\ket{\Phi_{\mathrm{biph,out}}})$.
\end{theorem}
A proof appears in Appendix \ref{app:1loccopt}.

A constructive proof for the existence of a $V_0$ satisfying Eq.~(\ref{eq:optconditioni2s_main}) has been established for finite dimensions~\citep{fillmore1969similarity,zhou2020saturating}.
It can be extended to trace-class~(and hence bounded and compact) operators on CV spaces,
including $\mathrm{Tr}_{\mathrm{S}}\,\ket{\Phi_{\mathrm{biph,out}}}\bra{\Phi_{\mathrm{biph,out}}^{\perp}}$ in Eq.~(\ref{eq:optconditioni2s_main}).

Following the upper bound in Eq.~(\ref{eq:cfiloccqfi_inequality}), the measurement-optimal 1-LOCC characterised by $V_0$ must correspond to the maximal CFI attainable.
Thus, the biphoton component of the QFI in Eq.~(\ref{eq:rhoPouttwinQFI}) may be attained in a measurement-optimal 1-LOCC scheme with $V_{\mathrm{opt}} = V_0$ --- that is, an unentangled measurement. The unitary $V_0$, in general,  depends on the the outgoing signal modes $\{\phi_n^{\mathrm{S}}\}$, which themselves change with the nature and strength of M-S  and M-E interactions.

\subsection{No advantage from entangled input probe}
\label{sec:noadvantage}

If $ \Gamma_{\perp}=0$, only the third term in Eq.~(\ref{eq:rhoPouttwinQFI})  survives. Then there always exists a single-photon Fock state
 \begin{equation}
    |\zeta^{\mathrm{opt},'}_{x_m}\rangle = \frac{1}{\sqrt{\langle\zeta^{\mathrm{opt}}_{x_m}|\zeta_{x_m}^{\mathrm{opt}}\rangle}}|\zeta_{x_m}^{\mathrm{opt}}\rangle,
\end{equation}
where
\begin{equation}
\ket{\zeta^{\mathrm{opt}}_{x_m}} = \frac{1}{\sqrt{1-N}}\,\sum_n\,r_n (V_{\mathrm{opt}})^{*}_{nx_m}\,\ket{\phi_n^{\mathrm{S}}}
\end{equation}
for some measurement outcome $x=x_m,$
which has at least as much QFI as the entangled input in Eq.~(\ref{eq:twophotonfreqstate}).
In principle, time-frequency entanglement of the input thus provides no advantage in this scenario. 

This follows from Theorem \ref{theorem1}, whereby the biphoton QFI can be written as the convex combination of signal-only QFIs as~(see Eqs.~(\ref{eq:prepnmeasureLOCC1}) and (\ref{eq:cfiLOCC_zero}) in Appendix \ref{app:1loccopt})
\begin{equation}\label{eq:convexsumoptQFI}
    \mathcal{Q}(\theta;\ket{\Phi_{\mathrm{biph,out}}}) = \sum_x\,\langle\zeta_x^{\mathrm{opt}}|\zeta_x^{\mathrm{opt}}\rangle\,\mathcal{Q}\left(\theta;\ket{\zeta_x^{\mathrm{opt},'}}\right)
\end{equation}
where $\ket{\zeta_x^{\mathrm{opt},'}} = \frac{1}{\sqrt{\langle\zeta_x^{\mathrm{opt}}|\zeta_x^{\mathrm{opt}}\rangle}}\ket{\zeta_x^{\mathrm{opt}}}$ are normalised conditional states. Equivalently, the biphoton QFI is always equal to the QFI of the following separable state
\begin{align}
&\mathcal{Q}(\theta;\ket{\Phi_{\mathrm{biph,out}}}) ~~~~~ \nonumber\noindent\\
&~~~=\mathcal{Q} \left( \theta;\sum_x\,\ket{\zeta^{\mathrm{opt}}_x}\bra{\zeta_x^{\mathrm{opt}}}\otimes\,V_{\mathrm{opt}}\ket{\xi_x^{\mathrm{I}}}\bra{\xi_x^{\mathrm{I}}}V^{\dag}_{\mathrm{opt}} \right).
\end{align}
This shows that it is always possible to engineer the incoming state of light so as to prepare deterministically  the product state component in the convex sum in Eq.~(\ref{eq:convexsumoptQFI})  with the maximal QFI $\max_x\,\mathcal{Q}(\theta;|\zeta^{\mathrm{opt,'}}_x\rangle)$~(which we will label by index $x=x_m$) so that the QFI of the separable state then yields at least as much precision as the entangled biphoton state.

Operationally, one need only start then with the (pre-conditioned)~single-photon signal state
\begin{equation}\label{eq:optimalproductstate}
    \ket{\psi_{x_m}^{\mathrm{opt}}} = \frac{1}{\sqrt{1-N}}\,\sum_n\,(V_{\mathrm{opt}})^{*}_{nx_m}\,\ket{\xi_n^{\mathrm{S}}},
\end{equation}
 which would yield the outgoing state $|\zeta^{\mathrm{opt,'}}_{x_m}\rangle$ whose QFI is always greater than, or equal to, the biphoton QFI $\mathcal{Q}(\theta;\ket{\Phi_{\mathrm{biph,out}}})$.

This subsection extends a similar conclusion in Ref.~\cite{Albarelli2022} for the restricted case of resonant 
$\Gamma$-estimation in TLS for $\Gamma_{\perp} = 0$ to an arbitrary Hamiltonian parameter $\theta$.

Our conclusion that an entangled input is not, in-principle, a resource can also be extended to scenarios with $\Gamma_{\perp} \neq 0$ when the biphoton state $\ket{\Phi_{\mathrm{biph,out}}}$ is post-selected,
because, in that case the first two terms in Eq.~(\ref{eq:rhoPouttwinQFI}) drop out.
The question of whether an entangled input is advantageous when all the three terms in Eq.~(\ref{eq:rhoPouttwinQFI}) contribute, however, remains open.

 \subsection{Lower Bound on Measurement-Optimal Protocols}

For any measurement-optimal 1-LOCC CFI,
\begin{align}\label{eq:i2sCFIlowerbound}
    \mathcal{C}_{\mathrm{max}}(\theta;V) &= \mathcal{Q}(\theta;\rho_{\mathrm{out}}'[V]) \nonumber\noindent\\
    &\geq  \mathcal{Q}(\theta;\sum_x\,\ket{\zeta_x}\bra{\zeta_x}) \nonumber \\
    &= \mathcal{Q}\bigg(\theta;\frac{1}{1-N}\sum_m\,|r_m|^2\ket{\phi_m^{\mathrm{S}}}\bra{\phi_m^{\mathrm{S}}}\bigg) \noindent\nonumber\\
    &= \mathcal{Q}(\,\theta;\mathrm{Tr}_{\mathrm{I}}\,\ket{\Phi_{\mathrm{biph,out}}}\,),
\end{align}
where the first line is true because maximisation of the CFI over $\{\Pi^{\mathrm{S}}_{y|x=x_m}\}$ in Eq.~(\ref{eq:cmax_defmaintext}) is precisely the maximisation that yields the Cram\'{e}r-Rao bound~\citep{braunstein1994statistical,paris2009quantum} for the conditional state $\rho_{\mathrm{out}}'[V]$~(see Eq.~(\ref{eq:measurementoptCFI}) Appendix \ref{app:1loccopt}). The second line is a consequence of the extended convexity of the QFI~\citep{alipour2015extended}.
The inequality is saturated iff $\langle\phi_m^{\mathrm{S}}|\partial_{\theta}\phi_n^{\mathrm{S}}\rangle = 0\,\,\forall\,\,m,n$. For a Schmidt basis $\{\ket{\phi_n^{\mathrm{S}}}\}$ that is complete on the signal Hilbert space, this is never true, and we get the stronger inequality
\begin{equation}\label{eq:lowerboundi2s}
    \mathcal{C}_{\mathrm{max}}(\theta;V) > \mathcal{Q}(\theta;\mathrm{Tr}_{\mathrm{I}}\,\ket{\Phi_{\mathrm{biph,out}}}).
\end{equation}
This shows that \textit{all} measurement-optimal 1-LOCC detection schemes yield  higher CFI than the QFI in signal photon obtained by tracing out the idler. Consequently, all measurement-optimal 1-LOCC detection schemes have a guaranteed metrological advantage over signal photon-only strategies.
This leads to the hierarchy
\begin{align}\label{eq:i2shierarchy}
    &\mathcal{Q}(\theta;\ket{\Phi_{\mathrm{biph,out}}}) = \mathcal{C}_{\mathrm{max}}(\theta;V_{\mathrm{opt}}) \geq \mathcal{C}_{\mathrm{max}}(\theta;V)\nonumber\noindent\\
    &~~~~~~~~~~~~~~~~~~~~~~~~> \mathcal{Q}(\theta;\mathrm{Tr}_{\mathrm{I}}\,\ket{\Phi_{\mathrm{biph,out}}}).
\end{align}

\subsection{Parameter-Independent Unitary $V = \mathds{1}^{\mathrm{I}}$}

Unlike $V = V_{\mathrm{opt}}$, $V = \mathds{1}^{\mathrm{I}}$ is independent of $\theta$ and $\{\phi_n^{\mathrm{S}}\}$, and presents a simpler
experimental scenario.
The corresponding CFI is 
\begin{align}\label{eq:maximalCFILOCCi2s_main}
    &\mathcal{C}_{\mathrm{max}}(\theta;V =\mathds{1}^{\mathrm{I}}) = \noindent \\
    &\frac{4}{1-N}\sum_n|r_n|^2\langle\partial_{\theta}\phi_n^{\mathrm{S}}|\phi_n^{\mathrm{S}}\rangle -\frac{4}{(1-N)^2}\sum_n|r_n|^2|\langle\phi_n^{\mathrm{S}}|\phi_n^{\mathrm{S}}\rangle|^2. \nonumber
\end{align}
For the special case of resonant $\Gamma$-estimation in a TLS with transition frequency $\omega_0$, it has been shown that~\citep[Section V A]{Albarelli2022})
\begin{equation}
    \mathcal{Q}(\Gamma;\ket{\Phi_{\mathrm{biph,out}}})\vert_{\Delta=0} = \mathcal{C}_{\mathrm{max}}(\Gamma;V=\mathds{1}^{\mathrm{I}});~\Delta = \omega_0-\bar{\omega}_{\mathrm{S}}.
\end{equation}
This implies that a 1-LOCC detection scheme with $V = \mathds{1}^{\mathrm{I}}$ attains the QFI for $\Gamma$-estimation at $\Delta=0$ in a TLS. 
While this conclusion no longer holds for the spectroscopy of more general systems and parameters, the measurement-optimal $V = \mathds{1}^{\mathrm{I}}$ 1-LOCC detection scheme continues to be an improvement over
simply tracing out the idler in the outgoing wavefunction, as per Eq.~\eqref{eq:i2shierarchy}.
We study its efficacy in spectroscopy with a PDC input probe numerically in Section \ref{sec6}.

\section{Spectroscopy Using PDC Light}
\label{sec6}

\begin{figure}[h!]
        \includegraphics[width=0.45\textwidth]{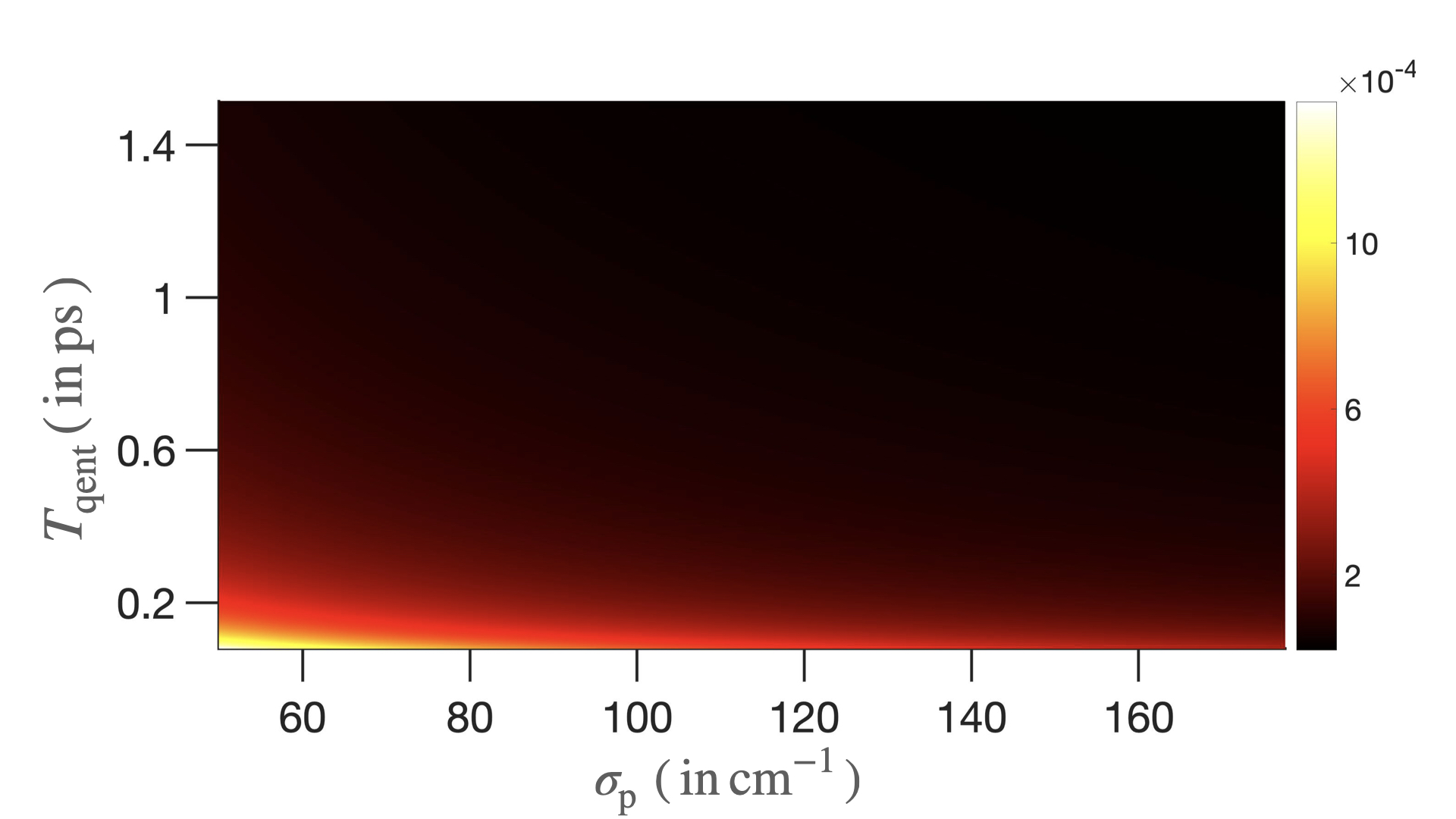}
        \label{entropyPDC_plot}
        \caption{  
        Plot of entanglement entropy $S$ in Eq.~\eqref{eq:entent} for PDC source characteristics - pump bandwidth ($\sigma_p$) and entanglement time 
        ($T_{\mathrm{qent}}$). The time-frequency entanglement is higher for short $T_{\mathrm{qent}}$, and small values of $\sigma_{\mathrm{p}}$. See \citep[Appendix H]{Albarelli2022} for definitions.
        \label{fig:entanglemententropyheatmap}}
\end{figure}

We now study pulsed quantum light spectroscopy using the experimentally ubiquitous entangled probes 
of weakly-downconverted PDC states~\citep[Appendix H]{Albarelli2022}
\begin{equation}\label{eq:PDCstate1}
    \ket{\Phi_{\mathrm{PDC}}} \approx  \frac{1}{N_{\mathrm{PDC}}^{1/2}} \left( \ket{0} + \sum_{n=0}^{\infty}\,r_{n,\mathrm{PDC}}\,\,\ket{h_n^{\mathrm{S}}}\ket{h_n^{\mathrm{I}}}    \right),
\end{equation}
where $\ket{h_n^{\mathrm{S}}}$ and $\ket{h_n^{\mathrm{I}}}$ are $n$-th Hermite-Gauss Schmidt basis modes for signal and idler spaces respectively, and $r_{n,\mathrm{PDC}}$ are corresponding Schmidt weights. The JTA function for the incoming PDC states may be obtained by inverting Eq.~(\ref{eq:JSAschmidtdecomp}) and recombining the Schmidt terms $\Phi_{\mathrm{PDC}}(t_{\mathrm{S}},t_{\mathrm{I}}) = \sum_n\,r_{n,\mathrm{PDC}}\,h_n(t_{\mathrm{S}})h_n(t_{\mathrm{I}})$. 
The time-frequency entanglement of PDC states can be quantified using the entanglement entropy function
\begin{equation}
\label{eq:entent}
S = -\sum_n |r_{n,\mathrm{PDC}}|^2\log |r_{n,\mathrm{PDC}}|^2,
\end{equation} 
which is plotted as a function of PDC characteristics of entanglement time $T_{\mathrm{qent}}$ and pumpwidth $\sigma_{\mathrm{p}}$~(see \citep[Appendix H]{Albarelli2022} for definitions) in Figure~  \ref{fig:entanglemententropyheatmap}.

We show that for these most practical of entangled probes, time-frequency entanglement provides a functional advantage in biphoton spectroscopy for asymptotically long detection times.
We also establish that it is possible to get close to the fundamental limits set by the corresponding QCRB using unentangled measurements that are independent of the true value of the parameter.
We thus provide a complete recipe for quantum-enhanced biphoton spectroscopy using PDC light probes and simple, unentangled measurements independent of the sample parameters.

\subsection{QFI Of Outgoing PDC State}
The outgoing S-I state for the PDC input in Eq.~(\ref{eq:PDCstate1}) has a structure similar to Eq.~(\ref{eq:outgoingrhoP_sum}), but with a modified normalisation
\begin{align}
    \rho_{\mathrm{PDC,out}}^{\mathrm{SI}} = \left( 1-\mathfrak{n} \right) \ket{\Phi_{\mathrm{PDC,out}}}\bra{\Phi_{\mathrm{PDC,out}}} + \mathfrak{n}\,\ket{0^{\mathrm{S}}}\bra{0^{\mathrm{S}}}\otimes \sigma^{\mathrm{I}},
\end{align}
where  $\sigma^{\mathrm{I}}$ is the same as in Eq.~(\ref{eq:sigmaIdef}), $\mathfrak{n} = N/N_{\mathrm{PDC}}$, 
\begin{align}
    &\ket{\Phi_{\mathrm{PDC,out}}}  = \\
    &\frac{1}{(N_{\mathrm{PDC}}(1-\mathfrak{n}))^{1/2}}\left( \ket{0} + \sum_n r_{n,\mathrm{PDC}}\ket{\phi_{n,\mathrm{PDC}}^{\mathrm{S}}}\ket{h_n^{\mathrm{I}}} \right),
    \nonumber\noindent
\end{align}
with
\begin{align}
&\ket{\phi_{n,\mathrm{PDC}}^{\mathrm{S}}} = \\
&\int_{-\infty}^{\infty} dt_{\mathrm{S}} \left( h_n(t_{\mathrm{S}}) -\Gamma\int_{-\infty}^{t_{\mathrm{S}}}d\tau f_{\mathrm{M}}(t_{\mathrm{S}}-\tau)h_n(t_{\mathrm{S}})    \right)  
	a_{\mathrm{S}}^{\dag}(t_{\mathrm{S}})\ket{0^{\mathrm{S}}}. \nonumber\noindent
\end{align}
The QFI of the outgoing PDC state has the familiar trinal contribution (cf. Eq.~\eqref{eq:rhoPouttwinQFI})
\begin{align}\label{eq:rhoPoutPDCQFI}
&\mathcal{Q}(\theta;\rho_{\mathrm{PDC,out}}^{\mathrm{SI}}) = \nonumber\noindent\\ 
&~\,\mathcal{C}\left(\mathfrak{n},1-\mathfrak{n}\right) 
+ \mathfrak{n}\mathcal{Q}(\theta;\sigma^{\mathrm{I}}) + \left(1-\mathfrak{n}\right)\,\mathcal{Q}(\theta;\ket{\Phi_{\mathrm{PDC,out}}})
\end{align}
where 
\begin{align}\label{eq:entangledPDCQFIexp_schmidt}     &\mathcal{Q}\left(\theta;\ket{\Phi_{\mathrm{PDC,out}}}\right) = \\
     &\frac{4}{N_{\mathrm{PDC}}(1-\mathfrak{n})} \sum_n\,|r_{n,\mathrm{PDC}}|^2\,\langle\partial_{\theta}\phi_{n,\mathrm{PDC}}^{\mathrm{S}}|\partial_{\theta}\phi_{n,\mathrm{PDC}}^{\mathrm{S}}\rangle  -\noindent\nonumber \\
      &\frac{4}{(N_{\mathrm{PDC}}(1-\mathfrak{n}))^2}\left|\sum_{n} |r_{n,\mathrm{PDC}}|^2 \langle\phi_{n,\mathrm{PDC}}^{\mathrm{S}}  |\partial_{\theta}\phi_{n,\mathrm{PDC}}^{\mathrm{S}}\rangle     \right|^2. \nonumber\noindent
\end{align}
The relative magnitudes of the three terms in the PDC QFI in Eq.~(\ref{eq:rhoPoutPDCQFI}) admit the same pattern with respect to the ratio 
$\Gamma_{\perp}/\Gamma$ as established for the biphoton QFI in Table \ref{table1},  with the only modification being that all the Fisher informations in Eq.~(\ref{eq:entangledPDCQFIexp_schmidt}) are scaled by the $\Gamma_{\perp}/\Gamma$-independent normalisation $N_{\mathrm{PDC}}$.

Also of interest is the CFI of
the measurement-optimal 1-LOCC mediated by preparation step unitary $V = \mathds{1}^{\mathrm{I}}$ for the PDC states~(see Eq.~(\ref{eq:maximalCFILOCCi2s_main}))
\begin{align}\label{eq:entangledPDCCFIloccv1}
     &\mathcal{C}_{\mathrm{max}}\left(\theta;V = 
     \mathds{1}^{\mathrm{I}}\right) = \\
     & \frac{4}{N_{\mathrm{PDC}}(1-\mathfrak{n})} \sum_n\,|r_{n,\mathrm{PDC}}|^2\,\langle\partial_{\theta}\phi^{\mathrm{S}}_{n,\mathrm{PDC}}|\partial_{\theta}\phi^{\mathrm{S}}_{n,\mathrm{PDC}}\rangle  -\noindent\nonumber \\
      & \frac{4}{(N_{\mathrm{PDC}}(1-\mathfrak{n}))^2}\sum_{n} |r_{n,\mathrm{PDC}}|^2 \left|\langle\phi^{\mathrm{S}}_{n,\mathrm{PDC}}  |\partial_{\theta}\phi^{\mathrm{S}}_{n,\mathrm{PDC}}\rangle\right|^2. \nonumber\noindent
\end{align}

Finally, some descriptions of the entangled PDC photons omit the vacuum term in Eq.~(\ref{eq:PDCstate1}), yielding just the biphoton state $\ket{\Phi_{\mathrm{biph}}}$ in Eq.~(\ref{eq:twophotonfreqstate}). 
This corresponds to the state post-selected for only successful detection of the two photons, and can be expressed in terms of the PDC JTA $\Phi_{\mathrm{PDC}}(t_{\mathrm{S}},t_{\mathrm{I}})$ as
\begin{equation}\label{eq:biphotonPDC_maintext}
    \ket{\Phi_{\mathrm{biph}}} = \frac{1}{\sqrt{\Lambda}}\,\int dt_{\mathrm{S}} \int dt_{\mathrm{I}} \,\Phi_{\mathrm{PDC}}(t_{\mathrm{S}},t_{\mathrm{I}})\,a_{\mathrm{S}}^{\dag}(t_{\mathrm{S}})a_{\mathrm{I}}^{\dag}(t_{\mathrm{I}})\ket{0},
\end{equation}
where $\Lambda = \int dt_{\mathrm{S}} \int dt_{\mathrm{I}}\, \Phi^*_{\mathrm{PDC}}(t_{\mathrm{S}},t_{\mathrm{I}})\Phi_{\mathrm{PDC}}(t_{\mathrm{S}},t_{\mathrm{I}})$ ensures unit norm.
The spectroscopic informations provided by states in Eqs.~(\ref{eq:PDCstate1}) and (\ref{eq:biphotonPDC_maintext}) are not, in general, a simple rescaling.
Rather,
\begin{align}\label{eq:PDCBPrelationsimplified_maintext}
    &\mathcal{Q}\left(\theta;\ket{\Phi_{\mathrm{PDC,out}}}\right)\approx \Lambda\,\mathcal{Q}\left(\theta;\ket{\Phi_{\mathrm{biph,out}}}\right)   \\
     &~~~+\frac{4}{\Lambda}\,\left|\int dt_{\mathrm{S}} \int dt_{\mathrm{I}}\, \partial_{\theta} \Phi_{\mathrm{PDC,out}}(t_{\mathrm{S}},t_{\mathrm{I}})^* \Phi_{\mathrm{PDC,out}}(t_{\mathrm{S}},t_{\mathrm{I}}) \right|^2,	\nonumber
\end{align}
assuming $N_{\mathrm{PDC}}\approx 1$ and $\Lambda\ll 1.$ See Appendix \ref{appendix:relationPDCbiphoton} for details.

We finally specialise our study of entangled quantum light spectroscopy to specific matter systems: 
in Section \ref{sec6b}, we address 1-site TLS~($P=1$ in Eq.~(\ref{eq:matterhamiltonian})) for which we will evaluate Fisher informations 
corresponding to pulse-matter coupling $\Gamma$, as well as level frequency $\omega_0$; 
in Section \ref{sec6c}, we address the 2-site CD systems~($P=2$ in Eq.~(\ref{eq:matterhamiltonian})), 
for which we will evaluate fundamental limits of inter-site coupling $J$ estimation. 
 As a first foray, we focus on the vibrationless Hamiltonian that does not include couplings to phonon baths.

We highlight aspects of engineering the source of PDC probes for spectroscopy while yielding tangible quantum enhancements.
We find larger time-frequency entanglement --- concomitantly shorter entanglement times and pump bandwidths --- to be beneficial.
Indeed, more entanglement in the PDC probe yields more spectroscopic information. Parameter-independent~(hence non-adaptive) unentangled detection, using the most entangled of PDC probes, also 
meaningfully outperform single-photon spectroscopies using the reduced signal state only.
Typically, these  simplified measurements capture between 60\% - 90\% of the spectroscopic information.

\subsection{TLS spectroscopy}
\label{sec6b}

For the TLS, $H_I^{\mathrm{M}} = \hbar\Delta\ket{e}\bra{e}$, $\Delta=\omega_0-\bar{\omega}_{\mathrm{S}}$ 
is the detuning between the carrier signal and TLS frequency $\omega_0$, and the characteristic response function for TLS takes the simple form
\begin{equation}
    f_{\mathrm{TLS}}(t) = \mathrm{exp} \left(-\left[\frac{\Gamma+\Gamma_{\perp}}{2} + i\Delta\right] t \right).
\end{equation}

\subsubsection{No Coupling to Environment $\mathrm{(E)}$: $\Gamma_{\perp} = 0$}

\begin{figure}
     \centering
    \includegraphics[width=0.49\textwidth]{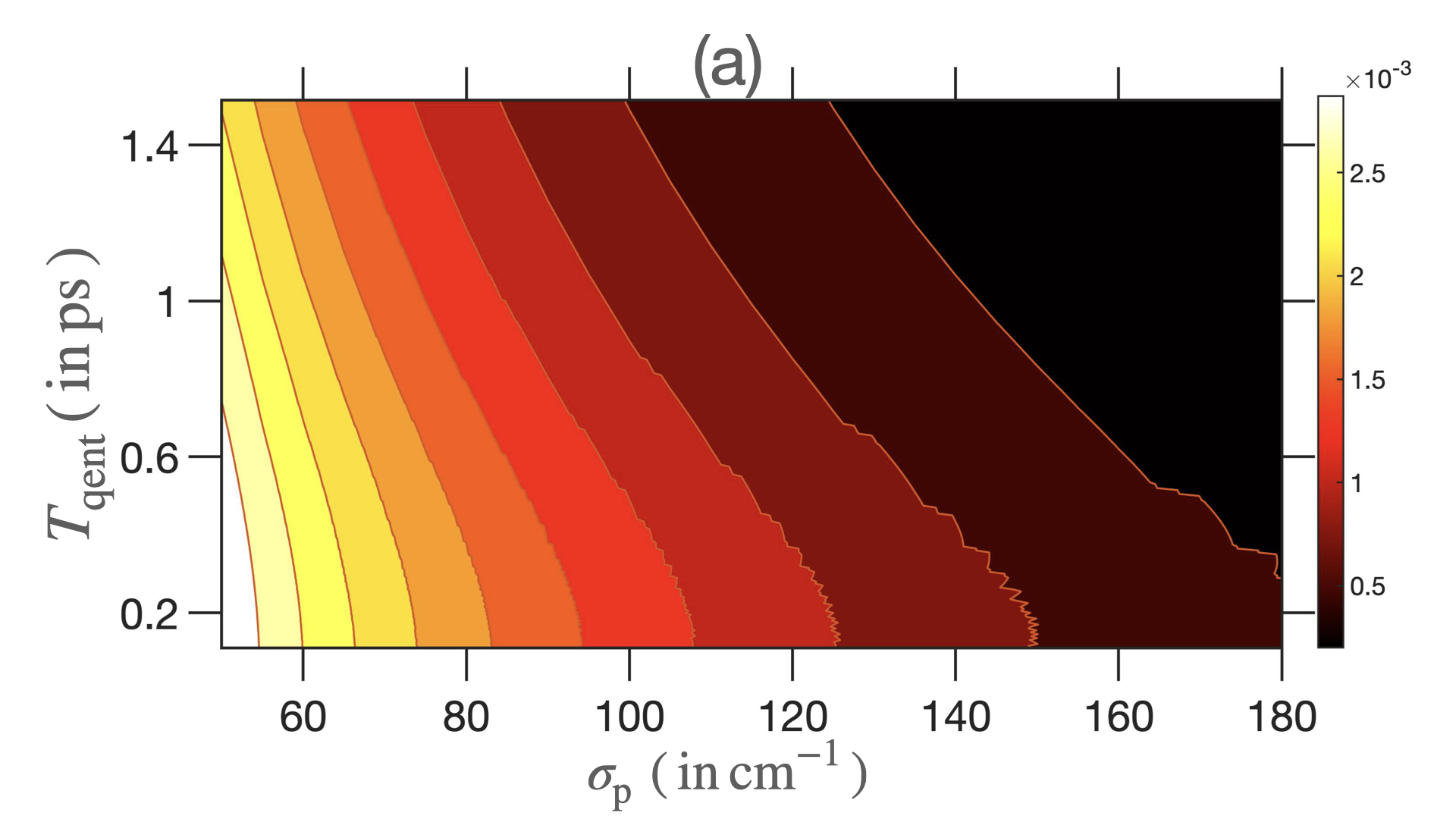}
    \includegraphics[width=0.49\textwidth]{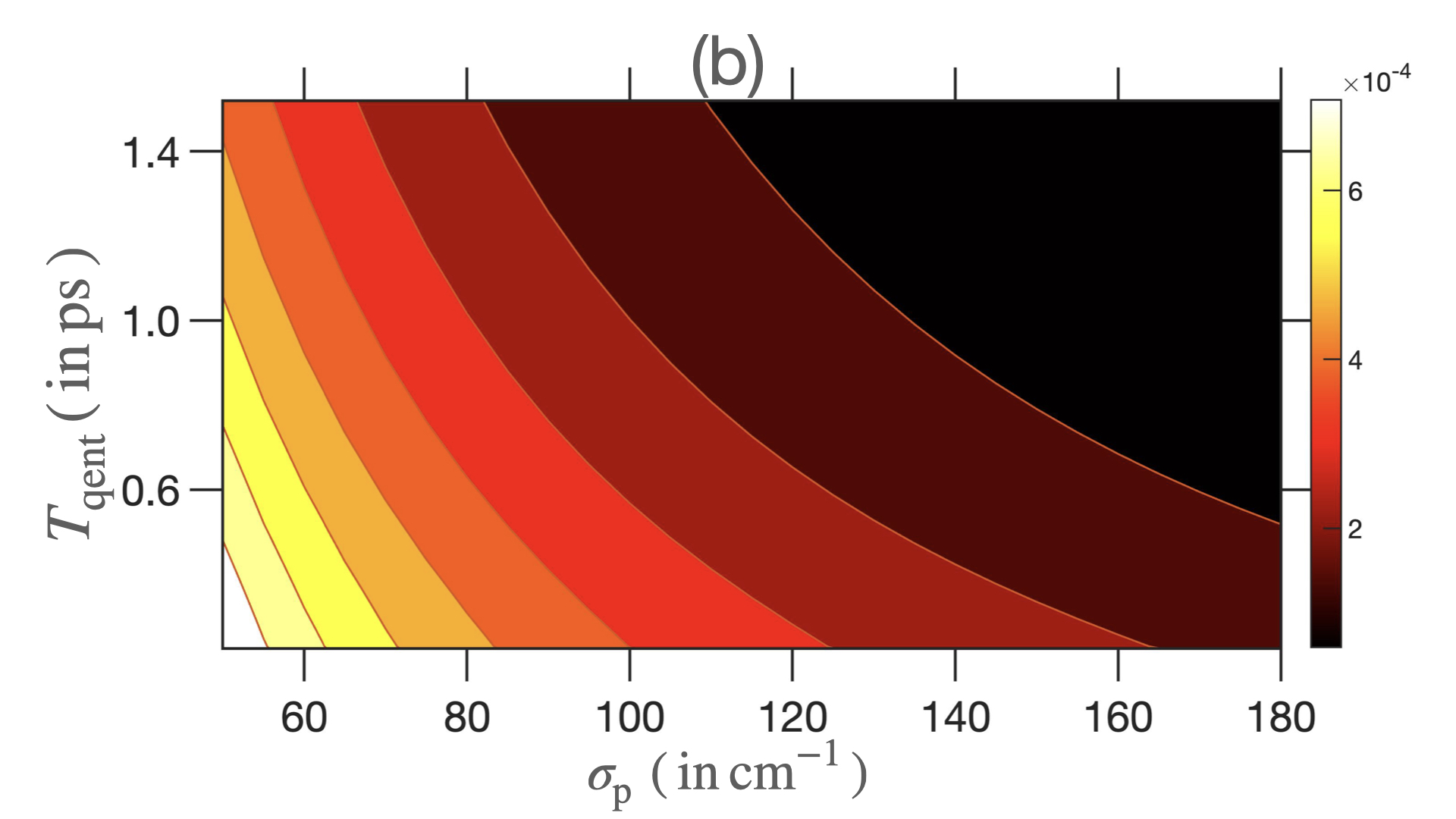}
\caption{QFI  $\mathcal{Q}(\theta;\ket{\Phi_{\mathrm{PDC,out}}})$, calculated numerically using Eq.~(\ref{eq:entangledPDCQFIexp_schmidt}), for varying PDC entanglement time $T_{\mathrm{qent}}$, and classical pumpwidth $\sigma_{\mathrm{p}}$, for TLS parameters 
(a) $\theta ~\equiv ~\Gamma$, and (b) $\theta ~\equiv~\omega_0$, and $\Gamma_{\perp} = 0 $.~($\Gamma = 0.15$\,THz\,,\,$\Delta = 0$\,THz).}  
\label{fig:PDCheatmap_TLS}
\end{figure}

In this case, only the last term in Eq.~\eqref{eq:rhoPoutPDCQFI} contributes. This QFI is plotted, 
for a grid of values of classical pumpwidths $\sigma_p$ and entanglement times $T_{\mathrm{qent}}$,
for the estimation of $\Gamma$ in Figure \ref{fig:PDCheatmap_TLS} (a), and for the $\omega_0$-parameter in Figure \ref{fig:PDCheatmap_TLS} (b). 
For both parameters, comparing Figs.~\ref{fig:entanglemententropyheatmap} and \ref{fig:PDCheatmap_TLS} shows that  more entanglement in the incoming PDC probe, as captured by entropy $S$ defined in Eq.~(\ref{eq:entent}), leads to a higher value for the outgoing QFI.

To uncover the role of entanglement in the incoming probe PDC state in this spectroscopy task more clearly, we also display scatter plots of the $\Gamma$- and $\omega_0$-QFIs as functions of the entanglement entropy $S$ in Appendix \ref{app:parametricplots}~(see Figure \ref{fig:scatterplots} (a)-(b)). 
These show that that a more entangled PDC input always yields more outgoing QFI.
However, the outgoing PDC QFI is not a one-to-one function of the incoming entanglement. Specifically, for the same amount of entanglement, incoming PDC states may yield outgoing states with different QFIs, depending on the experimental values of $T_{\mathrm{qent}}$ and $\sigma_{\mathrm{p}}$.

The apparent advantage conferred by time-frequency entanglement here  is not in contradiction with our earlier conclusion that entanglement provides no in-principle advantage in biphoton spectroscopy.
In Section \ref{sec:noadvantage} we showed that the outgoing QFI corresponding to \emph{any} incoming entangled state may be superseded by a 
suitably optimised product state in the S-I space.
In contrast, entanglement-enhanced sensing in Figure \ref{fig:PDCheatmap_TLS} is relative to PDC probe states \emph{only.}

For reference, in Figure \ref{fig:PDCheatmap_TLS}, the ratios of outgoing QFI corresponding to the most entangled PDC state~(bottom-left edge in either plot) to that of the least entangled (almost product) PDC state~(top-right edge) are 
\begin{equation}
\frac{\mathcal{Q}(\Gamma;\ket{\Phi_{\mathrm{PDC,out}}}) \vert_{T_{\mathrm{qent}} = 0.150\,\mathrm{ps},\sigma_p = 50\,\mathrm{cm}^{-1}}}{\mathcal{Q}(\Gamma;\ket{\Phi_{\mathrm{PDC,out}}}) \vert_{T_{\mathrm{qent}} = 1.995\,\mathrm{ps},\sigma_p = 180\,\mathrm{cm}^{-1}}} \approx 14.0391,
\end{equation}
and
\begin{equation}
\frac{\mathcal{Q}(\omega_0;\ket{\Phi_{\mathrm{PDC,out}}}) \vert_{T_{\mathrm{qent}} = 0.150\,\mathrm{ps},\sigma_p = 50\,\mathrm{cm}^{-1}}}{\mathcal{Q}(\omega_0;\ket{\Phi_{\mathrm{PDC,out}}}) \vert_{T_{\mathrm{qent}} = 1.995\,\mathrm{ps},\sigma_p = 180\,\mathrm{cm}^{-1}}} \approx 15.5665,
\end{equation}
showing that for
TLS spectroscopy, there is significant advantage to engineering the source to produce more entangled states --- within the class of PDC states.
Our result also indicates that, for the parameter regime considered, using entangled photons 
enables more precise spectroscopy than heralded single photon states~\cite{pittman2005heralding,mosley2008heralded,kaneda2016heralded}. 

To further underscore the subtle role of entanglement, we consider a family of time-frequency pulse mode~(TFM) states of the form~\citep{graffitti2020direct}
\begin{equation}\label{eq:TFMdef}
    \ket{\Phi_{\mathrm{TFM}}} = \frac{1}{N_{\mathrm{TFM}}^{1/2}}\,\left( \ket{0} + \frac{\alpha_{\mathrm{pump}}}{\hbar}\,\ket{\Phi^{\mathfrak{t}}_{\mathrm{TFM}}}   \right),
\end{equation}
where
\begin{equation}
\ket{\Phi^{\mathfrak{t}}_{\mathrm{TFM}}} = \cos\mathfrak{t}\ket{h_0^{\mathrm{S}}}\ket{h_1^{\mathrm{I}}} + \sin\mathfrak{t}\ket{h_1^{\mathrm{S}}}\ket{h_0^{\mathrm{I}}}, ~0\leq\mathfrak{t}\leq\pi,
\end{equation}
and $\alpha_{\mathrm{pump}}/\hbar = 0.01$~(same as for $\ket{\Phi_{\mathrm{PDC}}}$). 
Parametric plots of  the QFI of the outgoing TFM state against the entanglement of the incoming state in Figure \ref{fig:TFMscatterplots} in Appendix \ref{app:parametricplots}
show that for both TLS parameters, the maximal QFI in outgoing state corresponds to the product input $\ket{h_0^{\mathrm{S}}}\ket{h_1^{\mathrm{I}}}$.
Thus, for this family of biphoton states, time-frequency mode entanglement is not a useful resource.

As a final caveat, recall that our conclusions are for asymptotically large times.
They may not hold in general for finite-time evolutions.
As a matter of fact, an opposite behaviour was shown in a similar parameter region~\cite{Albarelli2022} for short detection times, where the problem essentially reduces to absorption estimation, and the perturbation induced on the signal photon wavefunction is not relevant.
This further highlights the intricacies in understanding the role of entanglement in quantum light spectroscopy.

To understand the measurements that attain a large fraction of the maximal spectroscopic information,
we now define two ratios to capture the efficacy of $\mathcal{C}_{\mathrm{max}}\left(\theta;V = 
     \mathds{1}^{\mathrm{I}}\right).$ They are motivated by the simpler parameter-independent $V=\mathds{1}^{\mathrm{I}}$ measurements enabling practical entangled light spectroscopy.    
The first is a ``degree of optimality"  defined as
\begin{equation}\label{eq:V1degreeoptimality}
    \varkappa(\theta) = \frac{\mathcal{C}_{\mathrm{max}}(\theta;V=\mathds{1}^{\mathrm{I}})}{\mathcal{Q}(\theta;\ket{\Phi_{\mathrm{PDC,out}}})},~~0 \leq \varkappa(\theta) \leq 1.
\end{equation}
A value closer to unity indicates proximity to the fundamental precision afforded by appropriately constructed estimators set by the PDC QFI, 
which for $\Gamma_{\perp} = 0$ is the biphoton QFI $\mathcal{Q}(\theta;\ket{\Phi_{\mathrm{PDC,out}}(\theta)})$ only.

The second is an ``enhancement factor" defined as
\begin{equation}\label{eq:V1enhancementfactor}
    \varsigma(\theta) = \frac{\mathcal{C}_{\mathrm{max}}(\theta;V=\mathds{1}^{\mathrm{I}})}{\mathcal{Q}(\theta;\mathrm{Tr}_{\mathrm{I}}\left[\ket{\Phi_{\mathrm{PDC,out}}}\right])},~~ \varsigma(\theta) > 1,
\end{equation}
where $\varsigma(\theta)>1$ indicates a spectroscopic advantage offered by 1-LOCC measurement with $V=\mathds{1}^{\mathrm{I}}$ over all single-photon strategies on the reduced signal-photon state $\mathrm{Tr}_{\mathrm{I}}\ket{\Phi_{\mathrm{PDC,out}}}$.
For completeness,
\begin{align}
&\mathcal{Q}\left(\theta;\mathrm{Tr}_{\mathrm{I}}\ket{\Phi_{\mathrm{PDC,out}}}\right) = \mathcal{C}_{\mathrm{max}}\left(\theta;V=\mathds{1}^{\mathrm{I}}\right) \;-\\
    &\frac{1}{(N_{\mathrm{PDC}}(1-\mathfrak{n}))^2} \sum_{n>m}\frac{16~|r_m|^2 |r_n|^2}{|r_m|^2 + |r_n|^2}|\langle\phi_{m,\mathrm{PDC}}^{\mathrm{S}}  \left|\partial_{\theta}\phi_{n,\mathrm{PDC}}^{\mathrm{S}}\rangle\right|^2.	\nonumber\noindent
\end{align}

\begin{figure}[t!]
     \centering
         \includegraphics[width=0.49\textwidth]{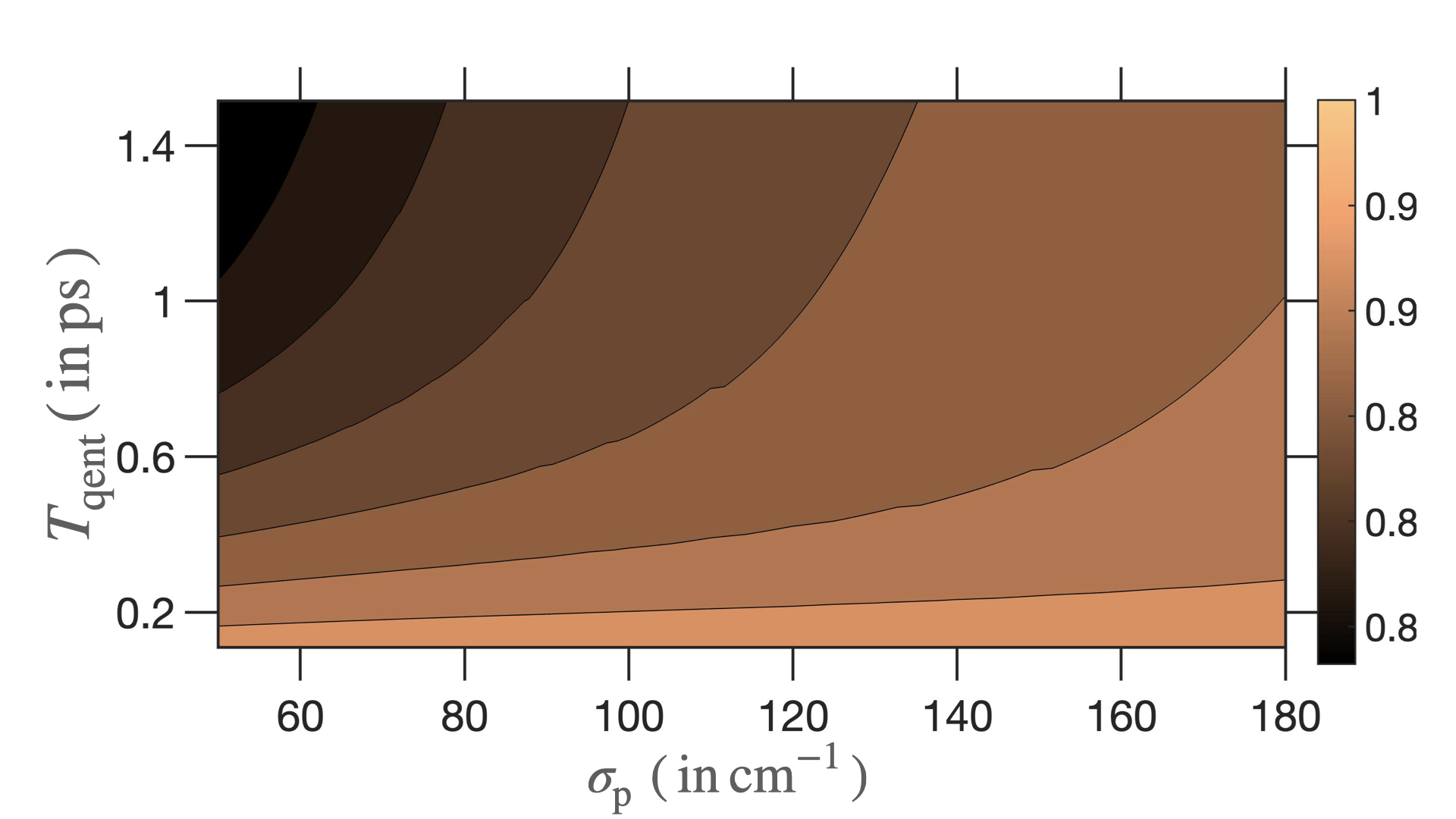}
\caption{Degree of optimality as per the ratio $\varkappa(\omega_0)$ using PDC light, for $\Gamma = 0.15$\,THz,\,$\Delta = 0$\,THz.}  
\label{fig:LOCCCFItoQFIratio}
\end{figure}

\begin{figure}
     \centering
         \includegraphics[width=0.49\textwidth]{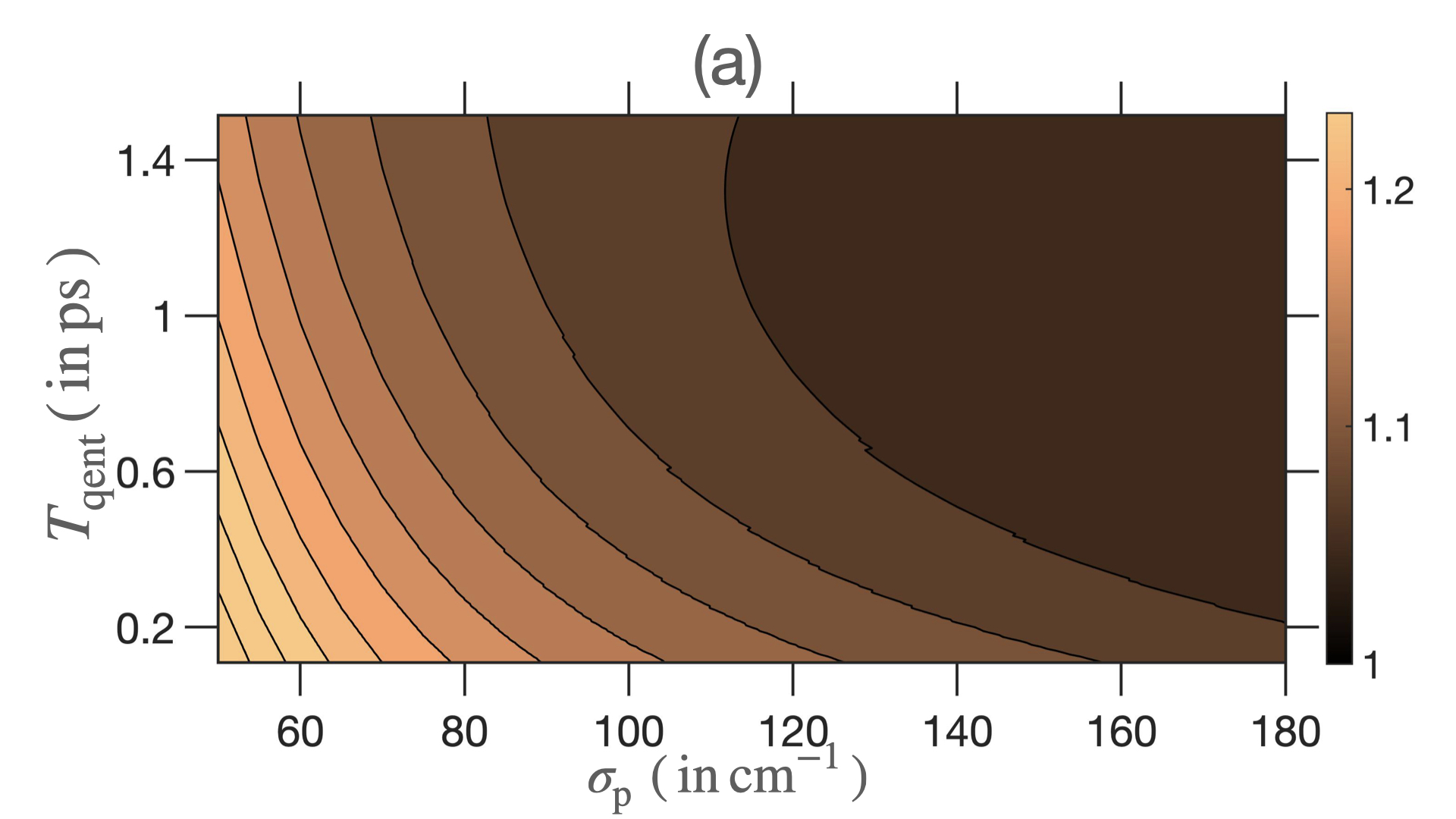}
         \includegraphics[width=0.49\textwidth]{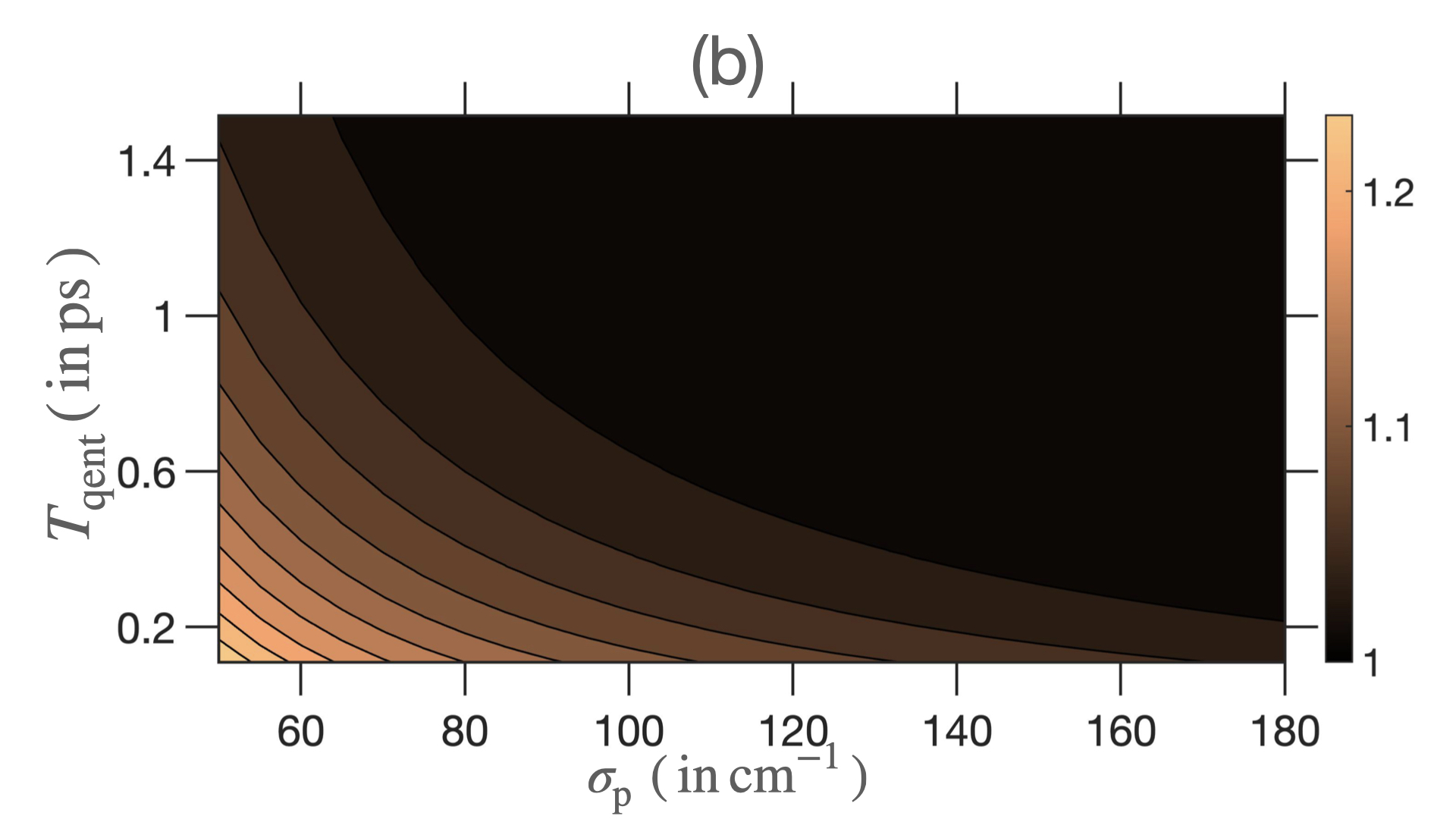}
\caption{Enhancement factor $\varsigma(\theta)$ for estimation of TLS parameters (a) $\Gamma$, and (b) $\omega_0$ using PDC light, for $\Gamma = 0.15$\,THz,\,$\Delta = 0$\,THz.}  
\label{fig:LOCCCFItotrQFIratio_enhancement}
\end{figure}

Figure~\ref{fig:LOCCCFItoQFIratio} displays the optimality ratio $\varkappa(\theta)$ for the TLS parameter $\omega_0$ when $\Delta=0$. 
It shows that more that 80\% of the QFI for a PDC probe is recovered by a parameter-independent 1-LOCC measurement.
For resonant $\Gamma$ estimation, $\varkappa(\Gamma)\vert_{\Delta=0} = 1$ for all points on the grid~\cite{Albarelli2022}. 
As the magnitude of the detuning $|\Delta|$ increases, $\langle\Phi_{\mathrm{PDC,out}}\,|\,\partial_{\Gamma}\Phi_{\mathrm{PDC,out}}\rangle\vert_{\Delta\neq 0}\neq 0$, 
and the degree of optimality drops below unity, as can be seen in Figure \ref{fig:LOCCCFItoQFIratio_Gamma_Deltaneq0} in Appendix \ref{appendix:TLSestimationDeltaneq0}. 

Lastly, Figure \ref{fig:LOCCCFItotrQFIratio_enhancement}  displays the enhancement factor $\varsigma(\theta)$ for TLS parameters $\omega_0$ and $\Gamma$.
From Eq.~(\ref{eq:lowerboundi2s}), we always expect $\varsigma(\theta) > 1$, which is substantiated in these plots. We also see that spectroscopy with only the most entangled of PDC states can meaningfully outperform single-photon spectroscopy using the reduced signal state $\mathrm{Tr}_{\mathrm{I}}[\ket{\Phi_{\mathrm{PDC,out}}}]$.
Interestingly, for $\omega_0$-estimation, a more entangled state yields a larger $\varsigma(\omega_0)$ as the detuning $|\Delta|$ increases, whereas the reverse is true for $\varsigma(\Gamma)$.
For the effects of non-zero $|\Delta|$ on these quantities, see Appendix \ref{appendix:TLSestimationDeltaneq0}. 
This has practical consequences for spectroscopy using PDC probes, as one may resort to the even simpler setup of single photon probes
 if the enhancement offered by the 1-LOCC scheme, as measured by $\varsigma(\theta)$ is not large enough.

\subsubsection{Coupling to Environment $\mathrm{(E)}$:  $\Gamma_{\perp} > 0$}
\label{subsec:gammaperpneq0}

\begin{figure}
         \centering
         \includegraphics[width=0.49\textwidth]{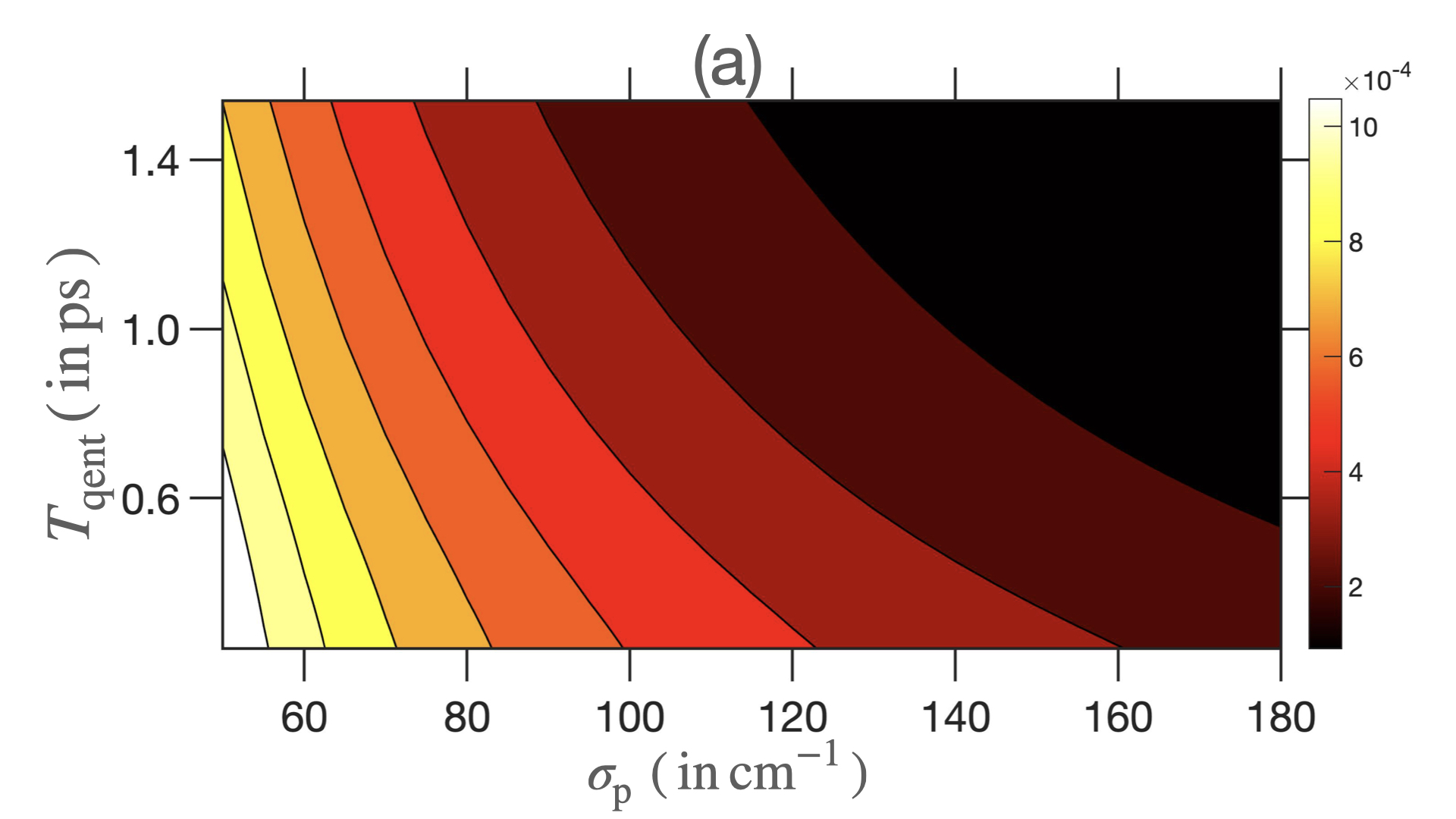}
         \includegraphics[width=0.49\textwidth]{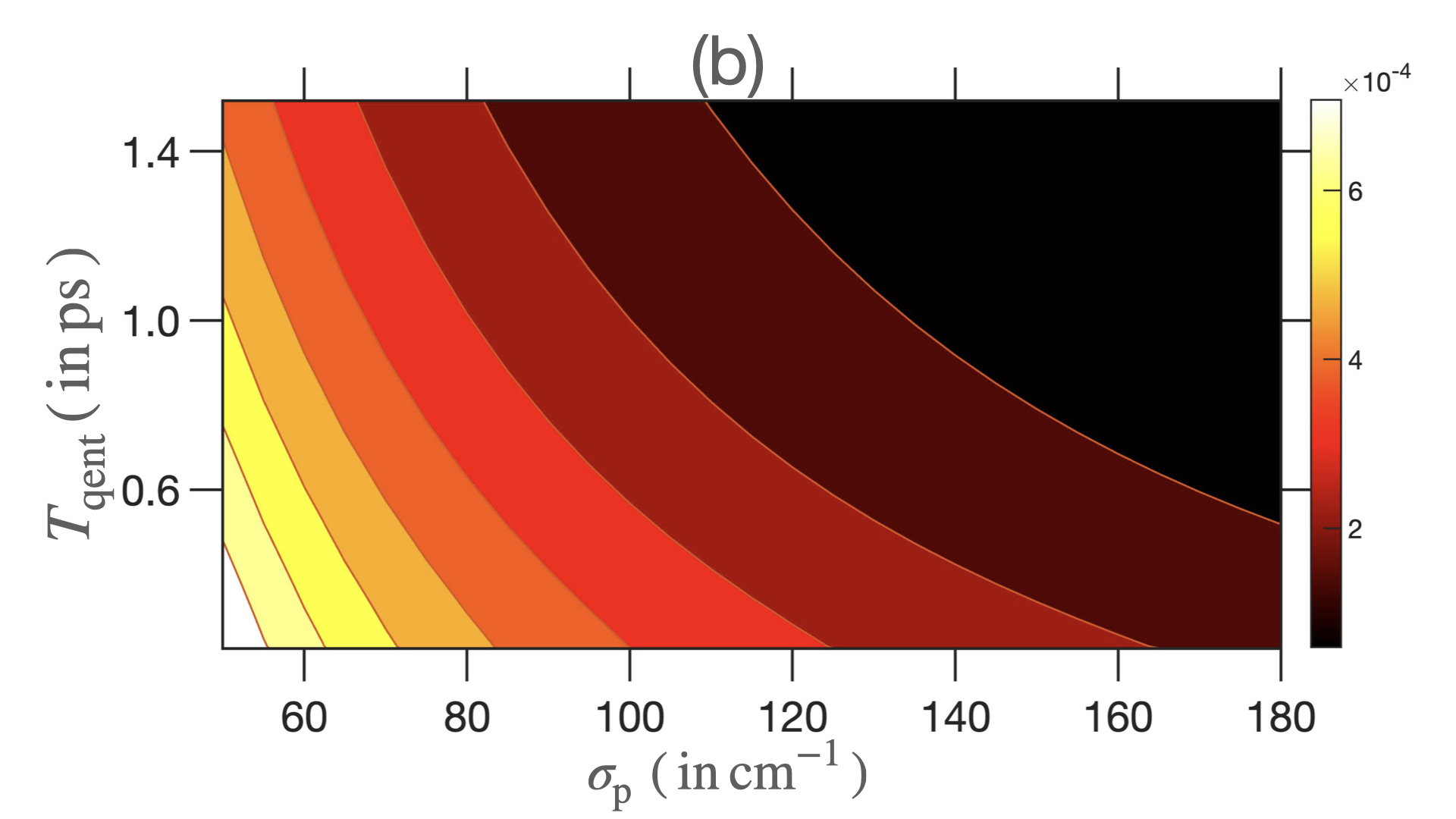}
\caption{ QFI $\mathcal{Q}(\Gamma;\ket{\Phi_{\mathrm{PDC,out}}})$ for varying PDC entanglement time $T_{\mathrm{qent}}$, and classical pumpwidth $\sigma_{\mathrm{p}}$, for TLS parameter $\Gamma$, and free space M-E coupling set to (a) $\Gamma_{\perp}/\Gamma = 0.5$, and (b) $\Gamma_{\perp}/\Gamma = 10.0$. Note the change in scale of QFI values between the two plots.~($\Gamma = 0.15$\,THz,\,$\Delta = 0$\,THz).}  
\label{fig:freespaceQFIGam}
\end{figure}

\begin{figure}
         \centering
         \includegraphics[width=0.49\textwidth]{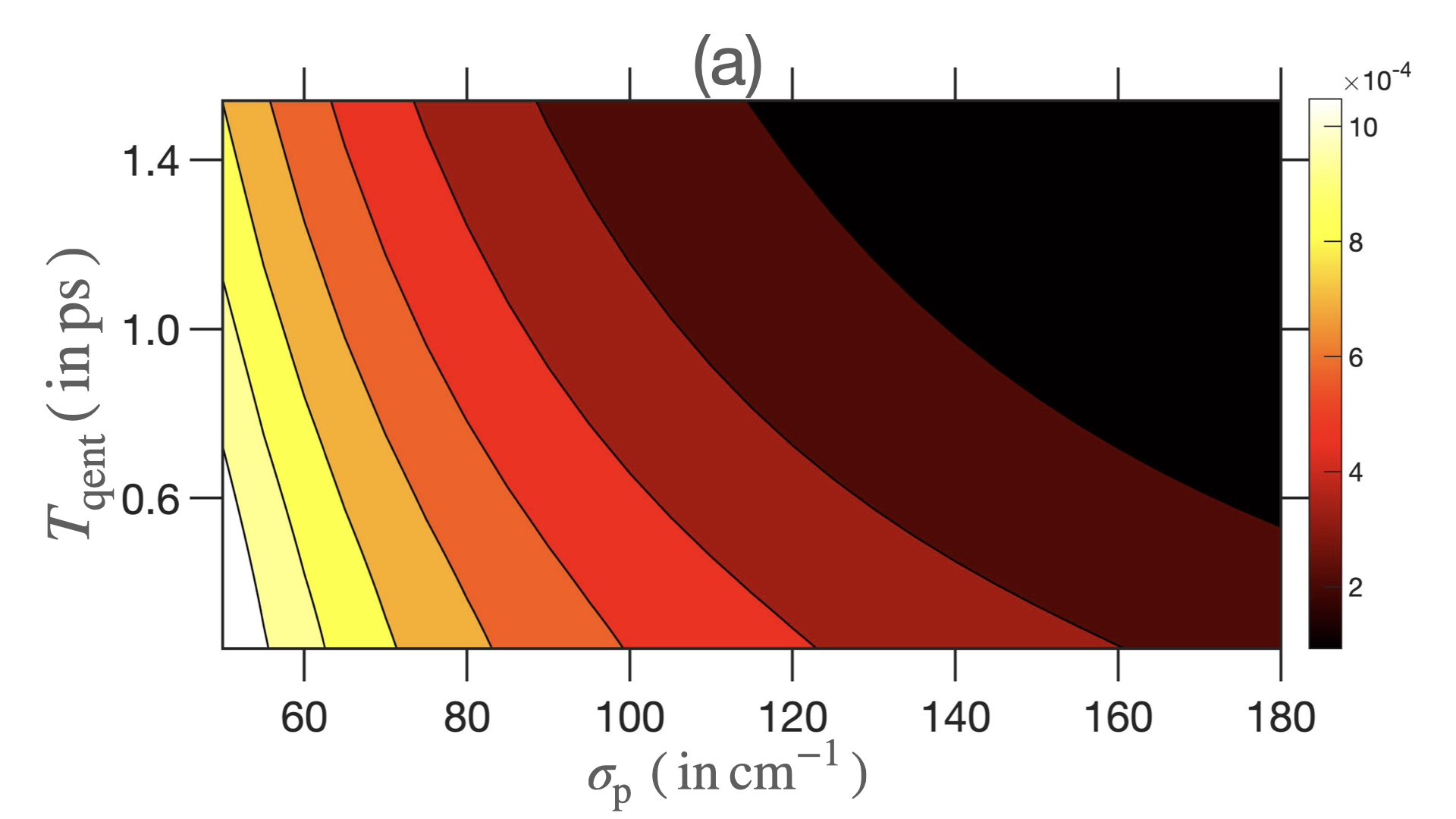}
         \includegraphics[width=0.49\textwidth]{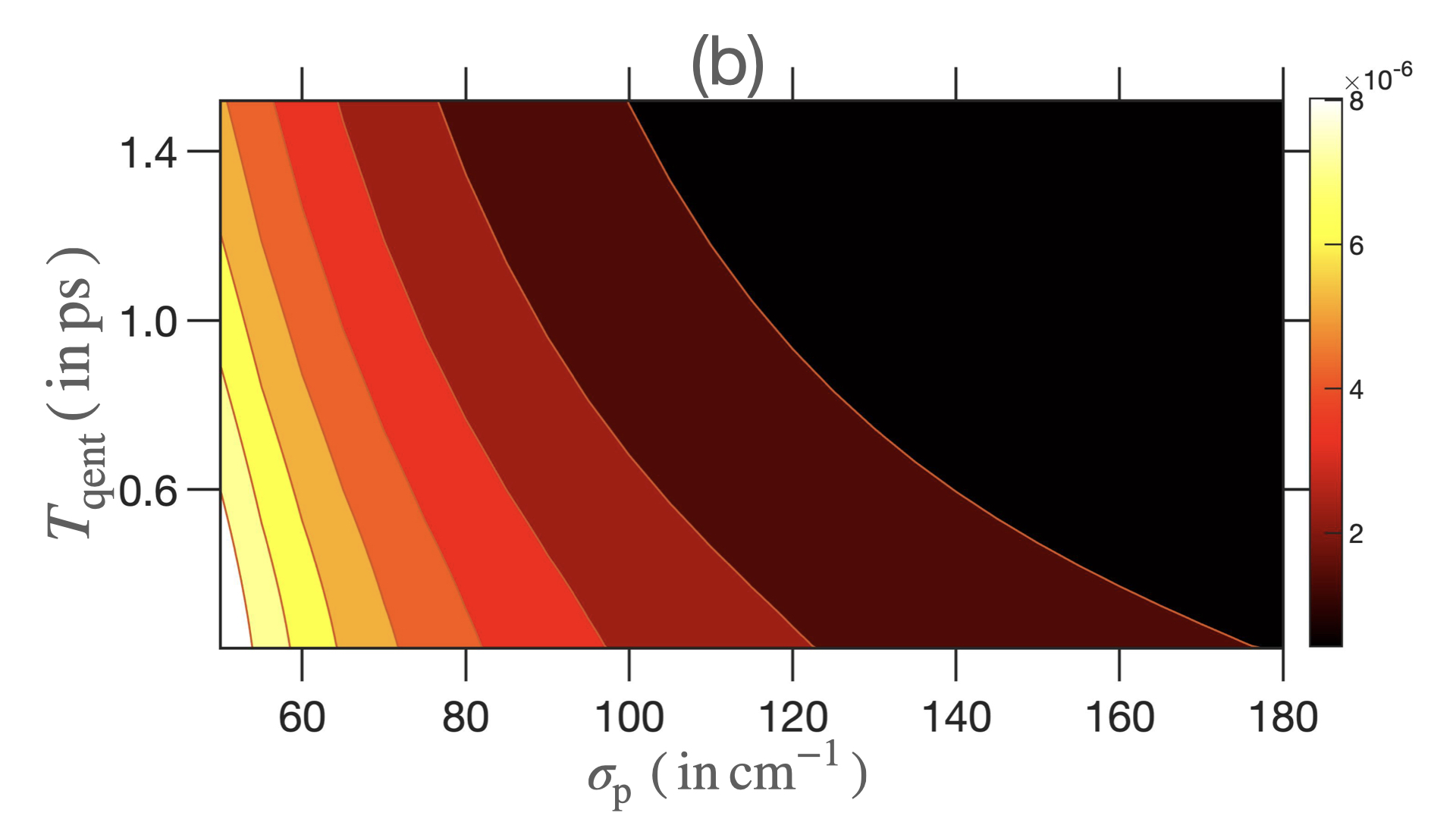}
\caption{QFI $\mathcal{Q}(\omega_0;\ket{\Phi_{\mathrm{PDC,out}}})$ for varying PDC entanglement time $T_{\mathrm{qent}}$, and classical pumpwidth $\sigma_{\mathrm{p}}$, for TLS parameter $\omega_0$, and free space M-E coupling set to (a) $\Gamma_{\perp}/\Gamma = 0.5$, and (b) $\Gamma_{\perp}/\Gamma = 10.0$. Note the change in scale of QFI values between the two plots.~($\Gamma = 0.15$\,THz,\,$\Delta = 0$\,THz).}  
\label{fig:freespaceQFIDel}
\end{figure}

For non-zero coupling to the environmental modes in E so that $\Gamma_{\perp}>0$, the outgoing two-photon QFI must now be evaluated using Eq.~(\ref{eq:rhoPoutPDCQFI}) which includes contributions from both single- and two-photon terms, in addition to the term corresponding to classical mixing. 
Figure \ref{fig:freespaceQFIGam} displays the outgoing $\Gamma$-QFI~(for the same grid of values as Figure \ref{fig:PDCheatmap_TLS} and $\Delta =0$) 
for two representative values of M-E coupling.
In panel (a), $\Gamma_{\perp} = 0.5\Gamma$ corresponds to comparable M-S and M-E couplings, 
whereas panel (b) corresponds to  $\Gamma_{\perp} = 10.0\Gamma$ such that the matter-environment coupling is much stronger than coupling to the incoming signal mode. Corresponding results are shown in Figure \ref{fig:freespaceQFIDel} for $\omega_0$-QFI. 
The effect of non-zero detuning on TLS parameter QFIs in the presence of an environment is studied in Appendix \ref{appendix:TLSestimationDeltaneq0}. 

Even in the presence of an environment, Figures~\ref{fig:freespaceQFIGam} and \ref{fig:freespaceQFIDel} show that time-frequency 
entanglement continues to be a useful resource for spectroscopy using PDC probes. The magnitudes of the QFI values for either TLS parameter are also diminished, an expected consequence of the M-E coupling which causes the matter sample to decay into the environmental modes, thus reducing the information content in the measured signal and idler modes.

\subsection{CD spectroscopy: $J$-estimation}
\label{sec6c}

The Hamiltonian of a coupled dimer~(CD)~($P=2$ in Eq.~(\ref{eq:matterhamiltonian})) is given by
\begin{equation}
    H^{\mathrm{CD}} = \sum_{j=a,b}\,\hbar\omega_j\ket{j}\bra{j} + \hbar(\omega_a +\omega_b)\ket{f}\bra{f} + J(\ket{a}\bra{b} + \ket{b}\bra{a}),
\end{equation}
where $J$ is the coupling strength between the two sites $a$ and $b$~(see Figure \ref{fig:biphotonspec_diagram}). Transforming to an appropriately chosen interaction frame and diagonalising the matter-only part~(details in Appendix \ref{appendix:CDdetails}), we can express the CD Hamiltonian in the delocalised excitonic basis as
\begin{equation}
    H^{\mathrm{CD}}_{I} = \sum_{i=\alpha,\beta}\hbar\Delta_i\ket{i}\bra{i} + \hbar(\Delta_{\alpha}+\Delta_{\beta})\ket{f}\bra{f},
\end{equation}
where $\Delta_i = \omega_i - \bar{\omega}_{\mathrm{S}}\,(i=\alpha,\beta)$ are the detunings from the central signal pulse frequency of the singly-excited manifold~(SEM) excitonic levels $\ket{\alpha}$ and $\ket{\beta}$. 

The explicit form for the characteristic CD function $f_{\mathrm{CD}}(t)$ appears in Appendix \ref{appendix:characteristicfnCD}.
It can be used to evaluate~(assuming no M-E coupling so that $\Gamma_{\perp} = 0$) the
 fundamental limits on the $J$ coupling parameter between the sites $a$ and $b$, using Eq.~(\ref{eq:rhoPoutPDCQFI}).

\begin{figure}
         \centering
         \includegraphics[width=0.49\textwidth]{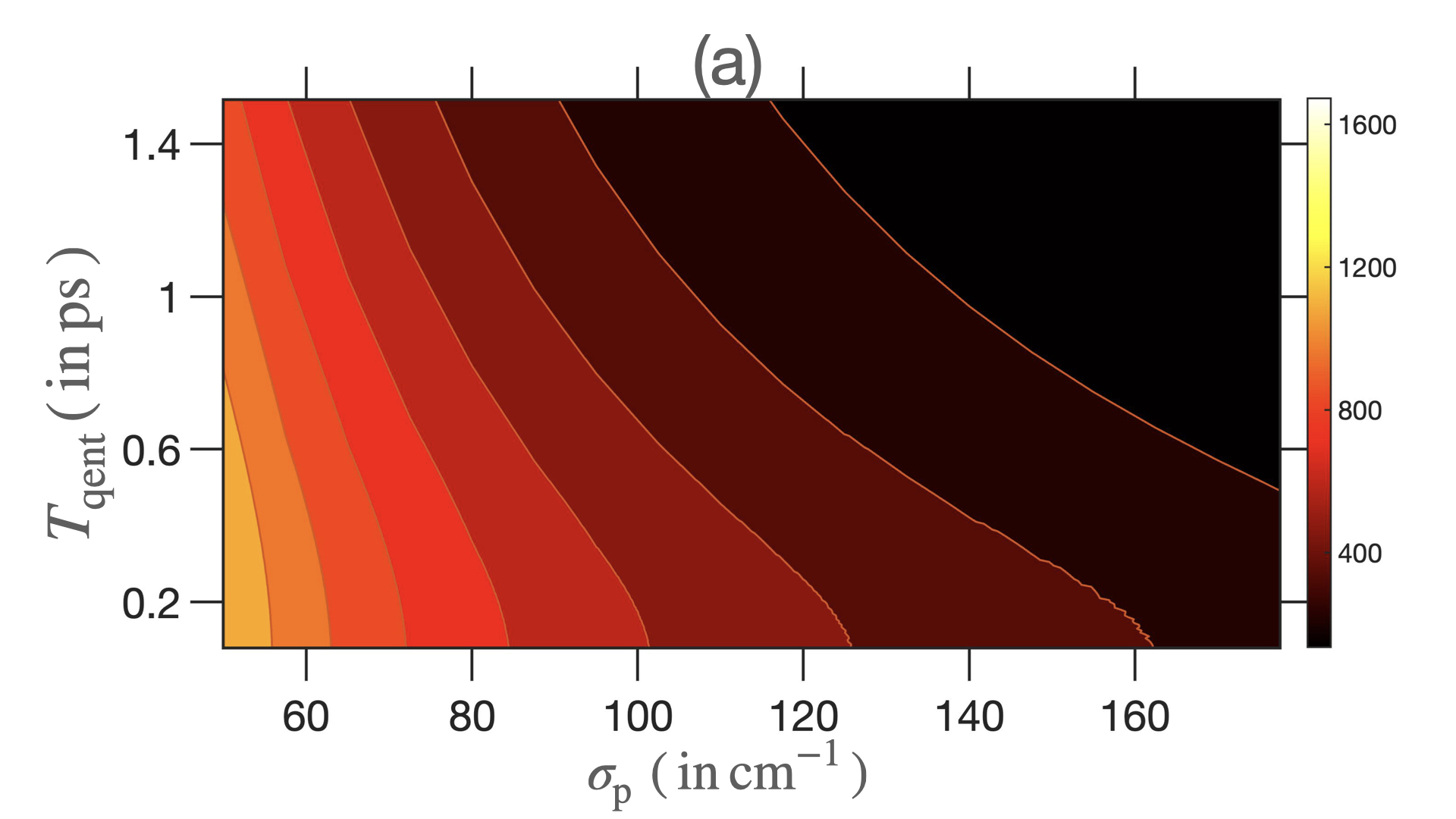}
         \centering
         \includegraphics[width=0.49\textwidth]{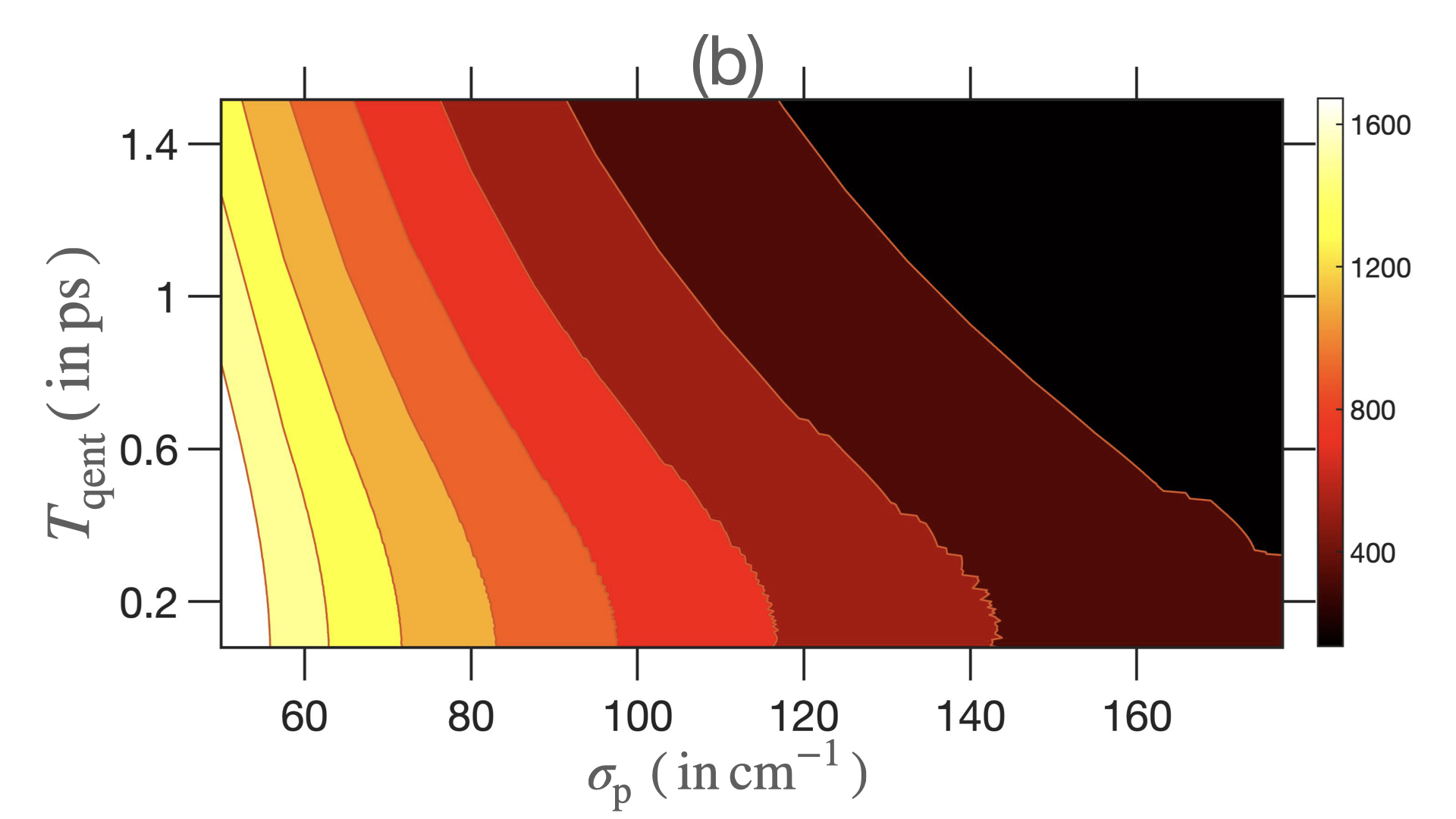}
        \caption{QFI  $\mathcal{Q}(J;\ket{\Phi_{\mathrm{PDC,out}}})$ for varying PDC entanglement time $T_{\mathrm{qent}}$, and classical pumpwidth $\sigma_{\mathrm{p}}$, for CD interstitial coupling $J$, and (a)\,$\bar{\omega}_{\mathrm{S}} = \omega_{\alpha}$, and (b)\,$\bar{\omega}_{\mathrm{S}} = \omega_{\beta}$. The CD parameters used are those of the allophycocyanin dimer~\citep{womick2009exciton} --- $\hbar\omega_a = 1.6$\,eV, $\hbar\omega_b = 1.8$\,eV, $J = -0.07$\,eV, $\mu_a = 1$\,Debye and $\mu_b = 1.5$\,Debye.~($\Gamma = 0.15$\,THz)}
        \label{fig:PDCheatmap_J}
\end{figure}
        
In Figure \ref{fig:PDCheatmap_J}, the $J$-QFI are plotted as heat maps, for identical ranges of $\sigma_p$ and $T_{\mathrm{qent}}$ values as for entanglement entropy $S$ 
in Figure \ref{fig:entanglemententropyheatmap}, for
signal carrier frequency $\bar{\omega}_{\mathrm{S}}$ resonant withe $\omega_{\alpha}$ in panel (a), and  $\omega_{\beta}$ in panel (b). 
Akin to TLS estimation, we find that a higher value of the entanglement entropy of the incoming PDC state, for the parameter ranges considered, yields a higher $J$-QFI.~(see Figure \ref{fig:scatterplots} (c)-(d) in Appendix \ref{app:parametricplots} for parametric $\mathcal{Q}(J;\ket{\Phi_{\mathrm{PDC,out}}})-S$ plots.)
This implies again that, within the set of PDC probes, more time-frequency entanglement enhances the spectroscopic performance of $J$-estimation.

We also see that the values of $J$-QFI for $\bar{\omega}_{\mathrm{S}} = \omega_{\alpha}$~(so that the signal beam is resonant with the $g$-$\alpha$ transition) are smaller than that for the choice $\bar{\omega}_{\mathrm{S}} = \omega_{\beta}$~(signal beam resonant with $g$-$\beta$ transition). This can be attributed to our choice of the small absolute value of $|\hbar(\omega_a-\omega_b)|$ relative to $\hbar\omega_a$ and $\hbar\omega_b$, meaning that the particular instance of the CD system that we are studying, 
for which $2|\omega_a-\omega_b|/(\omega_a+\omega_b) \approx 0.11$, is quite close to a homodimer for which the $g$-$\alpha$ transition is forbidden by the structure of the GSM-SEM dipole operator. Therefore, population transfer from $\ket{g}\rightarrow\ket{\alpha}$ for $\bar{\omega}_{\mathrm{S}} = \omega_{\alpha}$ is much smaller than for $\ket{g}\rightarrow\ket{\beta}$ when $\bar{\omega}_{\mathrm{S}} = \omega_{\beta}$, which accounts for the smaller values of the $J$-QFI.

\begin{figure}
         \centering
         \includegraphics[width=0.49\textwidth]{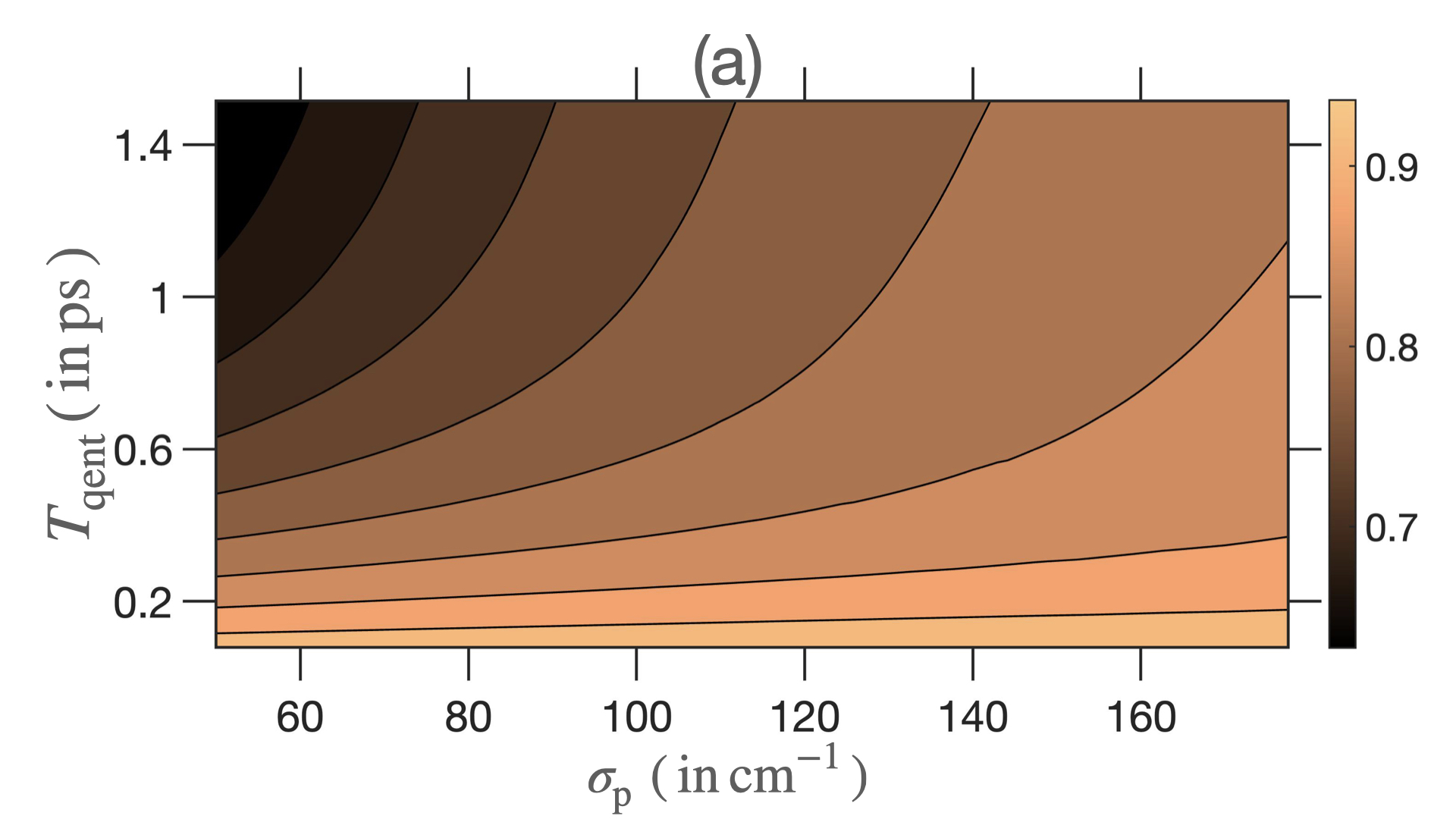}
         \includegraphics[width=0.49\textwidth]{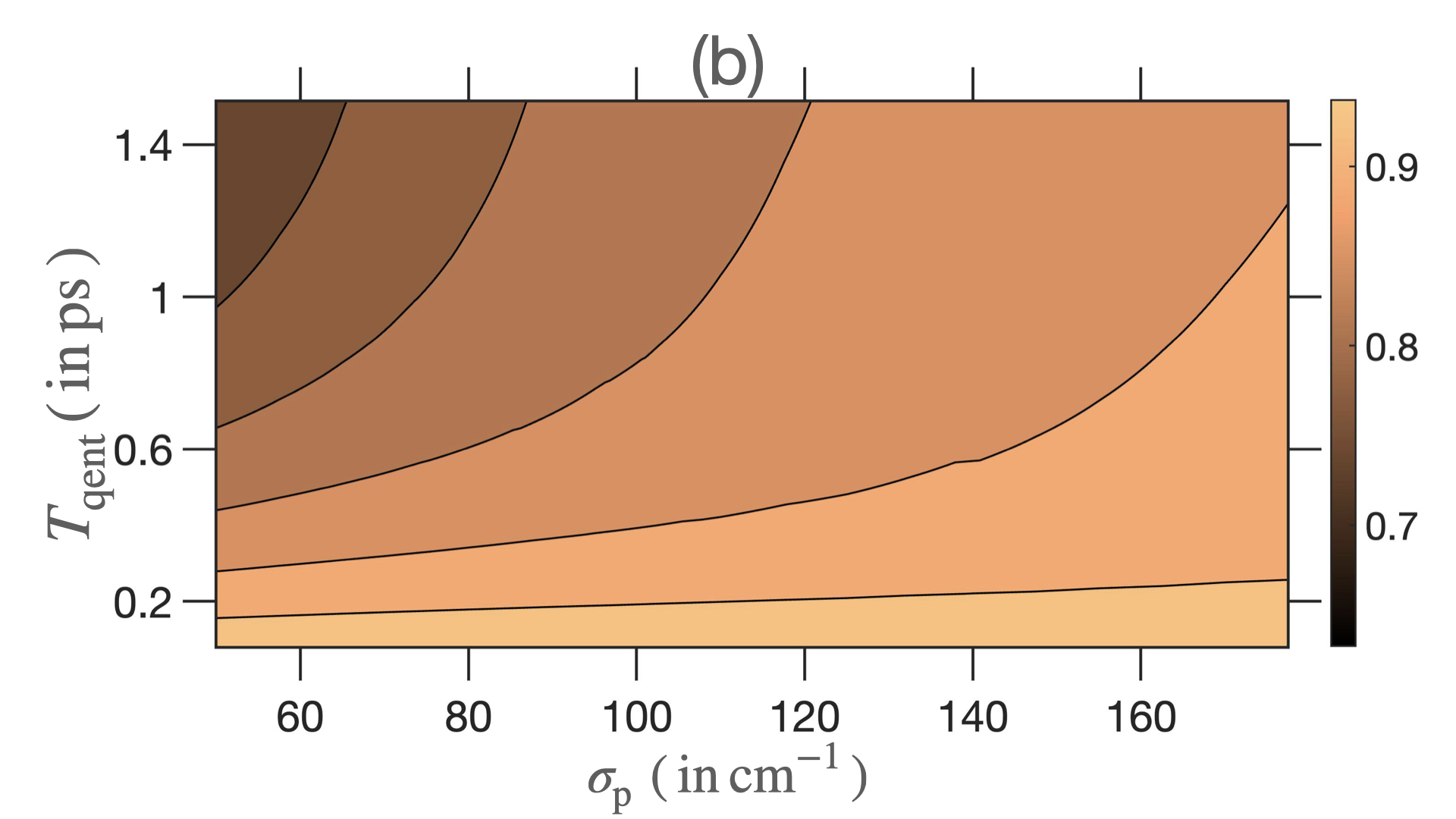}
    \caption{Degree of optimality $\varkappa(J)$ for estimation of CD parameter $J$, with (a)\,$\bar{\omega}_{\mathrm{S}} = \omega_{\alpha}$, and (b)\,$\bar{\omega}_{\mathrm{S}} = \omega_{\beta}$ using PDC light, for  $\Gamma=0.15$\,THz,\,$\Delta = 0$\,THz.}
       \label{fig:LOCCCFItoQFIratioJ}
\end{figure}

\begin{figure}[t!]
     \centering
         \includegraphics[width=0.49\textwidth]{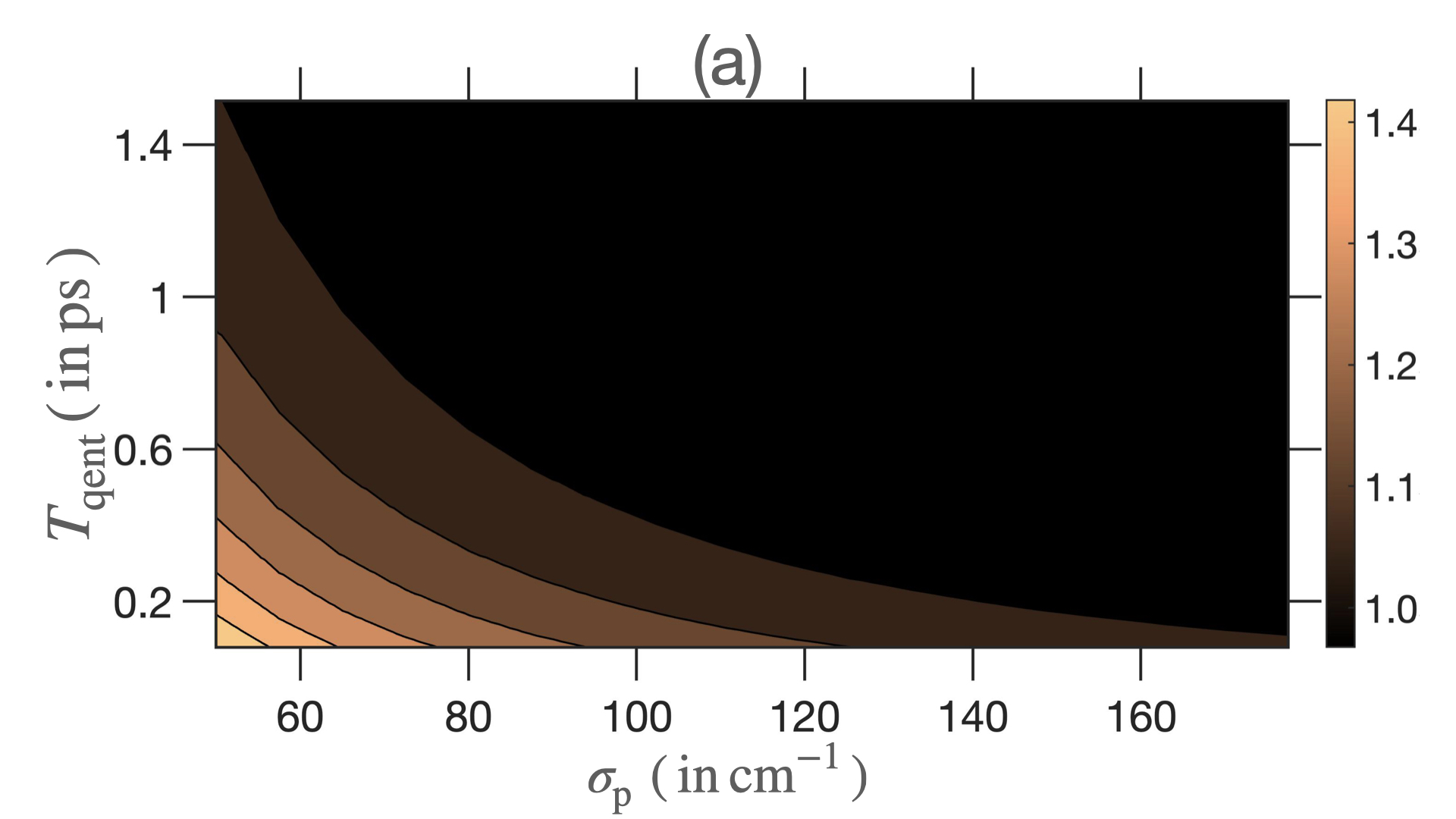}
         \label{fig:PDCZD_OmegaTLS}
         \includegraphics[width=0.49\textwidth]{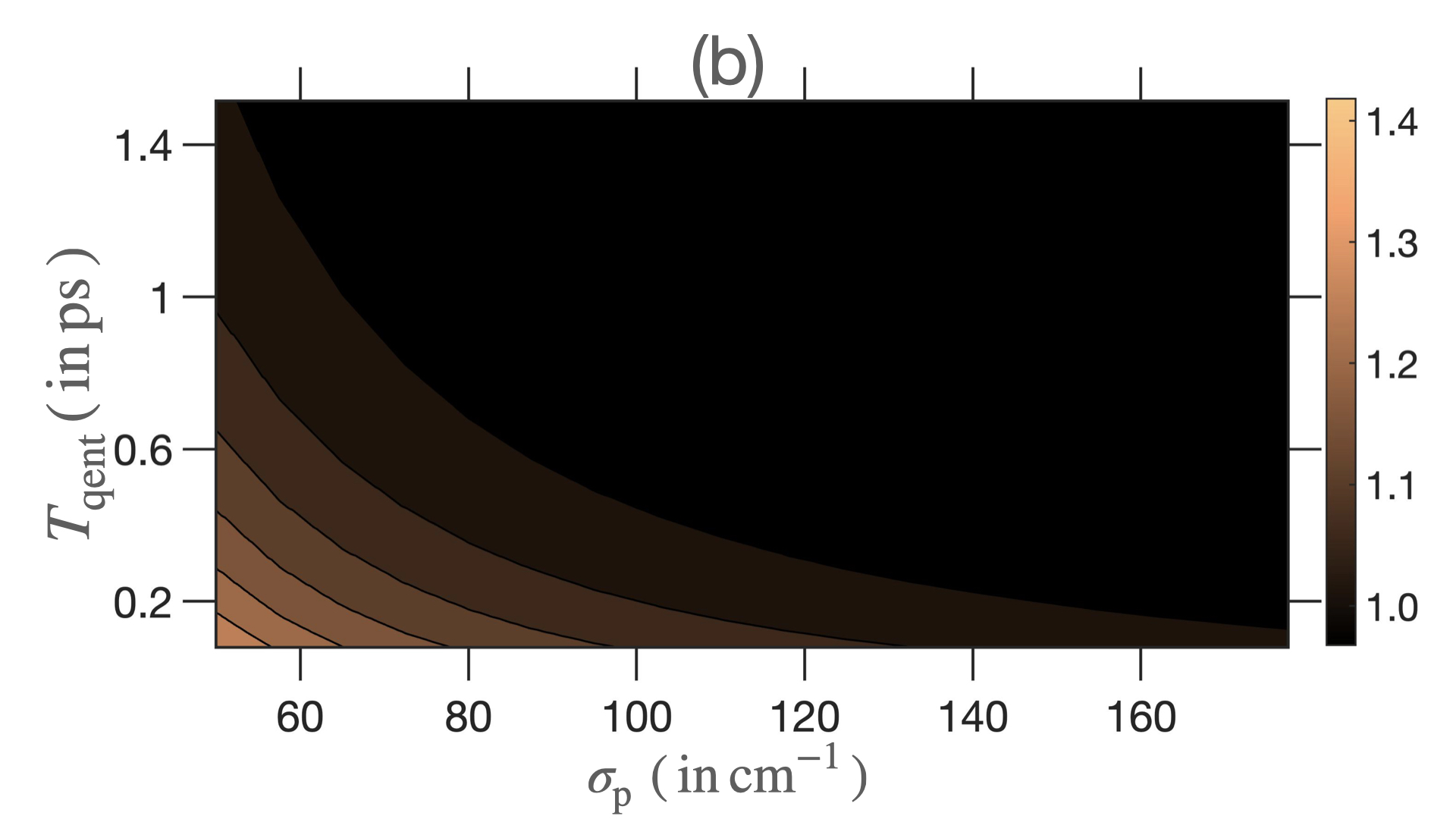}
        \caption{Enhancement factor $\varsigma(J)$ for CD parameter $J$, with (a) $\bar{\omega}_{\mathrm{S}} = \omega_{\alpha}$, and (b) $\bar{\omega}_{\mathrm{S}} = \omega_{\beta}$ using PDC light, for $\Gamma = 0.15$\,THz, $\Delta = 0$\,THz.}
       \label{fig:LOCCCFItotrQFIratioJ}
\end{figure}

Next, Figure \ref{fig:LOCCCFItoQFIratioJ} displays the
ratio $\varkappa(J)$
that captures the degree of optimality of the $V=\mathds{1}^{\mathrm{I}}$ measurement-optimal idler-to-signal scheme. 
For the central signal frequency $\bar{\omega}_{\mathrm{S}}$ set to either $\omega_{\alpha}$ or $\omega_{\beta}$, 
we see that the $V=\mathds{1}^{\mathrm{I}}$ idler-to-signal LOCC scheme recovers between 60\%-90\% of the QFI, especially for highly entangled states in the the bottom right corner.

Finally, Figure \ref{fig:LOCCCFItotrQFIratioJ} displays the enhancement factor $\varsigma(J)$.
For all values on the grid except for most highly entangled states with small values of $\sigma_p$ and $T_{\mathrm{qent}}$, 
$\varsigma(J)$ is close to 
unity.
As for TLS spectroscopy, CD spectroscopy with only the most entangled of PDC states can meaningfully outperform single-photon spectroscopy using the reduced state of the signal photon only, $\mathrm{Tr}_{\mathrm{I}}[\ket{\Phi_{\mathrm{PDC,out}}}]$.

\section{Conclusions}
\label{sec7}

Our quantum-information theoretic analysis of single-molecule biphoton spectroscopy of arbitrary quantum systems provides a characterisation of all the spectroscopic information that exists and can, in principle, be extracted using a biphoton probe in the long-time regime, when the excitation induced by the input pulse in the matter system has decayed.
This allows the design of simple unentangled measurements that can provide tangible quantum advantage in practice.

We provide a detailed analysis of the theoretical and experimental utility of time-frequency entanglement in single-molecule biphoton spectroscopy.
With the latter in mind, we compare the performance of biphoton spectroscopy to those with unentangled probes, especially single photons, and unentangled measurements. 
This reveals the subtle and intricate role entanglement can play in enhancing spectroscopic performance. 

Note added: A recent work~\citep{ko2023performing} examined the usefulness of quantum entangled light vis-\`{a}-vis classical coherent pulses in a non-linear generalisation of the biphoton setup in Figure \ref{fig:biphotonspec_diagram}. 
It focuses on the absorption signal~(hence small detection times) for ensemble systems, whereas we study fundamental limits of spectroscopic information for coherent single-molecule spectroscopies and asymptotically long detection times.

\begin{acknowledgments}
We thank Elnaz Darsheshdar and Sourav Das for fruitful discussions and feedback on the manuscript.
This work has been funded, in part, by an EPSRC New Horizons grant (EP/V04818X/1) and the UKRI (Reference Number: 10038209) under the UK Government's Horizon Europe Guarantee for the Research and Innovation Programme under agreement 101070700.
AK was supported, in part, by a Chancellor's International Scholarship from the University of Warwick.
Computing facilities were provided by the Scientific Computing Research Technology Platform of the University of Warwick.
\end{acknowledgments}

\bibliography{bibliography}

\appendix

\widetext
\section{Explicit Expressions For QFI of $\ket{\Phi_{\mathrm{biph,out}}}$}\label{appendix:twinQFIdetails}

\noindent The derivative of $\ket{\Phi_{\mathrm{biph,out}}}$ is 
\begin{equation}
    \ket{\partial_{\theta}\Phi_{\mathrm{biph,out}}} = -\frac{N_{\theta}}{2(1-N)^{3/2}}\,\sum_{m}\,r_m\ket{\phi_m^{\mathrm{S}}}\ket{\xi_m^{\mathrm{I}}} + \frac{1}{\sqrt{1-N}}\,\sum_m\,r_m\,\ket{\partial_{\theta}\phi_m^{\mathrm{S}}}\ket{\xi_m^{\mathrm{I}}}.
\end{equation}
Then we can get the two terms of the pure state QFI as 
\begin{equation}
    \langle \partial_{\theta}\Phi_{\mathrm{biph,out}}|\partial_{\theta}\Phi_{\mathrm{biph,out}}\rangle = \frac{N_{\theta}^2}{4(1-N)^2} + \frac{1}{1-N}\,\sum_m\,|r_m|^2\langle \partial_{\theta} \phi_m^{\mathrm{S}} | \partial_{\theta} \phi_m^{\mathrm{S}} \rangle - \frac{N_{\theta}}{(1-N)^2}\,\sum_m\,|r_m|^2\,\mathrm{Re}\langle \phi_m^{\mathrm{S}}|\partial_{\theta}\phi_m^{\mathrm{S}}\rangle,
\end{equation}
and
\begin{align}
    \langle \Phi_{\mathrm{biph,out}}|\partial_{\theta} \Phi_{\mathrm{biph,out}}\rangle &= -\frac{N_{\theta}}{2(1-N)^2}\,\sum_m\,|r_m|^2\,\langle\phi_m^{\mathrm{S}}|\phi_m^{\mathrm{S}}\rangle + \frac{1}{1-N}\,\sum_m\,|r_m|^2\,\langle\phi_m^{\mathrm{S}}|\partial_{\theta}\phi_m^{\mathrm{S}}\rangle \nonumber\noindent\\
    &= -\frac{N_{\theta}}{2(1-N)} +  \frac{1}{1-N}\,\sum_m\,|r_m|^2\,\langle\phi_m^{\mathrm{S}}|\partial_{\theta}\phi_m^{\mathrm{S}}\rangle .
\end{align}
The QFI, using Eq.~\eqref{eq:qfipure}, is 
\begin{align}\label{eq:explicittwophotonQFI}
    \mathcal{Q}(\theta;\ket{\Phi_{\mathrm{biph,out}}}) &= \cancel{\frac{N_{\theta}^2}{4(1-N)^2}} + \frac{1}{1-N}\,\sum_m\,|r_m|^2\,\langle \partial_{\theta} \phi_m^{\mathrm{S}} | \partial_{\theta} \phi_m^{\mathrm{S}} \rangle - \cancel{\frac{N_{\theta}}{(1-N)^2}\,\sum_m\,|r_m|^2\,\mathrm{Re}\langle \phi_m^{\mathrm{S}}|\partial_{\theta}\phi_m^{\mathrm{S}}\rangle}  \nonumber\noindent\\
    &- \cancel{\frac{N_{\theta}^2}{4(1-N)^2}} - \frac{1}{(1-N)^2}\left| \sum_m \, |r_m|^2\, \langle\phi_m^{\mathrm{S}}|\partial_{\theta}\phi_m^{\mathrm{S}}\rangle \right|^2 + \cancel{\frac{N_{\theta}}{(1-N)^2}\sum_m\,|r_m|^2\,\mathrm{Re}\langle\phi_m^{\mathrm{S}}|\partial_{\theta}\phi_m^{\mathrm{S}}\rangle} \nonumber\noindent\\
    &= \frac{1}{1-N}\,\sum_m\,|r_m|^2\,\langle \partial_{\theta} \phi_m^{\mathrm{S}} | \partial_{\theta} \phi_m^{\mathrm{S}} \rangle - \frac{1}{(1-N)^2}\left| \sum_m \, |r_m|^2\, \langle\phi_m^{\mathrm{S}}|\partial_{\theta}\phi_m^{\mathrm{S}}\rangle \right|^2.
\end{align}

\section{Relative Magnitudes of QFI Contributions in Table~\ref{table1}}
\label{app:relativemag_orders}

\subsection{$\Gamma_{\perp}\ll\Gamma$}
We can calculate orders of the various terms in Eq.~(\ref{eq:rhoPouttwinQFI}) by letting M-E coupling strength $\Gamma_{\perp}\rightarrow0$ while M-S coupling strength $\Gamma$ remains finite in this limit.

Expanding the matrix exponential in the characteristic response function of the molecule M, defined in Eq.~(\ref{eq:characteristicfn_main}), we have $f_{\mathrm{M}}(t) \propto O(1)$, in orders of $\Gamma$ as well as $\Gamma_{\perp}$, which in turn means $N\propto O(\Gamma\Gamma_{\perp})$. 

The orders of the parametric $N$-derivative, $N_{\theta} = \partial N/\partial \theta$ depend on the parameter of interest. For $\theta\equiv\Gamma$,  $N_{\Gamma} \propto O(\Gamma_{\perp})$, while for molecular Hamiltonian $H_{I}^{\mathrm{M}}$ parameters, it can be worked out that $N_{\theta} \propto O(\Gamma\Gamma_{\perp})$. This then gives, for $\Gamma$ parameter, $\mathcal{C}(N,1-N)\propto O(\Gamma_{\perp}^2/\Gamma\Gamma_{\perp})$, meaning $\lim_{\,\Gamma_{\perp}/\Gamma\rightarrow 0\,} \mathcal{C}(N,1-N) = 0$. For the molecular Hamiltonian parameters, we get similarly, $\mathcal{C}(N,1-N)\propto O(\Gamma^2\Gamma_{\perp}^2/\Gamma\Gamma_{\perp})$, which again yields $\lim_{\,\Gamma_{\perp}/\Gamma\rightarrow 0\,} \mathcal{C}(N,1-N) = 0$.

Moving on to the conditional idler state, we get from the unity orders of $f_{\mathrm{M}}(t)$ that $\sigma^{\mathrm{I}}\propto O(1)$, which yields, for both $\Gamma$ as well as molecular Hamiltonian parameters, $L_{\theta}\propto O(1)$, and $N\mathcal{Q}(\Gamma;\sigma^{\mathrm{I}})\propto O(\Gamma\Gamma_{\perp})$, yielding a vanishing contribution in the limit of $\Gamma_{\perp}/\Gamma \rightarrow 0$. 

\subsection{$\Gamma_{\perp}\gg\Gamma$}

Next, we establish the vanishingly small contribution of the biphoton QFI term $(1-N)\,\mathcal{Q}(\theta;\ket{\Phi_{\mathrm{biph,out}}})$ in Eq.~(\ref{eq:rhoPouttwinQFI}) in the limit of $\Gamma_{\perp}\gg\Gamma$. This limit can be interpreted, in turn, as $\Gamma_{\perp}\rightarrow 0$ for finite magnitudes of $\Gamma_{\perp}$, which gives $\ket{\phi_n^{\mathrm{S}}} \rightarrow \ket{\xi_n^{\mathrm{S}}}$, so that $\ket{\partial_{\theta}\phi_n^{\mathrm{S}}} \rightarrow 0$. Further, utilising again the fact that $f_{\mathrm{M}}(t) \propto O(1)$, we have $N\propto O(\Gamma\Gamma_{\perp})$. Putting these together, we have $(1-N)\,\mathcal{Q}(\theta;\ket{\Phi_{\mathrm{biph,out}}})\rightarrow 0$, following from Eq.~(\ref{eq:biphotonQFIexpr}).

\section{Signal-to-Idler 1-LOCC Scheme}\label{appendix:S2Iprepareandmeasure}

\begin{figure}[h!]
        \includegraphics[width=0.75\textwidth]{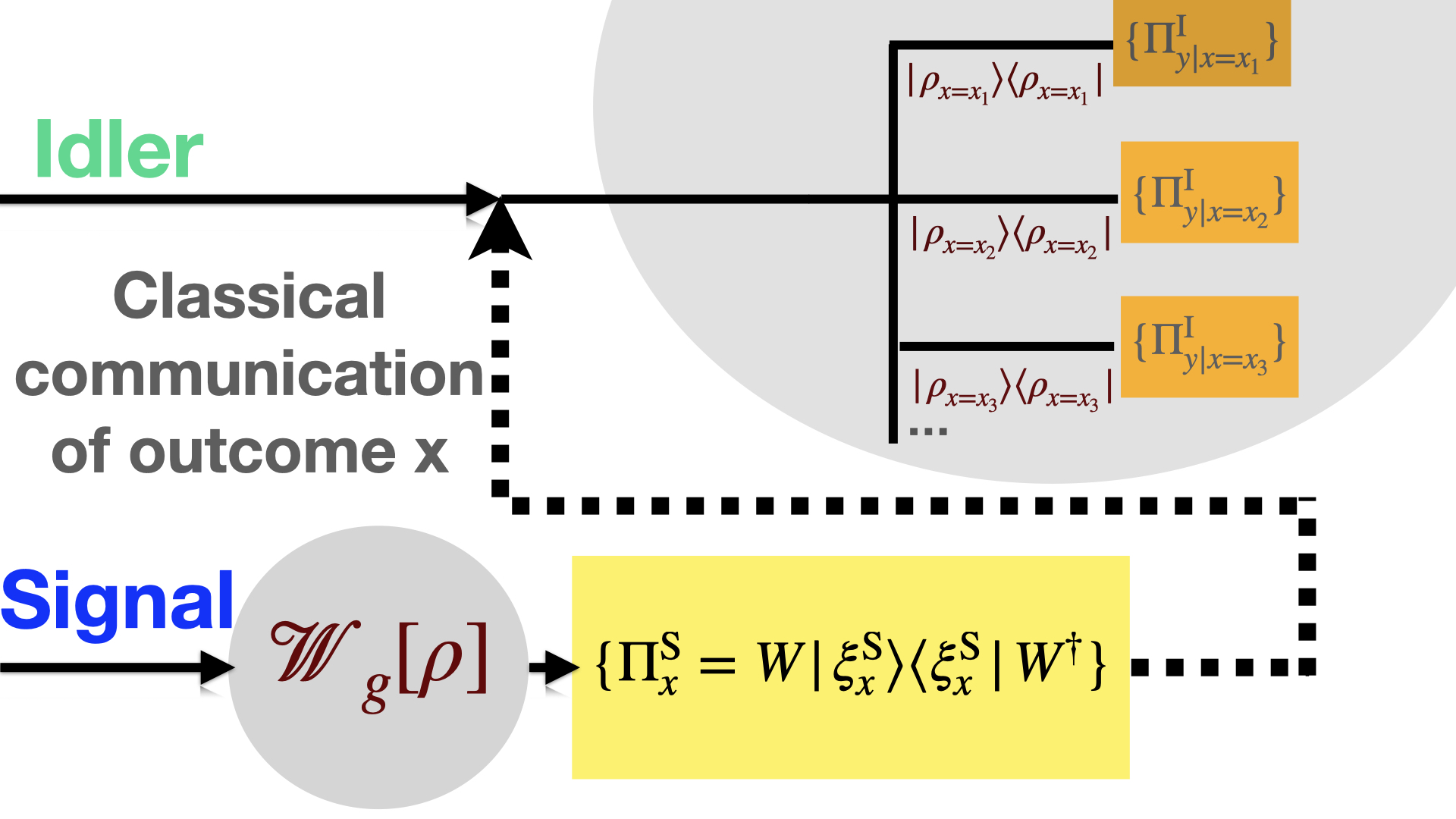}
        \caption[Signal-to-idler prepare-and-measure LOCC scheme.]{Signal-to-idler prepare-and-measure LOCC scheme. The two-photon Kraus operators for the overall LOCC scheme are given as $\{\Pi_{x,y} = \Pi_{x}^{\mathrm{S}}\otimes \Pi_{y|x}^{\mathrm{I}}$\}}
        \label{fig:signaltoidlerscheme}
\end{figure}

This class of signal-to-idler 1-LOCC protocol proceeds in the following three steps~(illustrated schematically in Figure \ref{fig:signaltoidlerscheme}):
\begin{itemize}
    \item (\textbf{Prepare}): Projectively measure the signal photon in arbitrary basis $\{W\ket{\xi_x^{\mathrm{S}}}\bra{\xi_x^{\mathrm{S}}}W^{\dag}\}$, where $W$ is a unitary operator on signal Hilbert space, \textit{after} the M-S interaction effected by superoperator $\mathcal{W}_g$ in Eq.~(\ref{eq:LMsuperoperator}). This amounts to the following transformation of the entangled two-photon state, given by Kraus elements $\Pi_x^{\mathrm{S}} = W\ket{\xi_x^{\mathrm{S}}}\bra{\xi_x^{\mathrm{S}}}W^{\dag}\otimes\mathds{1}^{\mathrm{I}}$,
    \begin{align}
        \rho_{\mathrm{out}}'[W] &= \sum_x\,\Pi_x^{\mathrm{S},\dag}\,\ket{\Phi_\mathrm{biph,out}}\bra{\Phi_{\mathrm{biph,out}}}~\Pi_x^{\mathrm{S}}\noindent\nonumber\\
        &=  \sum_x\, W\ket{\xi_x^{\mathrm{S}}}\bra{\xi_x^{\mathrm{S}}}W^{\dag}\otimes\ket{\rho_x}\bra{\rho_x},\,\ket{\rho_x} = \frac{1}{\sqrt{1-N}} \sum_n\,r_n \langle\xi_x^{\mathrm{S}}|W^{\dag}|\phi_n^{\mathrm{S}}\rangle\,\ket{\xi_n^{\mathrm{I}}},
    \end{align}
    where $\ket{\rho_x}$ are conditional states of the idler photon, the parametric dependence on the parameter $\theta$ entering through the coefficient $\langle\xi_x^{\mathrm{S}}|W^{\dag}|\phi_m^{\mathrm{S}}\rangle$, as well as the $\theta$-dependent quantity $N$.
    \item The outcomes of the projective measurement $\{\Pi_x^{\mathrm{S}}\}$ is classically communicated to the idler substation. 
    \item (\textbf{Measure}):The idler photon ensemble is partitioned into sub-ensembles~(labelled by the signal measurement outcome $x=x_m$) that are in the conditional states $\ket{\rho_{x=x_m}}\bra{\rho_{x=x_m}}$. These are now detected in measurement bases that depends on the preparation step outcome $\{\Pi^{\mathrm{I}}_{y|x=x_m}\}$. The joint signal-to-idler 1-LOCC projector is then $\Pi_{x,y} = \Pi_x^{\mathrm{S}}\otimes\Pi_{y|x}^{\mathrm{I}}$.
\end{itemize}
The above  prepare-and-measure 1-LOCC detection scheme is now one-way going signal-to-idler because the results of the signal measurement are communicated to the idler substation. An important point of difference from the idler-to-signal LOCC scheme, detailed in Section \ref{sec:idlertosignal}, is that the preparation-step quantum operation given by Kraus operators $\{\Pi_x^{\mathrm{S}}\}$ does not commute with the M-S interaction $\mathcal{W}_g$ as they happen in the same substation of the setup. It is meaningful then only to perform the preparation-step operation $\{\Pi_x^{\mathrm{S}}\}$ $\textit{after}$ the two-photon state has been encoded with information about the parameter $\theta$.

Analogous to the idler-to-signal LOCC scheme, we can find the optimal signal-to-idler LOCC measurements by maximizing the associated detection CFI in two steps --- first over all idler measurement strategies for the subensembles $\{\ket{\rho_x}\}$ for a given $W$, and subsequently over all preparation-step signal unitary operators $W$. Formally, we can state the maximisation analogously to idler-to-signal 1-LOCC in Eq.~(\ref{eq:optimise1LOCCCFI}) as:
\begin{align}\label{eq:optimise1LOCCCFI_s2i}
& \max_{\Pi_{y|x}^{\mathrm{I}},\Pi_x^{\mathrm{S}}}\,\mathcal{C}(\theta | \{ \Pi_{x,y} = \Pi_x^{\mathrm{S}}\otimes\Pi_{y|x}^{\mathrm{I}}\}) = \max_{\Pi_x^{\mathrm{S}}}\,\,\left(\max_{\Pi_{y|x}^{\mathrm{I}}}\,\,\mathcal{C}(\theta | \{ \Pi_{x,y} = \Pi_x^{\mathrm{S}}\otimes\Pi_{y|x}^{\mathrm{I}} \})\right) ~ \mathrm{s.t.}\,\,\sum_x\,\Pi_x^{\mathrm{S}} = \mathds{1}^{\mathrm{I}},\,\,\sum_{y}\Pi_{y|x}^{\mathrm{S}} = \mathds{1}^{\mathrm{I}}\,\forall\,\,x.
\end{align}

For a \textit{fixed} preparation step unitary $W$~(which fixes $\Pi_x^{\mathrm{S}}$), the maximal CFI of all measurement-step detection measurements on the idler photon is equal to the QFI of the intermediate state  $\rho_{\mathrm{out}}'[W]$ as maximisation over $\{\Pi_{y|x}^{\mathrm{I}}\}$ is precisely the maximisation that yields the quantum Cram\'{e}r-Rao bound~\citep{braunstein1994statistical,paris2009quantum} for the conditional state $\rho_{\mathrm{out}}'[W]$~(in analogy with the idler-to-signal measurement-optimal 1-LOCC):
\begin{equation}
    \mathcal{C}_{\mathrm{max}}(\theta;W) = \mathcal{Q}(\theta;\rho_{\mathrm{out}}'[W]) = \mathcal{Q}(\theta;\sum_x\,W\ket{\xi_x^{\mathrm{S}}}\bra{\xi_x^{\mathrm{S}}}W^{\dag}\otimes\ket{\rho_x}\bra{\rho_x}).
\end{equation}
where $\mathcal{C}_{\mathrm{max}}(\theta;W)$ is the CFI of measurement-optimal LOCC strategy for a given $W$, defined as the LOCC protocol that has maximal CFI for fixed $W$. By renormalizing the conditional states $\ket{\rho_x'} = \frac{1}{\sqrt{\langle\rho_x|\rho_x\rangle}}\ket{\rho_x}$ in order to express $\rho_{\mathrm{out}}'[W]$ in its spectral form, we can use Eq.~(\ref{eq:qfiarbitraryrank}) to evaluate the measurement-step-maximal CFI explicitly, 
\begin{align}\label{eq:prepnmeasureLOCC2}
    \mathcal{C}_{\mathrm{max}}(\theta;W) &=  \sum_x\,\frac{(\partial_{\theta}\langle\rho_x|\rho_x\rangle)^2}{\langle\rho_x|\rho_x\rangle}  + \frac{4}{1-N}\,\sum_m |r_m|^2 \langle\partial_{\theta}\phi^{\mathrm{S}}_m|\partial_{\theta}\phi^{\mathrm{S}}_m\rangle - \sum_x\, \langle\rho_x|\rho_x\rangle \,\left| \langle\rho_x'|\partial_{\theta}\rho_x'\rangle  \right|^2 \nonumber\noindent\\
    &=  \frac{4}{1-N}\,\sum_m |r_m|^2 \langle\partial_{\theta}\phi^{\mathrm{S}}_m|\partial_{\theta}\phi^{\mathrm{S}}_m\rangle  - \frac{4}{(1-N)^2}\sum_x\frac{1}{\langle\rho_x|\rho_x\rangle}\,\left( \mathrm{Im}\langle\rho_x|\partial_{\theta}\rho_x\rangle^2 - \mathrm{Re}\langle\rho_x|\partial_{\theta}\rho_x\rangle^2   \right) 
\end{align}
where 
\begin{equation}
\langle\rho_x|\rho_x\rangle = \frac{1}{1-N}\sum_m\,|r_m|^2\,|\langle\xi_x^{\mathrm{S}}|W^{\dag}|\phi_m^{\mathrm{S}}\rangle|^2
\end{equation}
has parametric dependence on $\theta$, just as for the idler-to-signal scheme. The overlap can be expressed in the more succint form
\begin{equation}
    \langle\rho_x|\partial_{\theta}\rho_x\rangle = \frac{1}{1-N}\langle\xi_x^{\mathrm{S}}|W^{\dag}\,Y\,W\ket{\xi_x^{\mathrm{S}}}
\end{equation}
where
\begin{equation}
    Y(\theta) = \sum_{m, n}\,|r_n|^2\,\langle\phi_m^{\mathrm{S}}|\partial_{\theta}\phi_n^{\mathrm{S}}\rangle\,\ket{\phi_m^{\mathrm{S}}}\bra{\phi_n^{\mathrm{S}}}
\end{equation}
is an operator on the signal Hilbert space. The measurement-optimal CFI for signal-to-idler LOCC scheme for given $W$ then has the following form:
\begin{align}
    \mathcal{C}_{\mathrm{max}}(\theta;W) &=  \frac{4}{1-N}\,\sum_m |r_m|^2 \langle\partial_{\theta}\phi^{\mathrm{S}}_m|\partial_{\theta}\phi^{\mathrm{S}}_m\rangle  \nonumber\noindent\\
 &- \frac{4}{(1-N)^2}\sum_x\,\frac{1}{\sum_m\,|r_m|^2\,|\langle\xi_x^{\mathrm{S}}|W^{\dag}|\phi_m^{\mathrm{S}}\rangle|^2}\,\,\left( \mathrm{Im}(\langle\xi_x^{\mathrm{S}}|W^{\dag}\,Y\,W\ket{\xi_x^{\mathrm{S}}})^2 -  \mathrm{Re}(\langle\xi_x^{\mathrm{S}}|W^{\dag}\,Y\,W\ket{\xi_x^{\mathrm{S}}})^2        \right)   
\end{align}
Note both the similarity of expressions to the idler-to-signal LOCC case in Eq.~(\ref{eq:prepnmeasureLOCC1}), as well as the differences -- the expressions for measurement-step-maximal CFI are not the same because the biphoton setup is not symmetrical, as we have remarked before.

We can now similarly obtain the overall optimal signal-to-idler LOCC measurement that fetches the maximal value of measurement CFI for \emph{any} $W$. This is done by maximizing the functional $\mathcal{C}_{\mathrm{max}}(\theta;W) $ over the set of all unitary matrices $W$, which amounts to a minimisation of the second term in Eq.~(\ref{eq:prepnmeasureLOCC2}). Defining the cost function as 
\begin{align}\label{eq:costfndef_appendix}
    \varpi(\theta;W) = \frac{4}{(1-N)^2}\sum_x\,\frac{1}{\sum_m\,|r_m|^2\,|\langle\xi_x^{\mathrm{S}}|W^{\dag}|\phi_m^{\mathrm{S}}\rangle|^2}\,\,\left( \mathrm{Im}(\langle\xi_x^{\mathrm{S}}|W^{\dag}\,Y\,W\ket{\xi_x^{\mathrm{S}}})^2 -  \mathrm{Re}(\langle\xi_x^{\mathrm{S}}|W^{\dag}\,Y\,W\ket{\xi_x^{\mathrm{S}}})^2        \right) ,
\end{align}
where $\varpi(\theta;W)\geq0$~(which follows from $\mathcal{Q}(\theta;\rho_{\mathrm{out}}'[W])\geq0$), and  the optimal unitary transformation $W_{\mathrm{opt}}$ is defined as:
\begin{align}\label{eq:optimizations2i}
    W_{\mathrm{opt}} : \varpi(\theta;W_{\mathrm{opt}}) \leq \varpi(\theta;W)~ \forall~ W\,\mathrm{s.t.}\,W^{\dag}W = \mathds{1}^{\mathrm{S}}.
\end{align}
The overall maximal CFI for idler-to-signal prepare-and-measure LOCC is then
\begin{equation}
    \mathcal{C}_{\mathrm{S}\rightarrow \mathrm{I}}(\theta) = \frac{4}{1-N}\,\sum_m\,|r_m|^2\,\langle\partial_{\theta}\phi_m^{\mathrm{S}}|\partial_{\theta}\phi_m^{\mathrm{S}}\rangle - \varpi(\theta;W_{\mathrm{opt}}).
\end{equation}

\subsubsection{Lower Bound}
Just like the idler-to-signal scenario, the measurement-optimal CFI $\mathcal{C}_{\mathrm{max}}(\theta;W)$ can be lower bounded using the convexity of the quantum Fisher information:
\begin{align}\label{eq:lowerbounds2i}
    \mathcal{C}_{\mathrm{max}}(\theta;W) &= \mathcal{Q}(\theta;\rho_{\mathrm{out}}'[W]) = \mathcal{Q}\bigg(\theta;\sum_x\,W\ket{\xi_x^{\mathrm{S}}}\bra{\xi_x^{\mathrm{S}}}W^{\dag}\otimes\ket{\rho_x}\bra{\rho_x}\bigg) \nonumber\noindent\\
    &\geq \mathcal{Q}(\theta;\mathrm{Tr}_{\mathrm{I}}\rho_{\mathrm{out}}'[W]) = \mathcal{Q}\left(\theta;\sum_x\,\langle\xi_x^{\mathrm{S}}|\,W^{\dag}\,\mathrm{Tr}_{\mathrm{I}}\ket{\Phi_{\mathrm{biph,out}}}\bra{\Phi_{\mathrm{biph,out}}}\,W|\xi_x^{\mathrm{S}}\rangle~W\ket{\xi_x^{\mathrm{S}}}\bra{\xi_x^{\mathrm{S}}}W^{\dag}\right) \nonumber\noindent\\
    &= \mathcal{C}(\theta;\langle\xi_x^{\mathrm{S}}|W^{\dag}\,\mathrm{Tr}_{\mathrm{I}}\ket{\Phi_{\mathrm{biph,out}}}\bra{\Phi_{\mathrm{biph,out}}}W|\xi_x^{\mathrm{S}}\rangle) \geq \mathcal{Q}(\theta;\mathrm{Tr}_{\mathrm{I}}\ket{\Phi_{\mathrm{biph,out}}})
\end{align}
\noindent which is identical to the lower bound in Eq.~(\ref{eq:i2sCFIlowerbound}) for the idler-to-signal protocol QFI. The second inequality in the above chain may always be saturated for single-parameter estimation with an appropriate choice for $W$. By the same reasoning as the idler-to-signal case, the first inequality is saturated iff $\langle\phi_m^{\mathrm{S}}|\partial_{\theta}\phi_n^{\mathrm{S}}\rangle = 0\,\,\forall\,\,m,n$~(see Eq.~(\ref{eq:qfiarbitraryrank})). For a Schmidt basis $\{\ket{\phi_n^{\mathrm{S}}}\}$ that is complete on the signal Hilbert space, this is never satisfied, and we get the stronger inequality
\begin{equation}
    \mathcal{C}_{\mathrm{max}}(\theta;W) > \mathcal{Q}(\theta;\mathrm{Tr}_{\mathrm{I}}\ket{\Phi_{\mathrm{biph,out}}}).
\end{equation}
This then leads to the following hierarchy of Fisher informations:
\begin{equation}\label{eq:s2ihierarchy}
    \mathcal{Q}(\theta;\ket{\Phi_{\mathrm{biph,out}}}) \geq \mathcal{C}_{\mathrm{S} \rightarrow \mathrm{I}}(\theta) \geq \mathcal{C}_{\mathrm{max}}(\theta;W) > \mathcal{Q}(\theta;\mathrm{Tr}_{\mathrm{I}}\ket{\Phi_{\mathrm{biph,out}}}).
\end{equation}
Therefore, the reduced signal state QFI $\mathcal{Q}(\theta;\mathrm{Tr}_{\mathrm{I}}\ket{\Phi_{\mathrm{biph,out}}})$ serves as a useful benchmark for the effectiveness of LOCC parameter estimation using entangled light. For the signal-to-idler scheme, practical difficulties for a setup requiring the preparation step to necessarily follow the light-matter interaction make this LOCC scheme much less attractive as a means for enhanced $\theta$-estimation compared to idler-to-signal scheme, where the idler photon is conditioned by preparation POVMs independent of interaction with the sample and subsequent measurement in the signal mode.

\section{Optimal 1-LOCC Measurement Projectors for Biphoton Setup}\label{app:1loccopt}

As indicated by the RHS of Eq.~(\ref{eq:optimise1LOCCCFI}), the maximisation of the 1-LOCC CFI function can proceed in two steps: first, for a fixed unitary transformation $V$, we maximise over all signal POVMs $\{\Pi^{\mathrm{S}}_{y|x=x_m}\}$; in the second step, the resulting quantity is then maximised over all choices of unitary preparation $V$.

\subsection{Optimisation over signal POVM $\{\Pi^{\mathrm{S}}_{y|x=x_m}\}$}

For a \textit{fixed} $V$, maximisation of the CFI over $\{\Pi^{\mathrm{S}}_{y|x=x_m}\}$ is precisely the maximisation that yields the Cram\'{e}r-Rao bound~\citep{braunstein1994statistical,paris2009quantum} for the conditional state $\rho_{\mathrm{out}}'[V]$, meaning that 
\begin{equation}\label{eq:measurementoptCFI}
    \mathcal{C}_{\mathrm{max}}(\theta;V) = \mathcal{Q}(\theta;\rho_{\mathrm{out}}'[V]).
\end{equation}
Abbreviating normalised conditional states  
\begin{equation}
\ket{\zeta_x'} = \frac{1}{\sqrt{\langle\zeta_x|\zeta_x\rangle}}\ket{\zeta_x},
\end{equation}
and using Eq.~(\ref{eq:qfiarbitraryrank}),
\begin{align}\label{eq:prepnmeasureLOCC1}
    &\mathcal{C}_{\mathrm{max}}(\theta;V) = \sum_x\,\langle\zeta_x|\zeta_x\rangle\,\mathcal{Q}(\theta;\ket{\zeta_x'}\bra{\zeta_x'})  + \mathcal{C}(\theta|\{\langle\zeta_x|\zeta_x\rangle\}),
\end{align}
where
\begin{align}\label{eq:prepnmeasurelocc2}
    &\sum_x\,\langle\zeta_x|\zeta_x\rangle\,\mathcal{Q}(\theta;\ket{\zeta_x'}\bra{\zeta_x'}) = \frac{4}{1-N}\,\sum_m |r_m|^2 \langle\partial_{\theta}\phi^{\mathrm{S}}_m|\partial_{\theta}\phi^{\mathrm{S}}_m\rangle - \frac{1}{(1-N)^2}\sum_x\,\frac{4}{\langle\zeta_x|\zeta_x\rangle} \,\left| \langle\xi_x^{\mathrm{I}}|V^{\dag}X(\theta)V|\xi_x^{\mathrm{I}}\rangle       \right|^2,
\end{align}
and 
\begin{equation}
    X(\theta) = \sum_{m,n} r_m r_n^*\, \langle\phi_n^{\mathrm{S}}|\partial_{\theta}\phi_m^{\mathrm{S}}\rangle\,\ket{\xi_m^{\mathrm{I}}}\bra{\xi_n^{\mathrm{I}}}
\end{equation}
is an operator on the idler I space.
The one other orthonormal basis vector~(besides $\ket{\Phi_{\mathrm{biph,out}}}$)
in the two-dimensional $\mathrm{Span}[\ket{\Phi_{\mathrm{biph,out}}},\ket{\partial_{\theta}\Phi_{\mathrm{biph,out}}}]$ can be constructed as~\cite{kurdzialek2023measurement}
\begin{equation}\label{eq:perpbiph_eqn}
    \ket{\Phi_{\mathrm{biph,out}}^{\perp}} \equiv \left(1-\ket{\Phi_{\mathrm{biph,out}}}\bra{\Phi_{\mathrm{biph,out}}}\right) \,\,\ket{\partial_{\theta}\Phi_{\mathrm{biph,out}}},
\end{equation}
in terms of which 
\begin{align}\label{eq:Xreln}
    X =&\,  (1-N)\,\mathrm{Tr}_{\mathrm{S}}\,\ket{\Phi_{\mathrm{biph,out}}}\bra{\Phi_{\mathrm{biph,out}}^{\perp}}  + \mathrm{Tr}(X)\,\mathrm{Tr}_{\mathrm{S}} \,\ket{\Phi_{\mathrm{biph,out}}}\bra{\Phi_{\mathrm{biph,out}}},
\end{align}
where 
\begin{equation}\label{eq:trXreln}
\mathrm{Tr}(X) = \left[(1-N)\langle \Phi_{\mathrm{biph,out}}|\partial_{\theta}\Phi_{\mathrm{biph,out}}\rangle - \frac{N_{\theta}}{2}\right].
\end{equation}
The relation in Eq.~(\ref{eq:Xreln}) can be obtained using Eq.~(\ref{eq:perpbiph_eqn}), and noting that 
\begin{align}
&\mathrm{Tr}_{\mathrm{S}}\,\ket{\partial_{\theta}\Phi_{\mathrm{biph,out}}}\bra{\Phi_{\mathrm{biph,out}}} = \frac{N_{\theta}}{2(1-N)}\,\mathrm{Tr}_{\mathrm{S}}\,\ket{\Phi_{\mathrm{biph,out}}}\bra{\Phi_{\mathrm{biph,out}}} + \frac{X}{1-N}.
\end{align}
Finally, we can also recast Eq.~(\ref{eq:biphotonQFIexpr}) in terms of $X$:
\begin{equation}\label{eq:qfibiphout_altern}
    \mathcal{Q}(\theta;\ket{\Phi_{\mathrm{biph,out}}}) = \frac{4}{1-N}\,\sum_{n}\,|r_n|^2\,\langle\phi_n^{\mathrm{S}}|\phi_n^{\mathrm{S}}\rangle - \frac{4}{(1-N)^2}|\mathrm{Tr}(X)|^2,
\end{equation}
where we have employed the explicit relation $\mathrm{Tr}(X) = \sum_{n}\,|r_n|^2\,\langle\phi_n^{\mathrm{S}}|\partial_{\theta}\phi_n^{\mathrm{S}}\rangle$.

\subsection{Optimal Choice of Preparation Unitary $V$}

Eq.~\eqref{eq:prepnmeasureLOCC1} is the Fisher information corresponding to the optimal 1-LOCC $\theta$-estimation strategy for a given unitary $V$, obtained by maximising the CFI functional over 
the set of measurement POVMs $\{\Pi_{y|x}^{\mathrm{S}}\}$ acting on the sub-ensemble states of the signal photon $\{\ket{\zeta_x'}\}$. 
To identify the optimal 1-LOCC detection scheme that fetches the maximum CFI, we now proceed to maximise $\mathcal{C}_{\mathrm{max}}(\theta;V)$ over all $V$. 
As only the second term on the RHS of Eq.~(\ref{eq:prepnmeasurelocc2}) depends on $V$, we can see that the maximisation is equivalent to minimisation of the following cost function:
\begin{align}\label{eq:costfndef}
    \vartheta(\theta;V) = \frac{1}{(1-N)^2}\sum_x\frac{4}{\langle\zeta_x|\zeta_x\rangle} \,\left| \langle\xi_x^{\mathrm{I}}|V^{\dag}X(\theta)V|\xi_x^{\mathrm{I}}\rangle       \right|^2 - \mathcal{C}(\theta;\{\langle\zeta_x|\zeta_x\rangle\}).
\end{align}
Using the inequality chain
\begin{equation}
\mathcal{C}_{\mathrm{max}}(\theta;V) \leq \mathcal{C}_{\mathrm{max}}(\theta;V_0) < \mathcal{Q}(\theta,\ket{\Phi_{\mathrm{biph,out}}}),
\end{equation}
we have for the cost function 
\begin{equation}\label{eq:hierarchyCFIQFI}
    \vartheta(\theta;V) \geq \vartheta(\theta;V_{\mathrm{opt}}) > \frac{4}{(1-N)^2}\,|\mathrm{Tr}(X)|^2,
\end{equation}
where $V_{\mathrm{opt}}$ is the optimal unitary, defined formally as
\begin{align}\label{eq:optimizationi2s}
    V_{\mathrm{opt}} : \vartheta(\theta;V_{\mathrm{opt}}) \leq \vartheta(\theta;V)~ \forall~ V\,\mathrm{s.t.}\,V^{\dag}V = \mathds{1}^{\mathrm{I}},
\end{align}
and the second inequality in Eq.~(\ref{eq:hierarchyCFIQFI}) follows from Eq.~(\ref{eq:qfibiphout_altern}).

Our strategy in the following will be to construct a preparation unitary that saturates the latter of the inequalities in Eq.~(\ref{eq:hierarchyCFIQFI}). While the existence of such a unitary~(and hence a 1-LOCC detection) is not guaranteed for general multipartite scenarios, showing that there exists such a unitary preparation for which the inequality is saturated is a \emph{sufficient} condition for maximisation. This follows from the fact that measurement CFIs can never exceed the QFI. 

\subsection{Proof of Theorem \ref{theorem1}}

For the traceless~(in I space) compact bounded operator $\mathrm{Tr}_{\mathrm{S}}\,|\Phi_{\mathrm{biph,out}}\rangle\langle\Phi_{\mathrm{biph,out}}^{\perp}|$, we can always construct~\cite{fillmore1969similarity,zhou2020saturating} a preparation-step unitary $V_0$ such that 
\begin{equation}\label{eq:optconditioni2s}
\bra{\xi_m^{\mathrm{I}}}V_0^{\dag}\,\mathrm{Tr}_{\mathrm{S}}\,\ket{\Phi_{\mathrm{biph,out}}}\bra{\Phi_{\mathrm{biph,out}}^{\perp}}\,V_0\ket{\xi_m^{\mathrm{I}}} = 0\,\,\, \forall m,
\end{equation}
A short calculation then reveals that~(which we will establish separately below)
\begin{equation}\label{eq:saturation}
    \vartheta(\theta;V_0) = \frac{1}{(1-N)^2}|\mathrm{Tr}(X)|^2,
\end{equation}
from which $\mathcal{C}_{\mathrm{max}}(\theta,V_0) = \mathcal{Q}(\theta;\ket{\Phi_{\mathrm{biph,out}}})$ follows, establishing Theorem \ref{theorem1}.
Thus for all unitary preparations $V_0$ that admit the condition in Eq.~(\ref{eq:saturation}), the corresponding measurement-optimal 1-LOCC saturates the utimate quantum Cr\'{a}mer-Rao bound.

\subsubsection{$V=V_0$ saturates cost function $\vartheta(\theta;V)$ bound}\label{sub:CV0iszero}

In order to establish the relation in Eq.~(\ref{eq:saturation}), we first note that
\begin{align}
    \langle \xi_x| V_0^{\dag} X V_0 |\xi_x \rangle &= \cancel{\langle \xi_x | V_0^{\dag}\,(1-N)\mathrm{Tr}_{\mathrm{S}}\,\ket{\Phi^{\perp}_{\mathrm{biph,out}}}\bra{\Phi_{\mathrm{biph,out}}}V_0|\xi_x\rangle} + \mathrm{Tr}(X)\,\langle\xi_x|V_0^{\dag} \mathrm{Tr}_{\mathrm{S}} \,\ket{\Phi_{\mathrm{biph,out}}}\bra{\Phi_{\mathrm{biph,out}}}V_0|\xi_x\rangle \nonumber\noindent\\
    \nonumber\noindent\\
    &=\mathrm{Tr}(X)\,\langle\zeta_x|\zeta_x\rangle.
\end{align}
where we have used the optimality relation for $V_0$ in Eq.~(\ref{eq:optconditioni2s}) in the first line, and the second line utilises the relation
\begin{equation}
    \langle\xi_x|V_0^{\dag}\,\mathrm{Tr}_{\mathrm{S}}\,\ket{\Phi_{\mathrm{biph,out}}}\bra{\Phi_{\mathrm{biph,out}}}V_0|\xi_x\rangle = \langle\zeta_x|\zeta_x\rangle
\end{equation}
which can be easily worked out explicitly. The cost function in Eq.~(\ref{eq:costfndef}) then becomes 
\begin{align}
    \vartheta(\theta;V_0) &= \frac{1}{(1-N)^2}\,\sum_x\,\frac{1}{\langle\zeta_x|\zeta_x\rangle}\,\langle\zeta_x|\zeta_x\rangle^2\,|\mathrm{Tr}(X)|^2 + \mathcal{C}(\theta;\langle\zeta_x|\zeta_x\rangle) \nonumber\noindent\\
    &= \frac{1}{(1-N)^2}\,|\mathrm{Tr}(X)|^2\,\sum_x\,\frac{1}{1-N}\,r_m^*r_n (V_0)_{mx}(V_0)_{nx}^*\,\langle\phi_m|\phi_n\rangle + \mathcal{C}(\theta;\langle\zeta_x|\zeta_x\rangle) = \frac{|\mathrm{Tr}(X)|^2}{(1-N)^2} + \mathcal{C}(\theta;\langle\zeta_x|\zeta_x\rangle)
\end{align}
where we have used the unitarity of the $V_0$ matrix : $\sum_x\,(V_{0})_{mx}(V_0)_{nx}^* = \delta_{mn}$.
The second step will be to establish that the classical Fisher information $\mathcal{C}(\theta;\langle\zeta_x|\zeta_x\rangle)$ corresponding to mixing of the subensemble states $\{\ket{\zeta_x'}\}$ vanishes for unitary $V_0$. In order to do this, we first look at the structure of the inner product
\begin{align}
    \langle\zeta_x|\partial_{\theta}\zeta_x\rangle &= \frac{N_{\theta}}{2(1-N)}\,\langle\zeta_x|\zeta_x\rangle + \frac{1}{1-N}\,\langle\xi_x|V_0^{\dag}XV_0|\xi_x\rangle = \left(\frac{N_{\theta}}{2(1-N)} + \mathrm{Tr}(X)\right)\,\langle\zeta_x|\zeta_x\rangle \nonumber\noindent\\
    &= \langle \Phi_{\mathrm{biph,out}}|\partial_{\theta} \Phi_{\mathrm{biph,out}}\rangle\,\,\langle\zeta_x|\zeta_x\rangle
\end{align}
where we have used the relation $\langle\xi_x|V_0^{\dag}XV_0|\xi_x\rangle = \mathrm{Tr}(X)\,\langle\zeta_x|\zeta_x\rangle$ in the first line, and Eq.~(\ref{eq:trXreln}) in the second line. Now, keeping in mind that the outgoing biphoton state $\ket{\Phi_{\mathrm{biph,out}}}$ is a normalised quantum state, we have $\langle\Phi_{\mathrm{biph,out}}|\Phi_{\mathrm{biph,out}}\rangle = 1$, and thence $\mathrm{Re}\,\,\langle\Phi_{\mathrm{biph,out}}|\partial_{\theta}\Phi_{\mathrm{biph,out}}\rangle = 0$. This then gives
\begin{equation}\label{eq:cfiLOCC_zero}
    \mathcal{C}(\theta;\{\langle\zeta_x|\zeta_x\}) = \sum_x 4\,\frac{\langle\zeta_x|\partial_{\theta}\zeta_x\rangle^2}{\langle\zeta_x|\zeta_x\rangle} =  
 \sum_x\,4\,\langle\zeta_x|\zeta_x\rangle\,\mathrm{Re}\langle\Phi_{\mathrm{biph,out}}|\partial_{\theta}\Phi_{\mathrm{biph,out}}\rangle^2 = 0,
\end{equation}
which proves Eq.~(\ref{eq:saturation}).

Briefly, we also note that a similar result was established recently for multipartite pure and rank $2$ states in finite-dimensional Hilbert spaces~\citep{zhou2020saturating}. 
However, the recipe can not be generalised to our infinite-dimensional CV case due to its dependence on the total system dimension.
We have thus employed a different and more direct approach here that establishes existence of 1-LOCC schemes whose CFI is shown to equal the QFI for the outgoing state.

\section{Parametric Relation Between PDC Entanglement and Outgoing QFI}\label{app:parametricplots}

\begin{figure}[h!]
     \centering  
         \includegraphics[width=\textwidth]{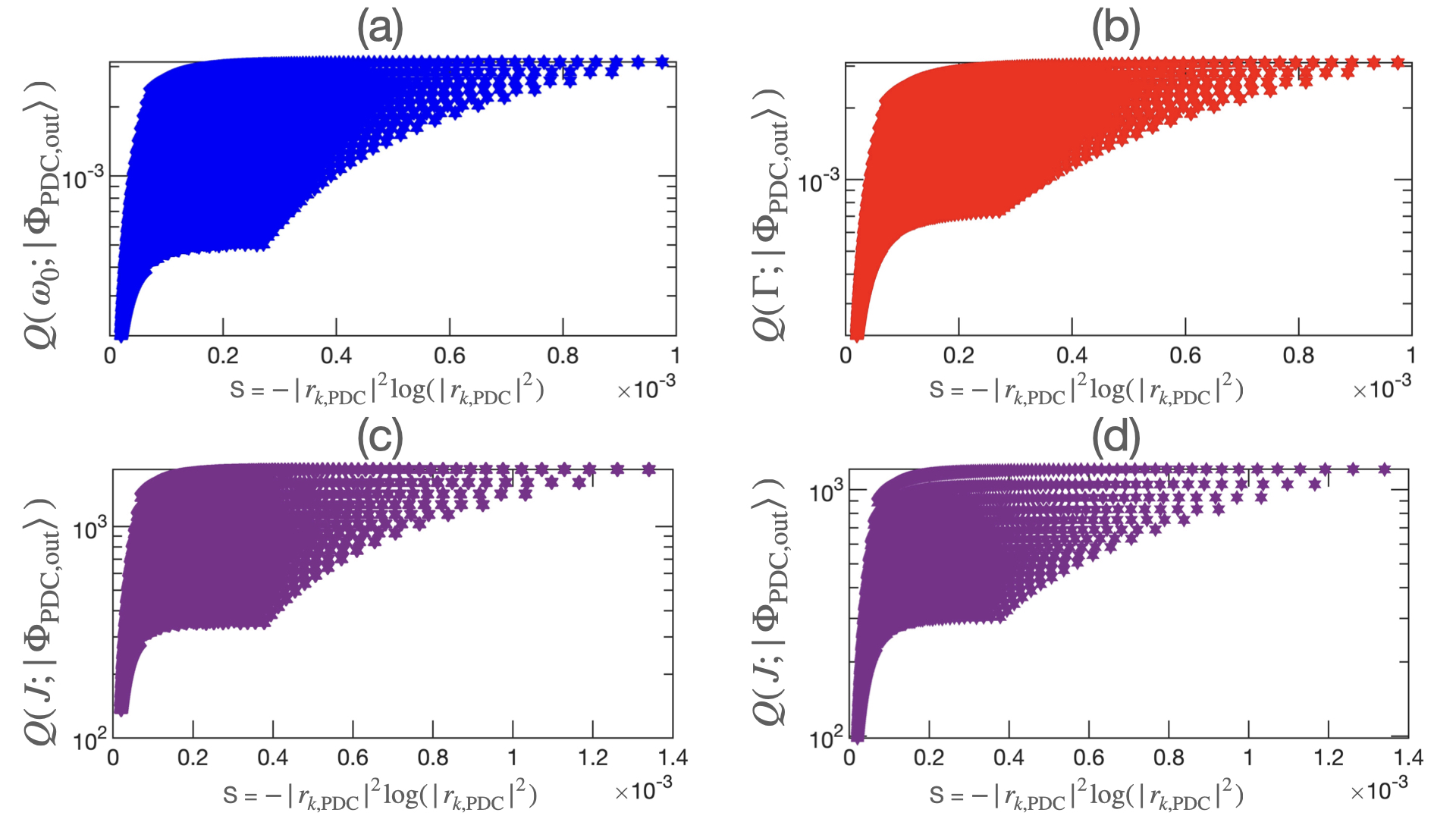}
\caption{Scatter plots of outgoing entangled beam QFI  $\mathcal{Q}(\theta;\ket{\Phi_{\mathrm{PDC,out}}})$, plotted as a function of entanglement entropy $S = -|r_{k,\mathrm{PDC}}|^2\mathrm{log}(|r_{k,\mathrm{PDC}}|^2)$ in the incoming state $\ket{\Phi_{\mathrm{PDC}}}$, for (a) TLS parameter $\theta\equiv\omega_0$~(corresponding to Figure \ref{fig:PDCheatmap_TLS}(a)), (b) TLS parameter $\theta\equiv\Gamma$ (corresponding to Figure \ref{fig:PDCheatmap_TLS}(b)), (c) CD parameter $J$ for $\bar{\omega}_{\mathrm{S}} = \omega_{\alpha}$~(corresponding to Figure \ref{fig:PDCheatmap_J}(a)), and (d) CD parameter $J$ for $\bar{\omega}_{\mathrm{S}} = \omega_{\beta}$~(corresponding to Figure \ref{fig:PDCheatmap_J}(b)).~($\Gamma = 0.15$\,THz\,\,$\Gamma_{\perp} = 0$)}  
\label{fig:scatterplots}
\end{figure}

\begin{figure}[h!]
     \centering   
         \includegraphics[width=0.49\textwidth]{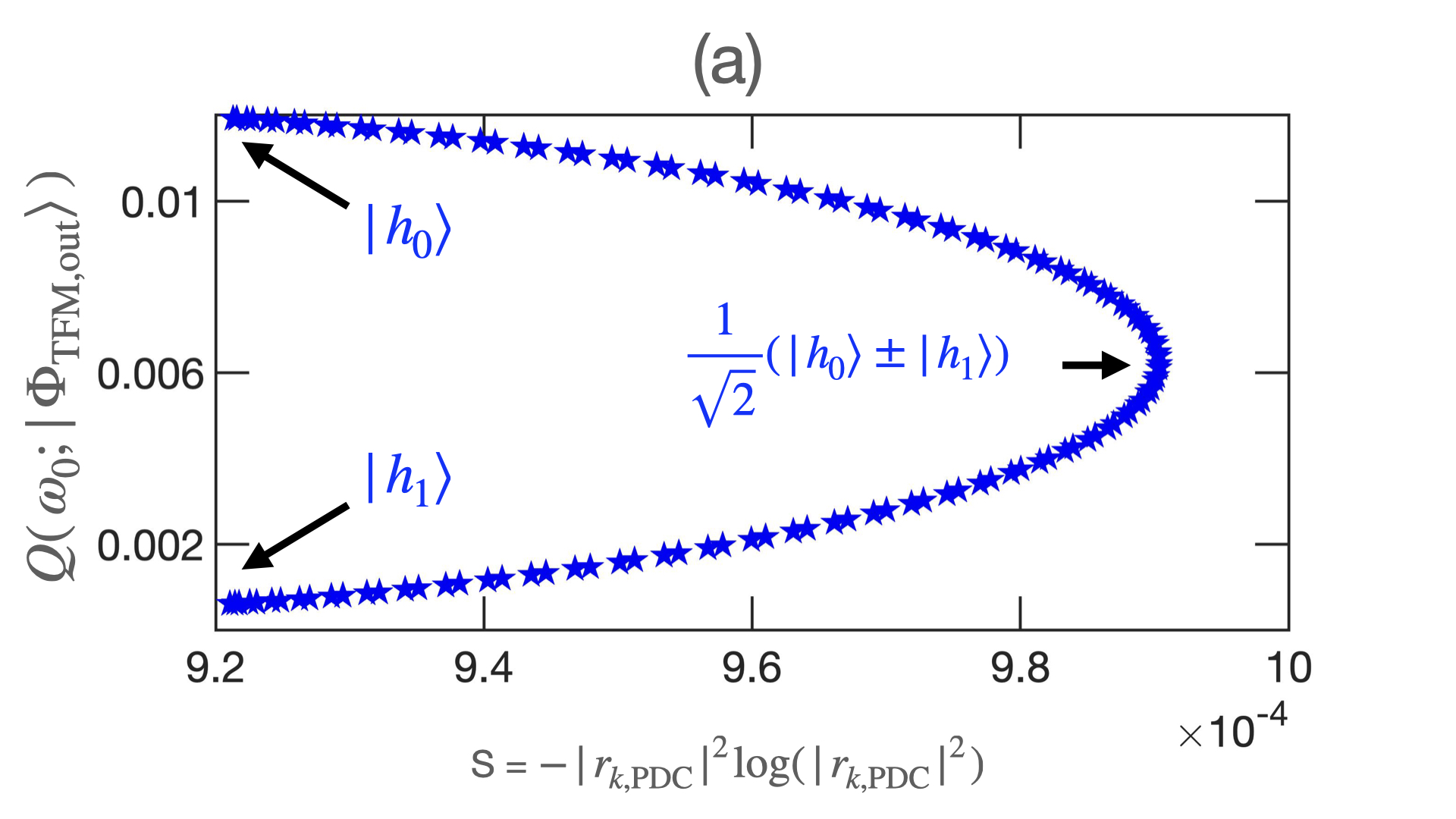}
         \label{fig:PDC20_gamma}      
         \includegraphics[width=0.49\textwidth]{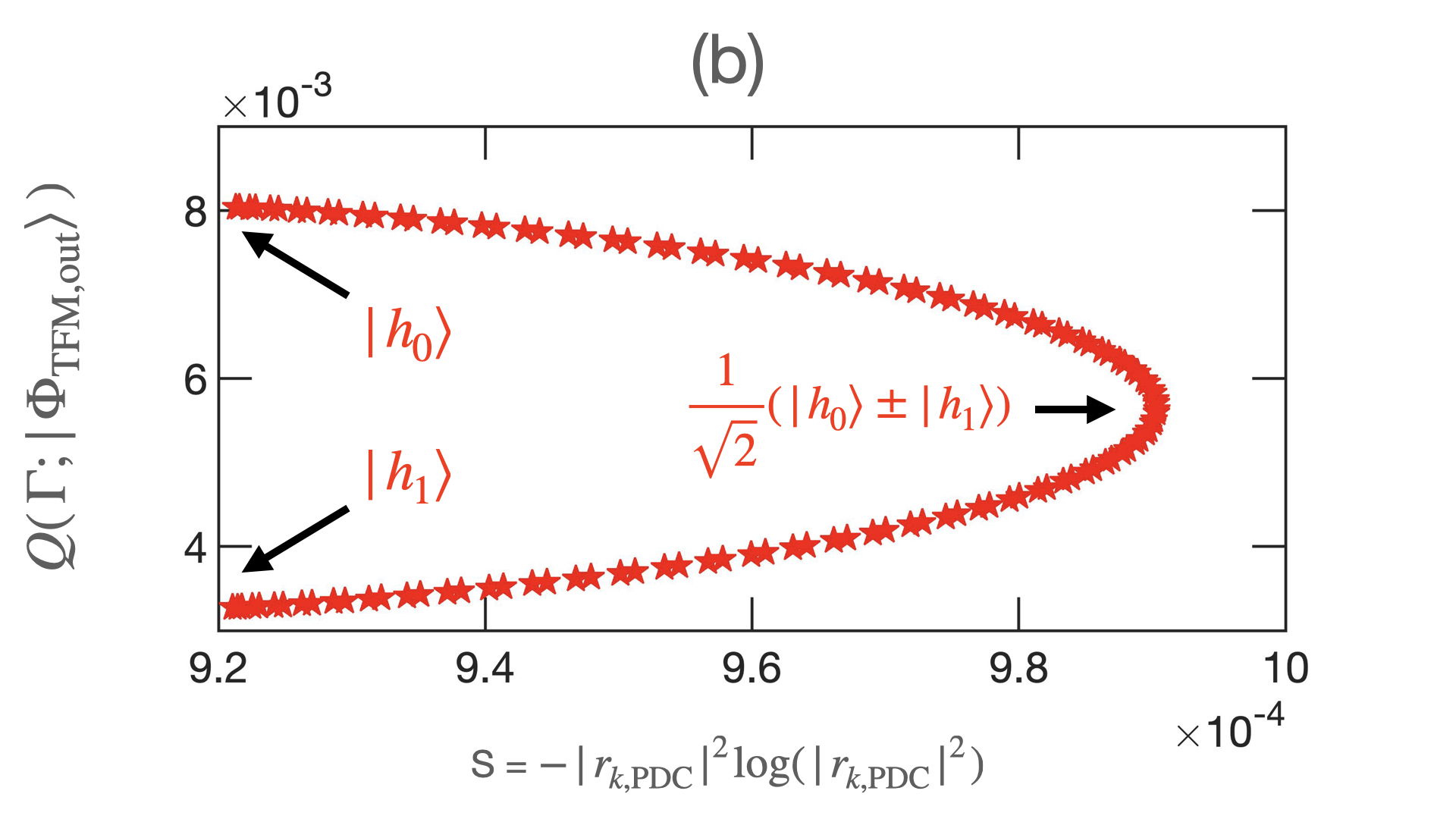}
         \label{fig:PDC100_gamma}
\caption{Parametric plot of outgoing TFM state QFI $\mathcal{Q}(\theta;\ket{\Phi_{\mathrm{TFM,out}}})$, plotted as a function of entanglement entropy $S = -|r_{k,\mathrm{PDC}}|^2\,\mathrm{log}(|r_{k,\mathrm{PDC}}|^2)$ in the incoming state $\ket{\Phi_{\mathrm{TFM}}}$~(defined in Eq.~(\ref{eq:TFMdef})), for (a) TLS parameter $\theta\equiv\omega_0$, and (b) TLS parameter $\theta\equiv\Gamma$.~($k_1 = k_2 = 1.3\,\mathrm{ps}, \Gamma = 0.15\,\mathrm{THz},\,\Gamma_{\perp} = 0\,\mathrm{THz},\,\Delta=0$.) }  
\label{fig:TFMscatterplots}
\end{figure}

This appendix contains parametric plots depicting the relationship between entanglement of input time-frequency entangled states, as captured by entropy of entanglement defined in Eq.~(\ref{eq:entent}), and outgoing biphoton QFI $\mathcal{Q}(\theta;\ket{\Phi_{\mathrm{PDC,out}}})$. Figure \ref{fig:scatterplots} (a)-(b) shows these plots for TLS parameters, while (c)-(d) display the same relationship for CD parameter $J$, all for PDC input states in Eq.~(\ref{eq:PDCstate1}). Figure \ref{fig:TFMscatterplots} shows analogous plots for TFM states defined in Eq.~(\ref{eq:TFMdef}).

\section{TLS Estimation Using PDC Light with $\Delta>0$}\label{appendix:TLSestimationDeltaneq0}
The following series of plots reproduce all quantities presented in Section  \ref{sec6b} for TLS parameter estimation using PDC light, but now for nonzero detuning between the central paraxial frequency $\bar{\omega}_{\mathrm{S}}$, and TLS frequency $\omega_0$.

\subsection{Perfect Coupling}

\begin{figure}[h!]
     \centering   
     \textbf{Outgoing PDC state QFI for TLS parameter $\Gamma$ for $\Delta\neq0$}\par\medskip
         \includegraphics[width=0.49\textwidth]{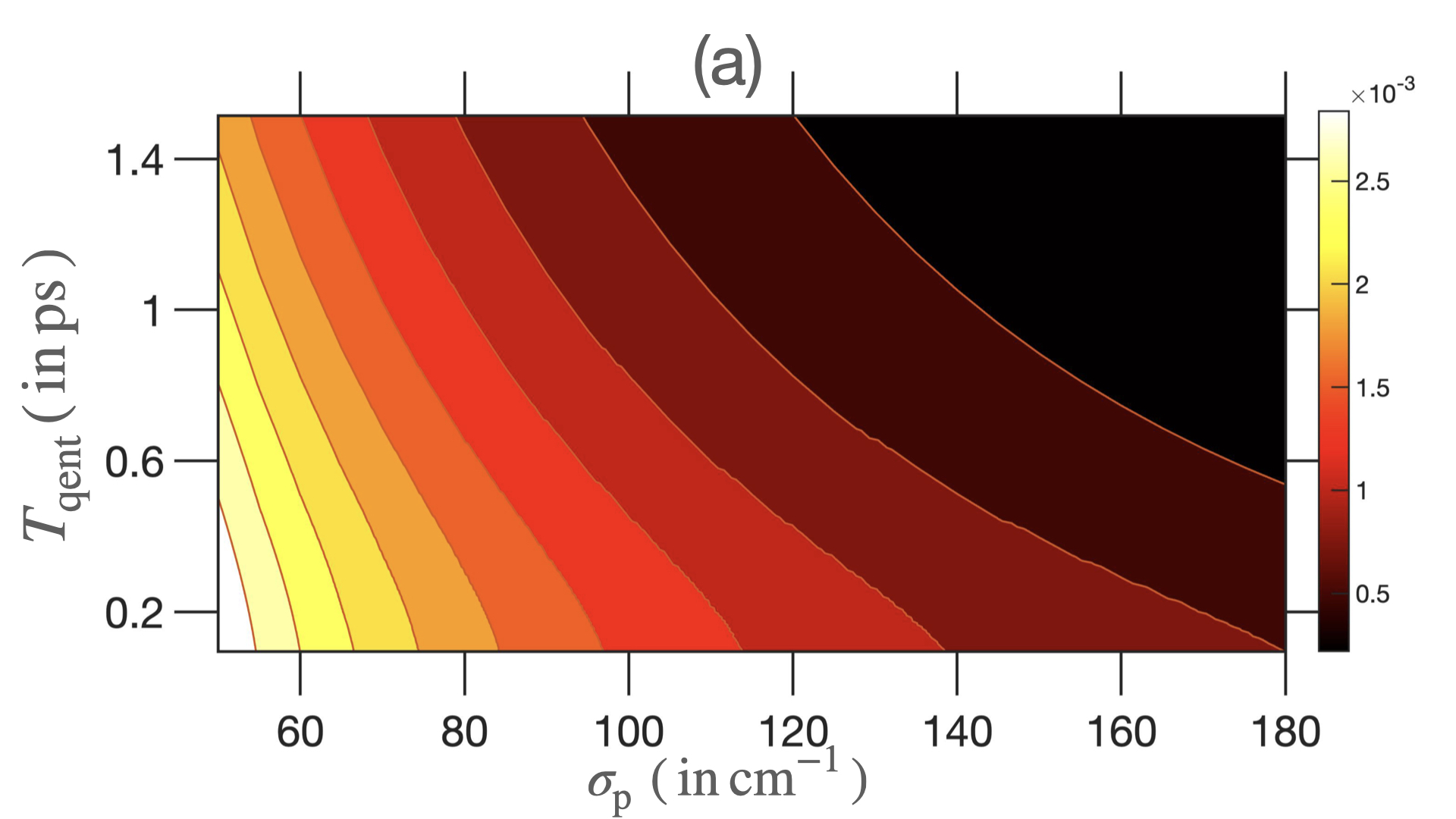}     
         \includegraphics[width=0.49\textwidth]{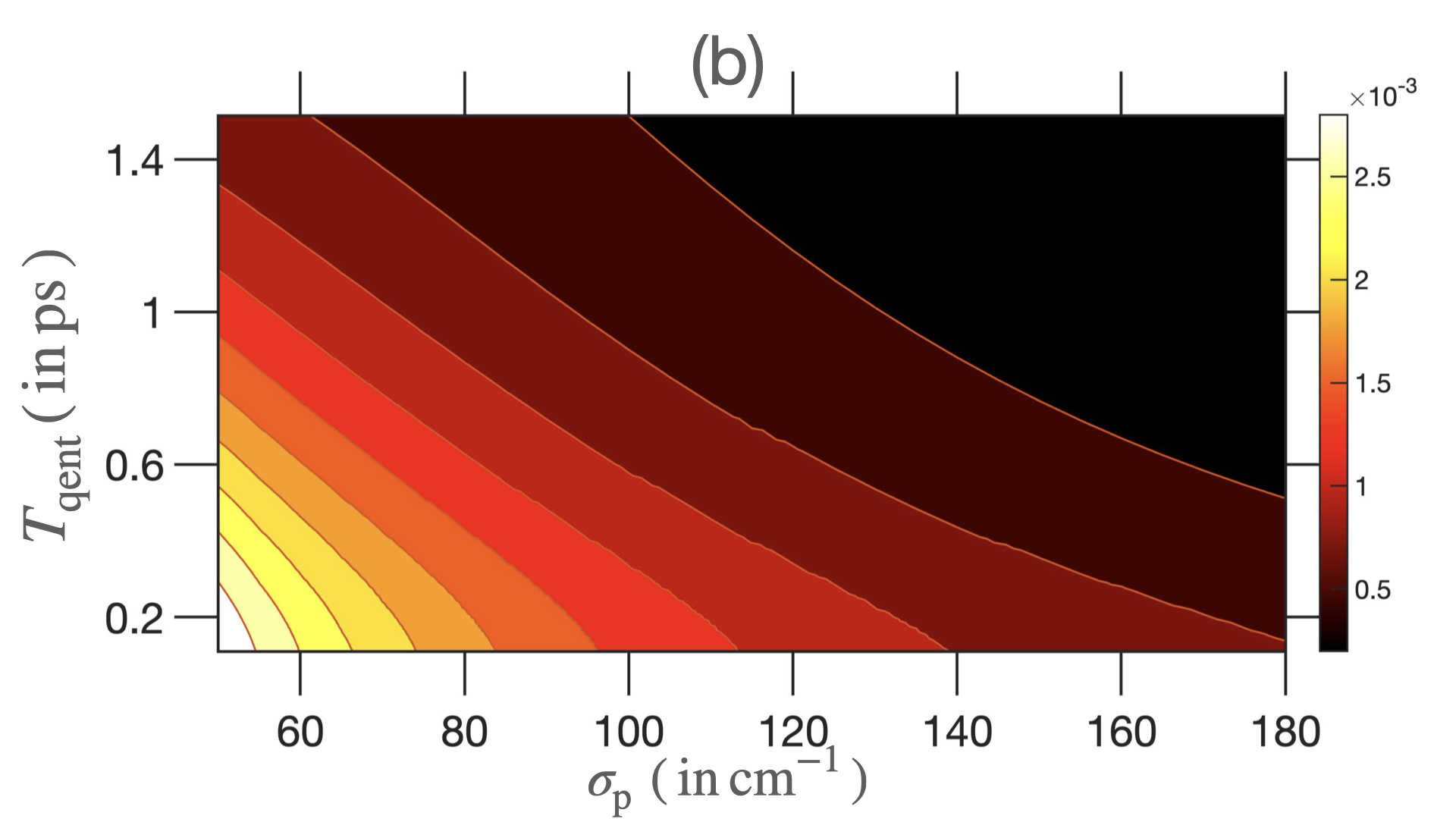}
\caption{Outgoing entangled beam QFI  $\mathcal{Q}(\Gamma;\ket{\Phi_{\mathrm{PDC,out}}})$, calculated numerically using Eq.~(\ref{eq:entangledPDCQFIexp_schmidt}), for varying entanglement time $T_{\mathrm{qent}}$, and classical pumpwidth $\sigma_{\mathrm{p}}$, for TLS parameter $\Gamma$ and detunings (a) $\Delta = 20\,$THz, and (b) $\Delta = 100\,$THz. ~($\Gamma = 0.15$\,THz,\,\,$\Gamma_{\perp} = 0$\,THz.)}  
\label{fig:PDCheatmap_TLS_GammaDeltaneq0}
\end{figure}

\begin{figure}[h!]
     \centering   
     \textbf{Outgoing PDC state QFI for TLS parameter $\omega_0$ for $\Delta\neq0$}\par\medskip
\includegraphics[width=0.49\textwidth]{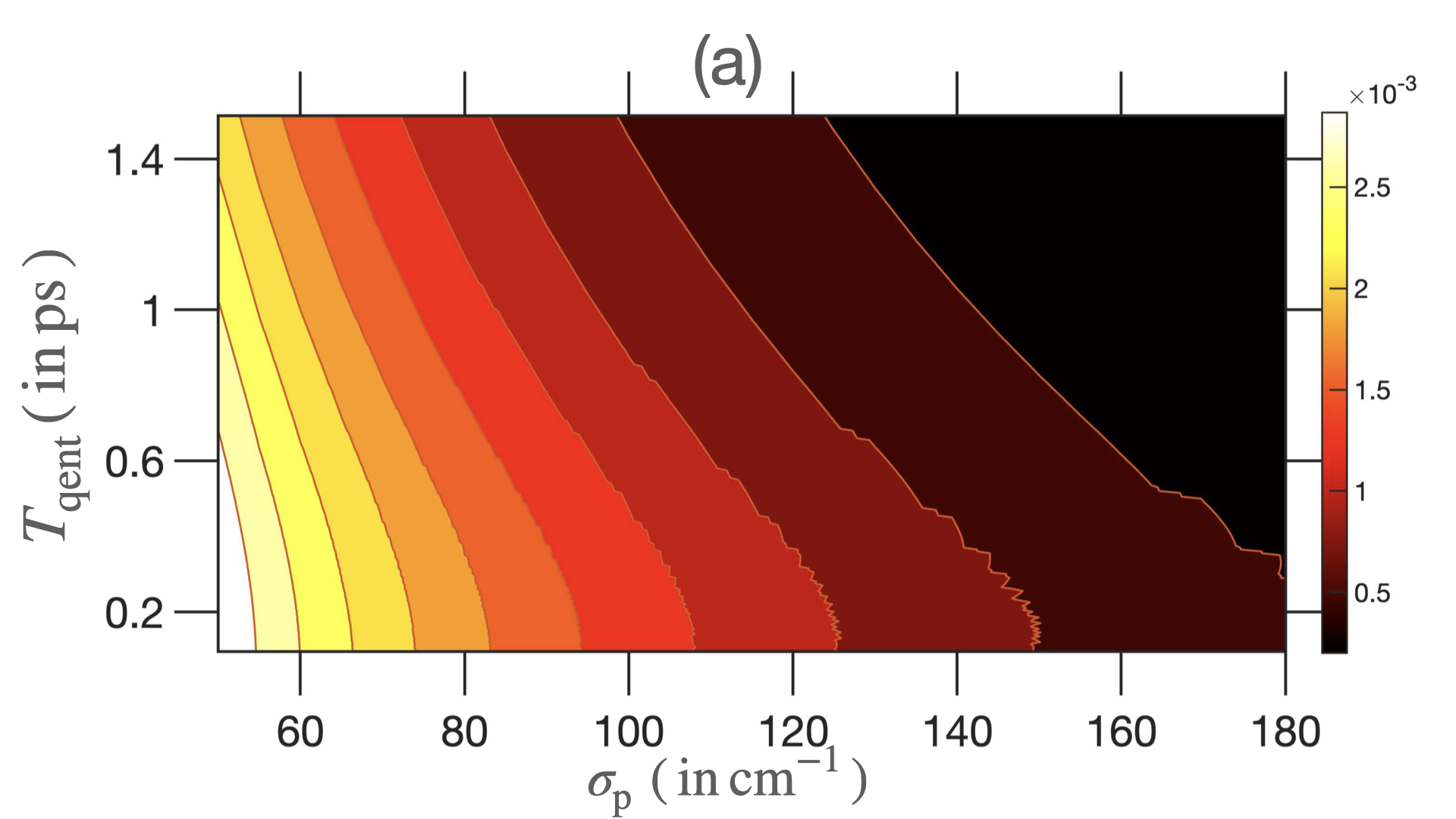}
         \label{fig:PDC20_omega0}      
         \includegraphics[width=0.49\textwidth]{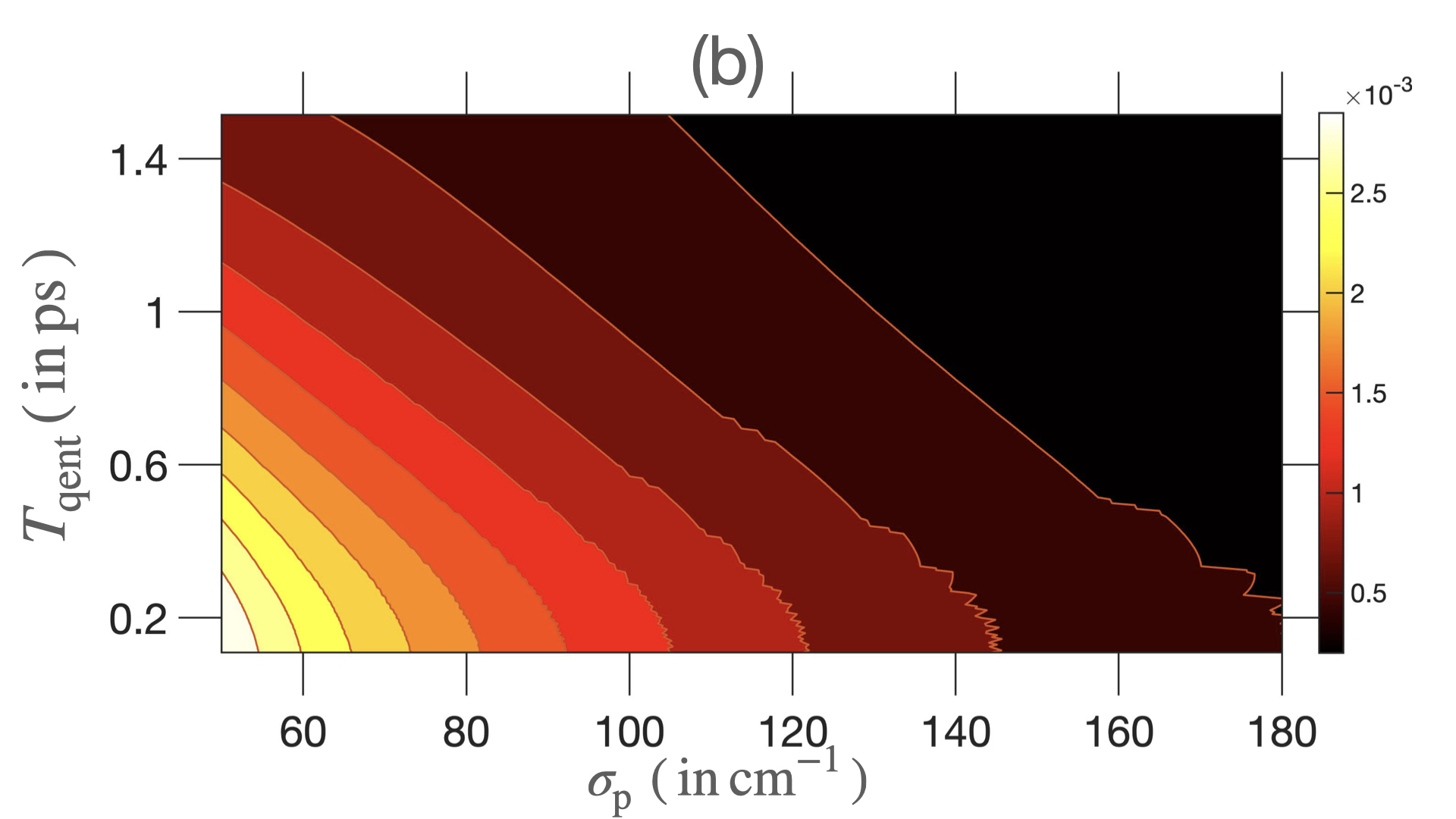}
         \label{fig:PDC100_omega0}
\caption{Outgoing entangled beam QFI  $\mathcal{Q}(\omega_0;\ket{\Phi_{\mathrm{PDC,out}}})$, calculated numerically using Eq.~(\ref{eq:entangledPDCQFIexp_schmidt}), for varying entanglement time $T_{\mathrm{qent}}$, and classical pumpwidth $\sigma_{\mathrm{p}}$, for TLS parameter $\omega_0$ and detunings (a) $\Delta = 20\,$THz, and (b) $\Delta = 100\,$THz. ~($\Gamma = 0.15$\,THz,\,\,$\Gamma_{\perp} = 0$\,THz.)}  
\label{fig:PDCheatmap_TLS_omega0Deltaneq0}
\end{figure}

In Figures \ref{fig:PDCheatmap_TLS_GammaDeltaneq0} and  \ref{fig:PDCheatmap_TLS_omega0Deltaneq0}, we see that the QFI admits the same trend with respect to $\sigma_p$ and $T_{\mathrm{qent}}$ as Figure \ref{fig:PDCheatmap_TLS} for either TLS paramters $\Gamma$ or $\omega_0$ as the magnitude of detuning $|\Delta|$ changes, with the only noticeable effect of the detuning manifesting in the diminished magnitudes for the QFIs.

\begin{figure}[h!]
     \centering
     \textbf{Degree of optimality $\varkappa(\Gamma)$ for TLS parameter $\Gamma$ for $\Delta\neq0$}\par\medskip
         \includegraphics[width=0.49\textwidth]{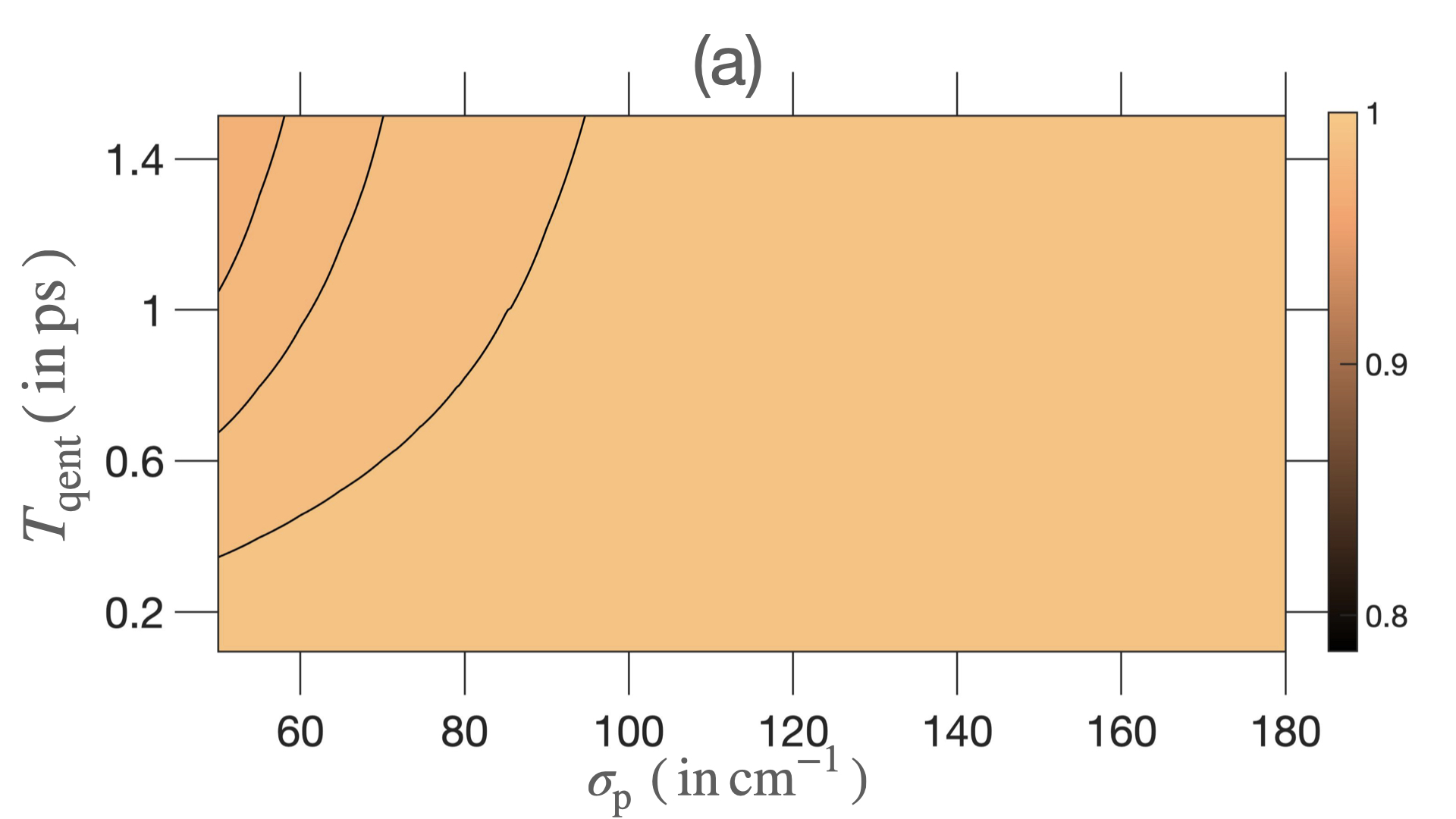}
         \includegraphics[width=0.49\textwidth]{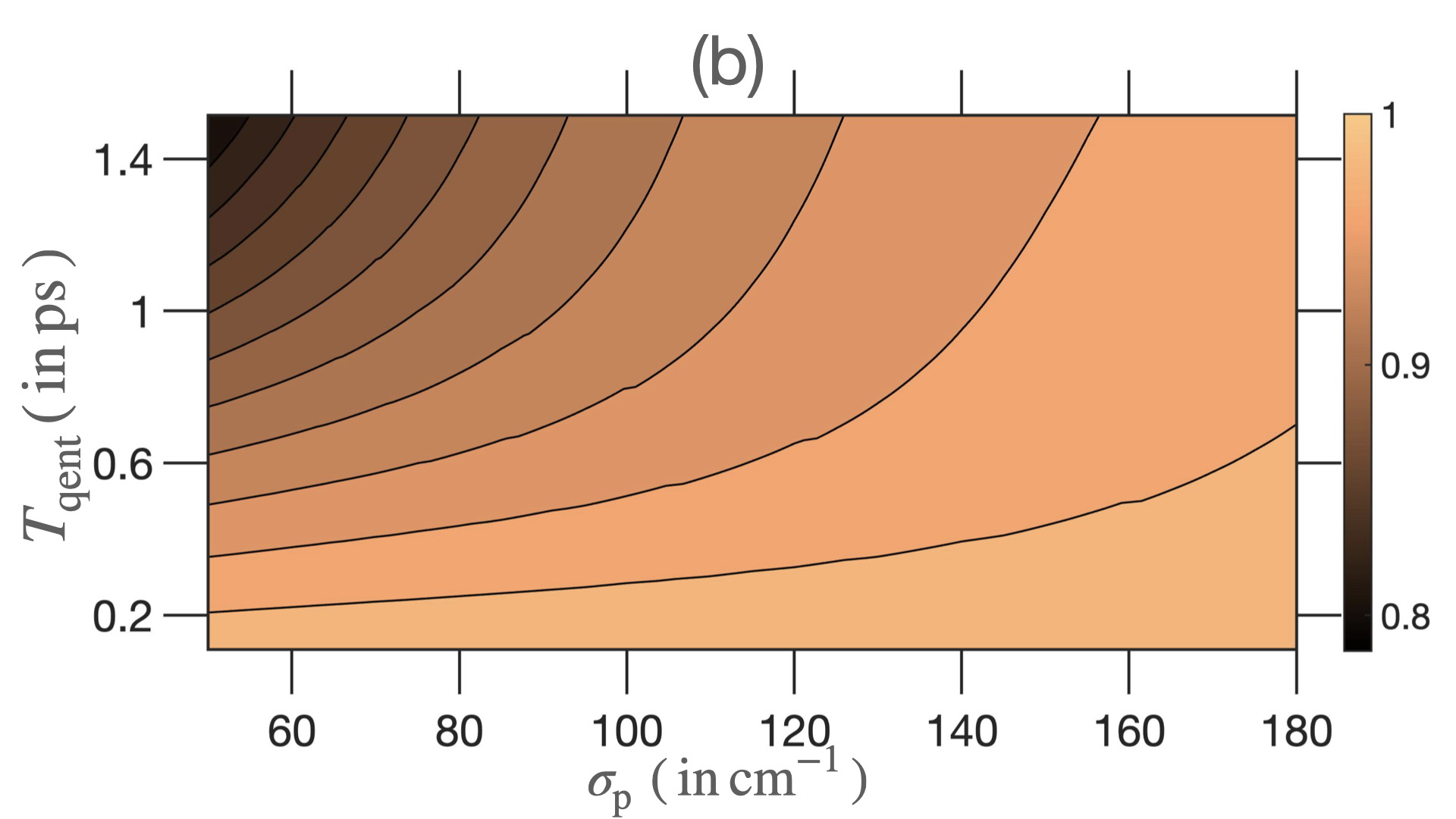}
\caption{Degree of optimality of measurement-optimal $V = \mathds{1}^{\mathrm{I}}$ LOCC schemes, measured as the ratio $\varkappa(\Gamma)$~(defined in Eq.~(\ref{eq:V1degreeoptimality})), for estimation of TLS parameter \,$\Gamma$, using PDC light for detunings (a) $\Delta = 20$\,THz, and (b) $\Delta = 100$\,THz.~($\Gamma = 0.15$\,THz)    }  
\label{fig:LOCCCFItoQFIratio_Gamma_Deltaneq0}
\end{figure}

\begin{figure}[h!]
     \centering
     \textbf{Degree of optimality $\varkappa(\omega_0)$ for TLS parameter $\omega_0$ for $\Delta\neq0$}\par\medskip
         \includegraphics[width=0.49\textwidth]{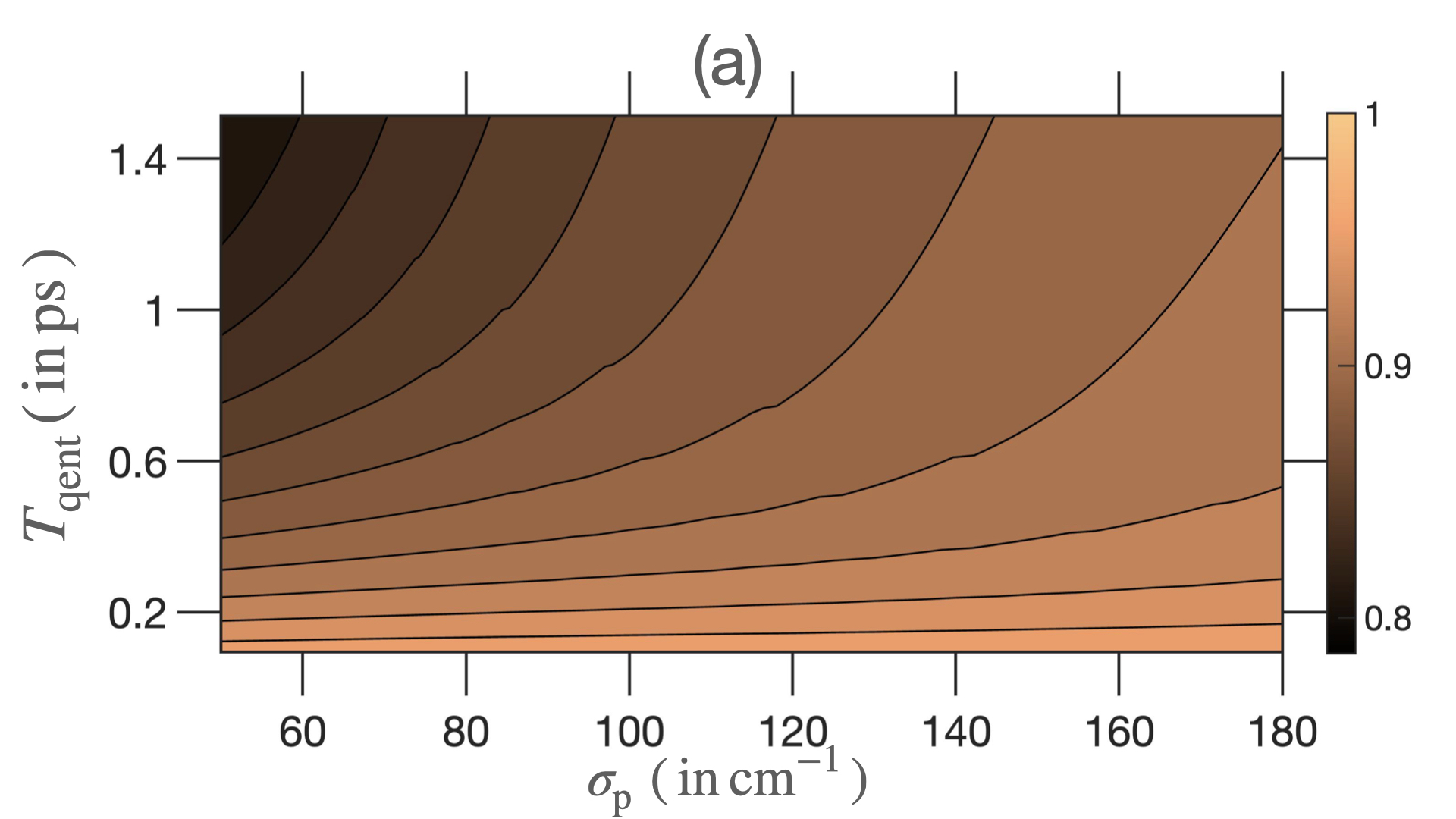}
         \includegraphics[width=0.49\textwidth]{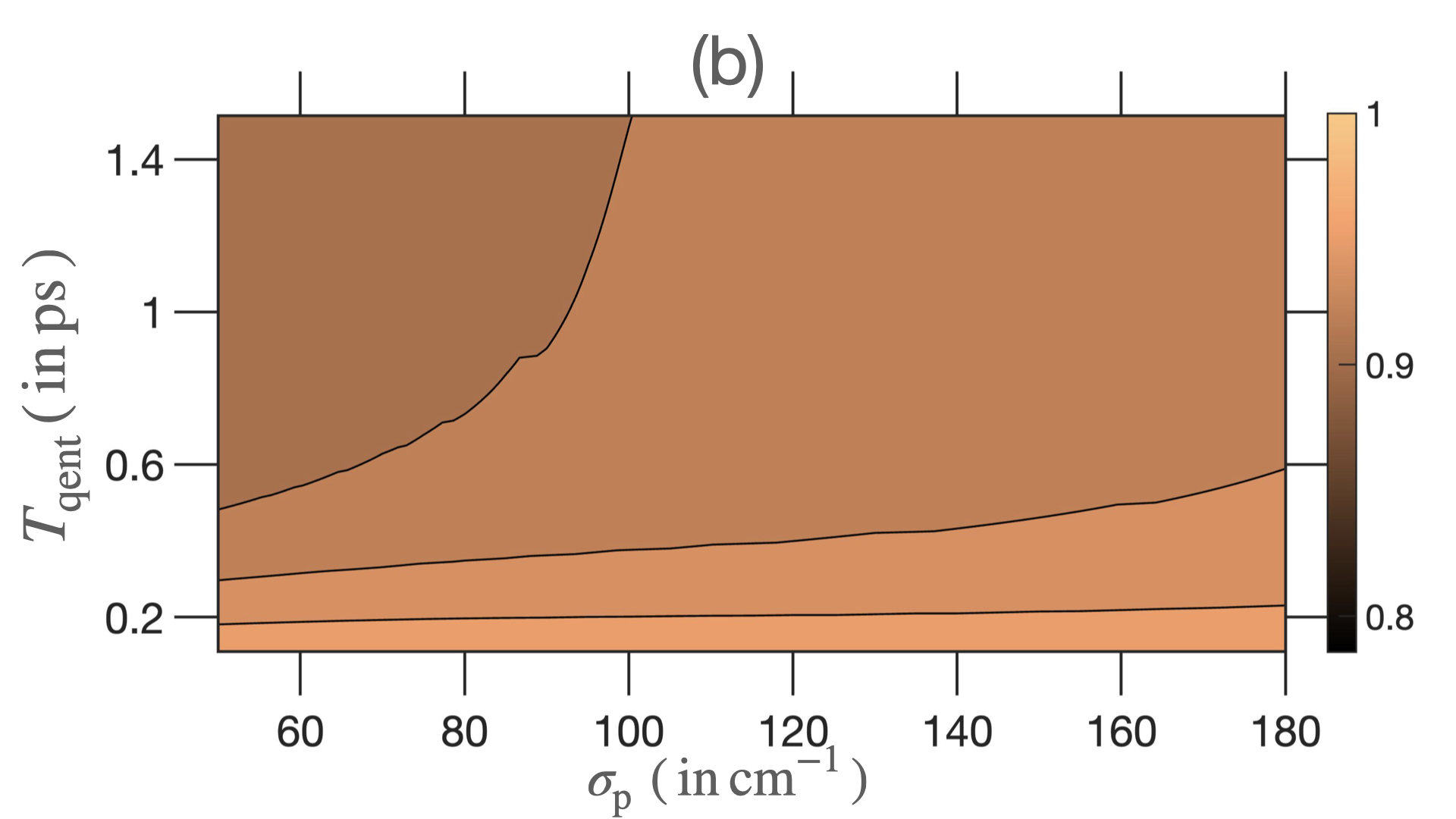}
\caption{Degree of optimality of measurement-optimal $V = \mathds{1}^{\mathrm{I}}$ LOCC schemes, measured as the ratio $\varkappa(\omega_0)$~(defined in Eq.~(\ref{eq:V1degreeoptimality})), for estimation of TLS parameter \,$\omega_0$, using PDC light for detunings (a) $\Delta = 20$\,THz, and (b) $\Delta = 100$\,THz.~($\Gamma = 0.15$\,THz)    }  
\label{fig:LOCCCFItoQFIratio_omega0_Deltaneq0}
\end{figure}

Next, in Figures \ref{fig:LOCCCFItoQFIratio_Gamma_Deltaneq0} and \ref{fig:LOCCCFItoQFIratio_omega0_Deltaneq0} we plot the degree of optimality of measurement-optimal $V = \mathds{1}^{\mathrm{I}}$ LOCC, $\varkappa(\theta)$, for TLS parameters $\Gamma$ and $\omega_0$ respectively. Figure \ref{fig:LOCCCFItoQFIratio_Gamma_Deltaneq0} displays the departure from the forecasted value of unity for $\varkappa(\Gamma)$ when there is zero detuning between $\omega_0$ and $\bar{\omega}_{\mathrm{S}}$, which can be attributed to the increasing value of the quantity $|\langle\Phi_{\mathrm{PDC,out}}|\partial_{\Gamma}\Phi_{\mathrm{PDC,out}}\rangle|$ as $|\Delta|$ increases. In contrast, Figure \ref{fig:LOCCCFItoQFIratio_omega0_Deltaneq0} shows that for certain incoming PDC states~(see upper left corner of the ($\sigma_p,T_{\mathrm{qent}}$) grid) a higher magnitude of detuning can actually enhance the optimality of the  measurement-optimal $V = \mathds{1}^{\mathrm{I}}$ LOCC detection scheme. This shows that, even though the overall QFI values decrease with increasing $|\Delta|$ for either parameter~(rendering therefore all possible estimators less precise), the effect of increasing detuning on the degree of optimality of measurement schemes is less certain, and may depend on the parameter of interest.

\begin{figure}[h!]
     \centering
     \textbf{Enhancement factor $\varsigma(\Gamma)$ for TLS parameter $\Gamma$ for $\Delta\neq0$}\par\medskip
         \includegraphics[width=0.49\textwidth]{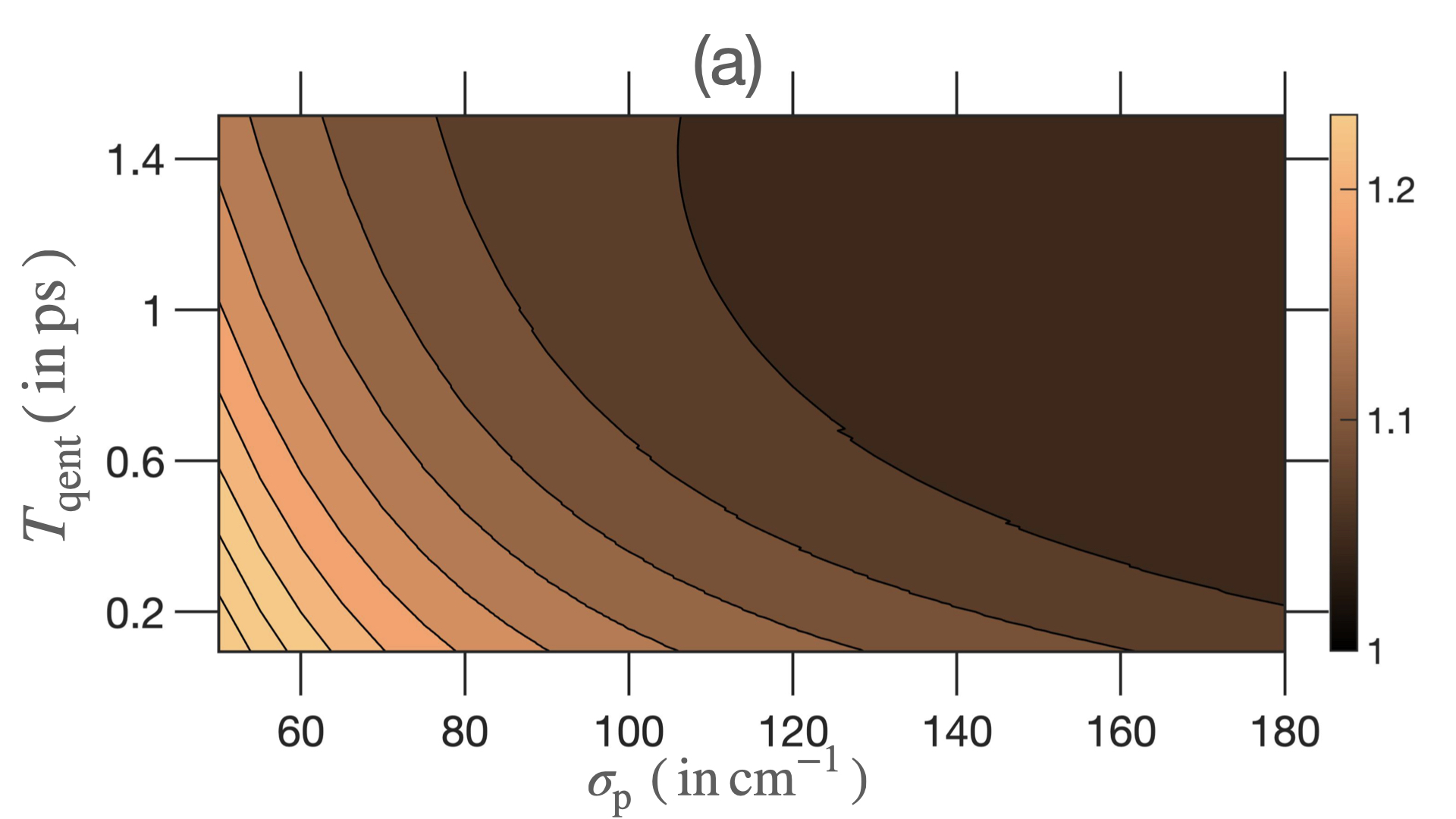}
         \includegraphics[width=0.49\textwidth]{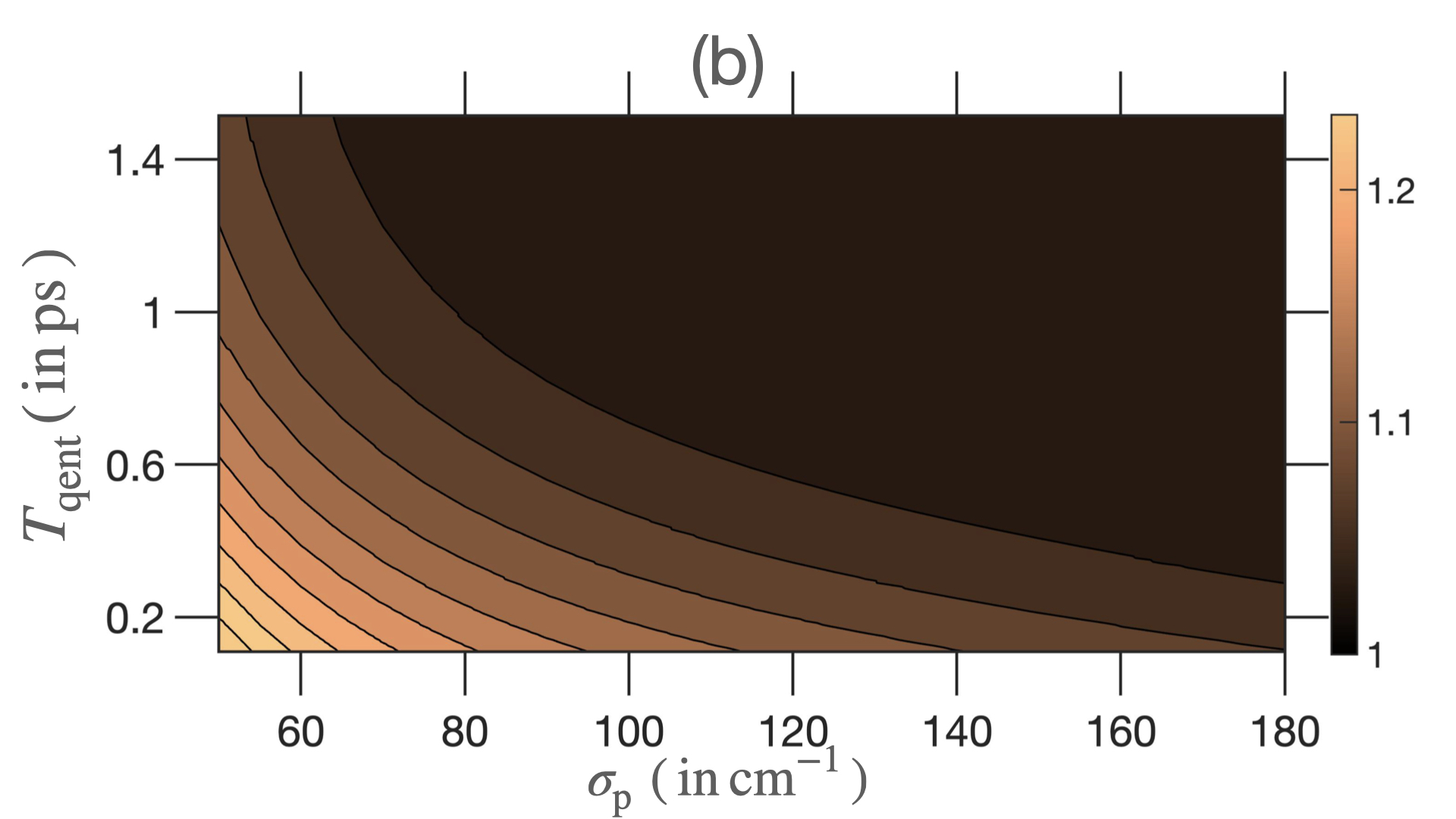}
\caption{Enhancement factor of measurement-optimal $V = \mathds{1}^{\mathrm{I}}$ LOCC schemes, measured as the ratio $\varsigma(\Gamma)$~(defined in Eq.~(\ref{eq:V1enhancementfactor})), for estimation of TLS parameter \,$\Gamma$, using PDC light for detunings (a) $\Delta = 20$\,THz, and (b) $\Delta = 100$\,THz.~($\Gamma = 0.15$\,THz)    }  
\label{fig:LOCCCFItoCFIratio_Gamma_Deltaneq0}
\end{figure}

\begin{figure}[h!]
     \centering
     \textbf{Enhancement factor $\varsigma(\omega_0)$ for TLS parameter $\omega_0$ for $\Delta\neq0$}\par\medskip
         \includegraphics[width=0.49\textwidth]{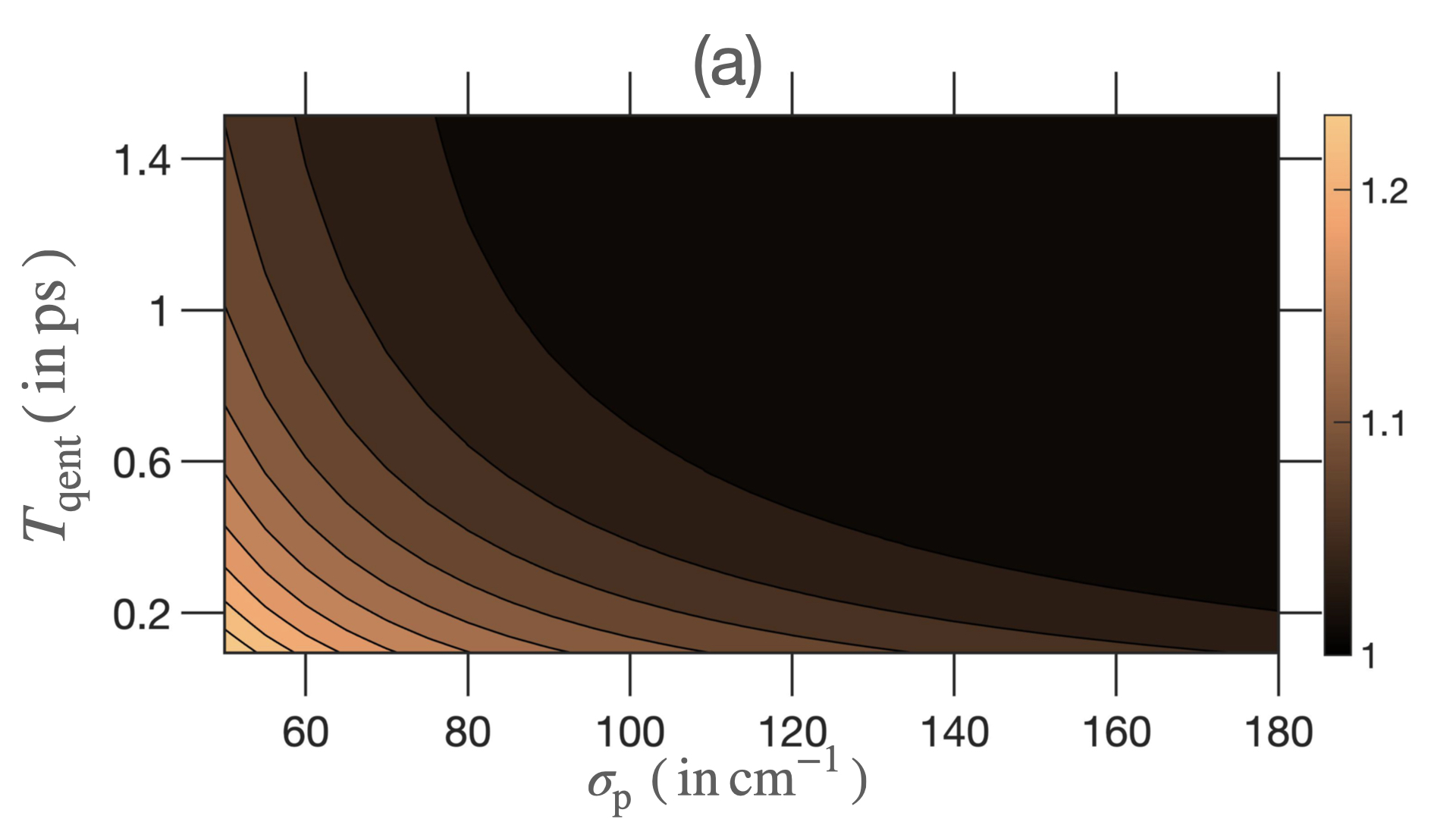}
         \includegraphics[width=0.49\textwidth]{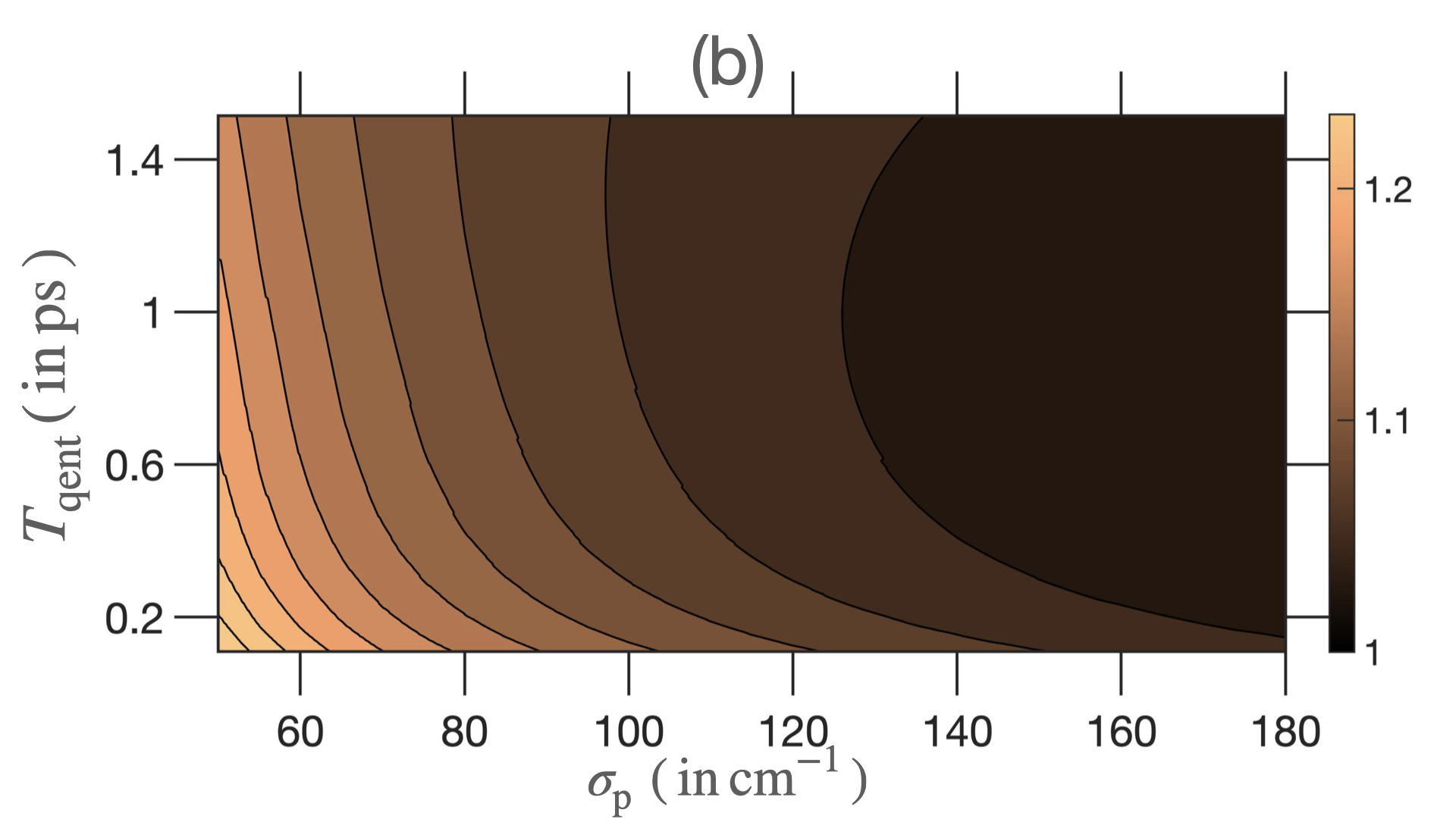}
\caption{Enhancement factor of measurement-optimal $V = \mathds{1}^{\mathrm{I}}$ LOCC schemes, measured as the ratio $\varsigma(\omega_0)$~(defined in Eq.~(\ref{eq:V1enhancementfactor})), for estimation of TLS parameter \,$\omega_0$, using PDC light for detunings (a) $\Delta = 20$\,THz, and (b) $\Delta = 100$\,THz.~($\Gamma = 0.15$\,THz)    }  
\label{fig:LOCCCFItoCFIratio_omega0_Deltaneq0}
\end{figure}

A similar behaviour is observed in Figures \ref{fig:LOCCCFItoCFIratio_Gamma_Deltaneq0} and \ref{fig:LOCCCFItoCFIratio_omega0_Deltaneq0} --- for $\Gamma$-estimation using PDC photons, the enhancement $\varsigma(\Gamma)$ of $V = \mathds{1}^{\mathrm{I}}$ over all single-mode schemes decreases as detuning increases~(Figure \ref{fig:LOCCCFItoCFIratio_Gamma_Deltaneq0}), whereas for $\omega_0$-estimation, a larger enhancement $\varsigma(\omega_0)$ can be obtained using PDC states with signal photon far detuned from the TLS frequency $\omega_0$~(Figure \ref{fig:LOCCCFItoCFIratio_omega0_Deltaneq0}).

\subsection{Free Space $\Gamma_{\perp}>0$ }
\begin{figure}
     \centering   
     \textbf{Outgoing PDC state QFI for TLS parameter $\Gamma$ for $\Delta\neq0~(\Gamma_{\perp}/\Gamma = 0.5)$}\par\medskip
         \includegraphics[width=0.49\textwidth]{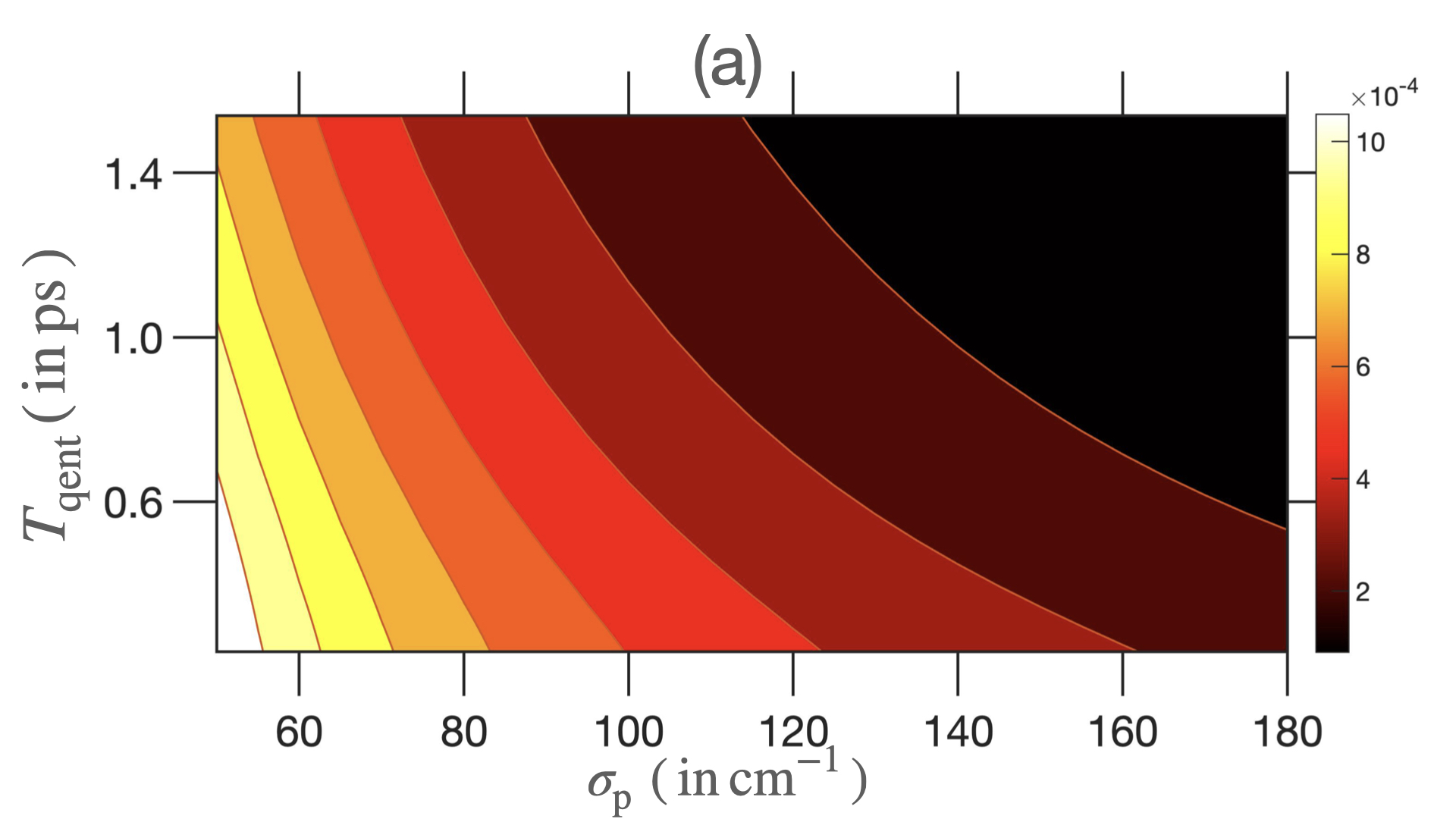}    
         \includegraphics[width=0.49\textwidth]{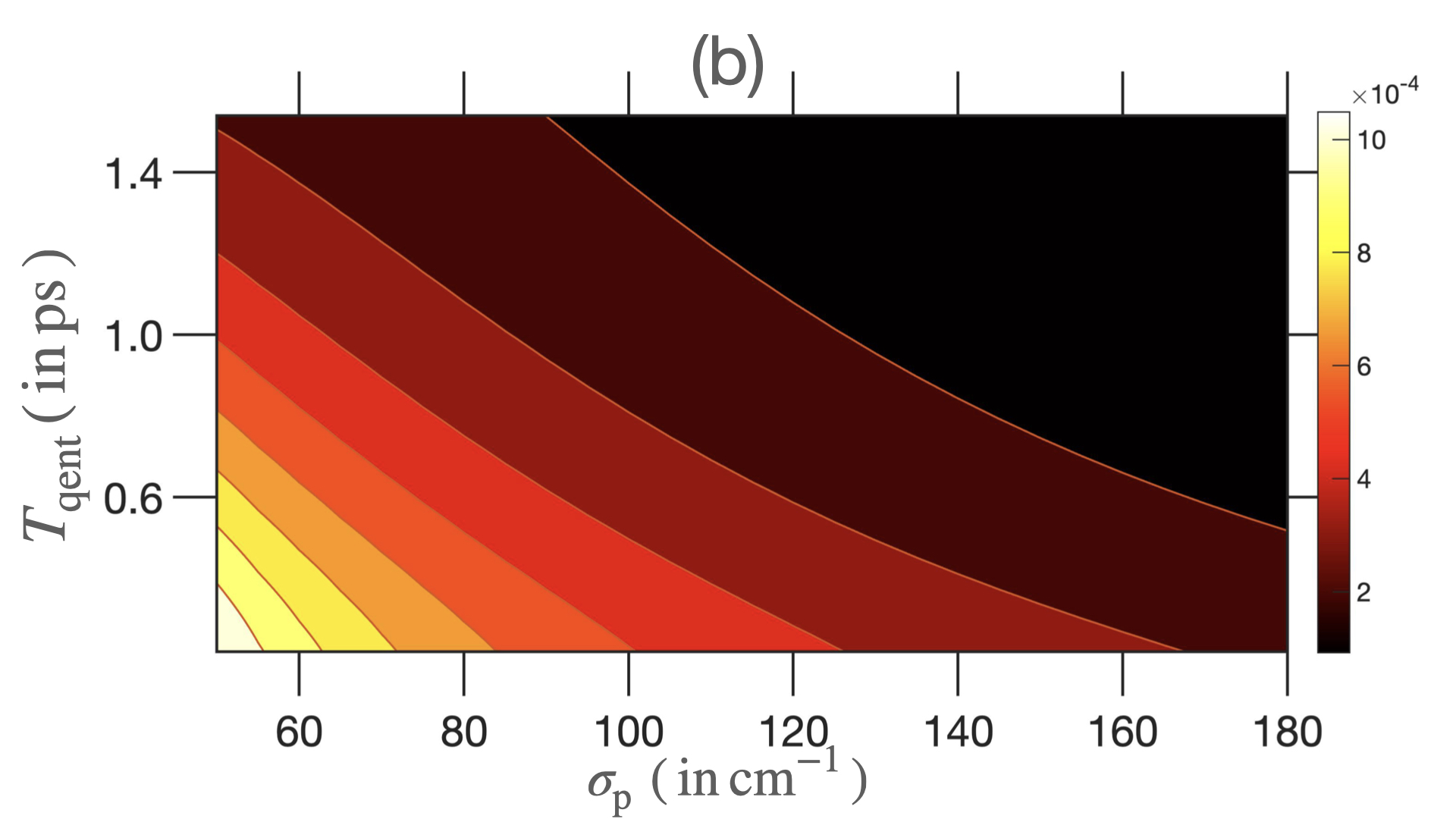}
\caption{Outgoing entangled beam QFI  $\mathcal{Q}(\Gamma;\ket{\Phi_{\mathrm{PDC,out}}})$, calculated numerically using Eq.~(\ref{eq:rhoPouttwinQFI}), for varying entanglement time $T_{\mathrm{qent}}$, and classical pumpwidth $\sigma_{\mathrm{p}}$, for TLS parameter $\Gamma$ and detunings (a) $\Delta = 20\,$THz, and (b) $\Delta = 100\,$THz. ~($\Gamma = 0.15$\,THz,\,$\Gamma_{\perp}/\Gamma = 0.5.$)}  
\label{fig:PDCheatmap_TLS_GammaDeltaneq0_gammperp5}
\end{figure}

\begin{figure}
     \centering   
     \textbf{Outgoing PDC state QFI for TLS parameter $\Gamma$ for $\Delta\neq0~(\Gamma_{\perp}/\Gamma = 10.0)$}\par\medskip
         \includegraphics[width=0.49\textwidth]{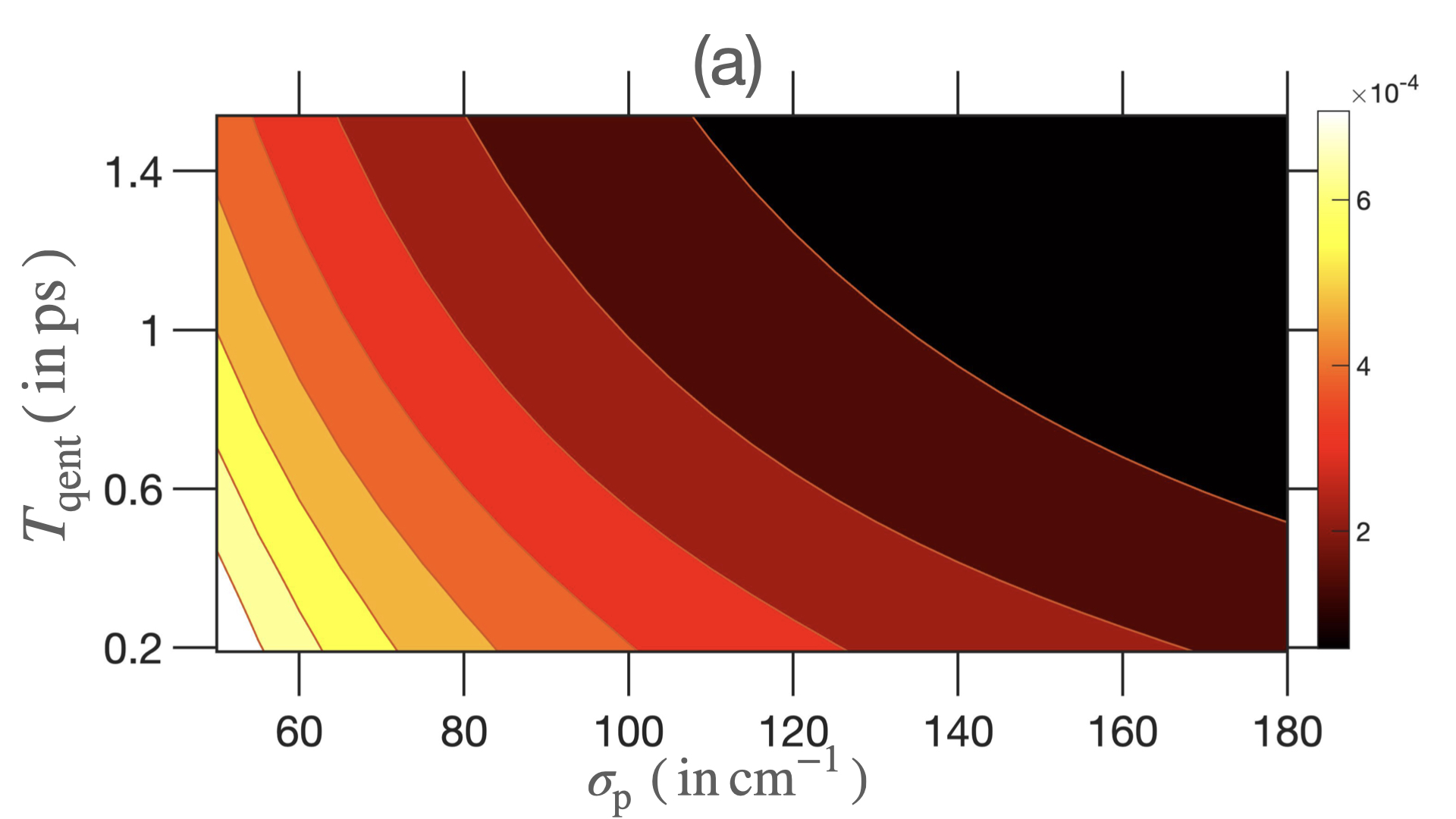}    
         \includegraphics[width=0.49\textwidth]{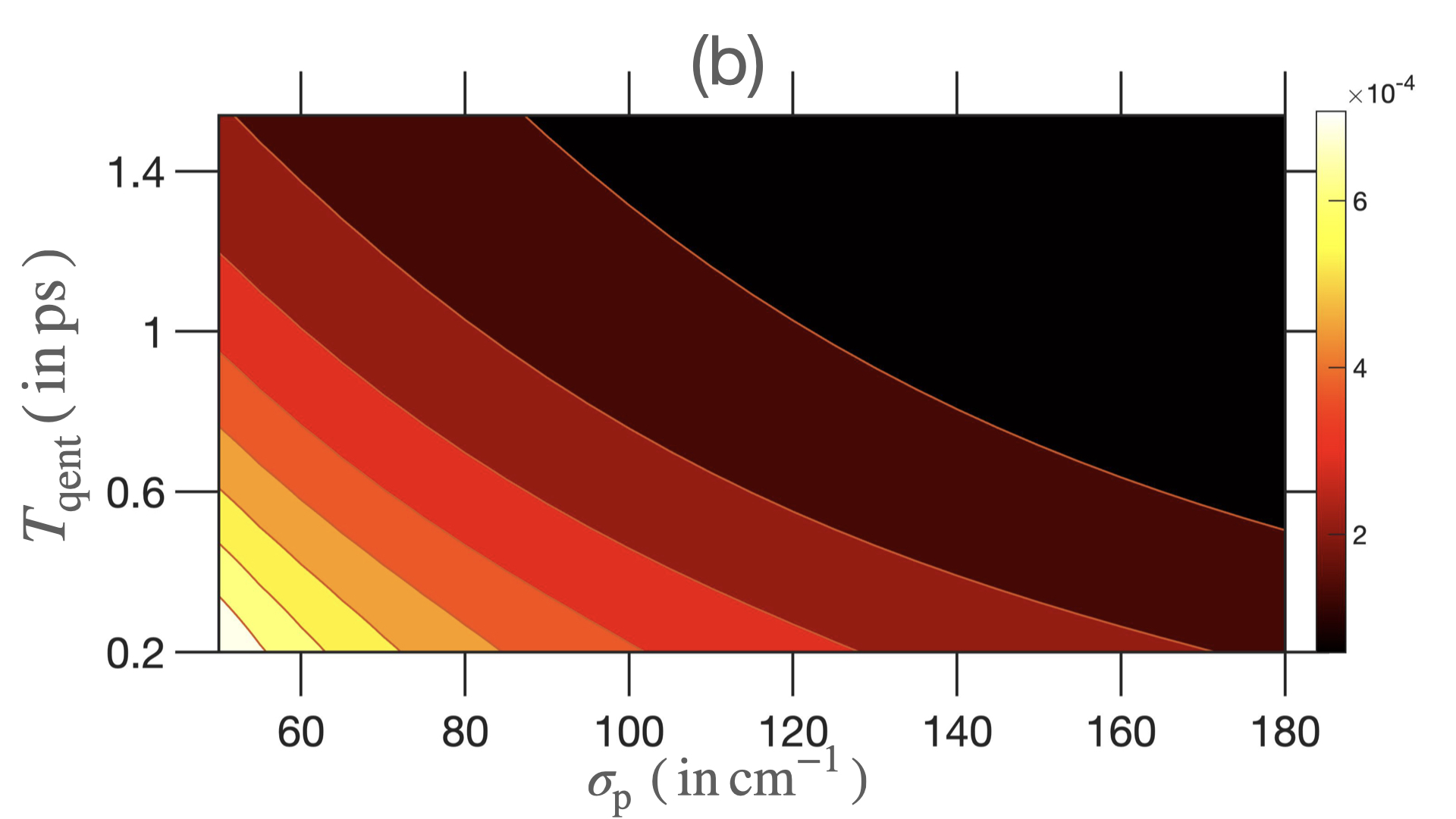}
\caption{Outgoing entangled beam QFI  $\mathcal{Q}(\Gamma;\ket{\Phi_{\mathrm{PDC,out}}})$, calculated numerically using Eq.~(\ref{eq:rhoPouttwinQFI}), for varying entanglement time $T_{\mathrm{qent}}$, and classical pumpwidth $\sigma_{\mathrm{p}}$, for TLS parameter $\Gamma$ and detunings (a) $\Delta = 20\,$THz, and (b) $\Delta = 100\,$THz. ~($\Gamma = 0.15$\,THz,\,$\Gamma_{\perp}/\Gamma = 10.0.$)}  
\label{fig:PDCheatmap_TLS_GammaDeltaneq0_gammperp10}
\end{figure}

\begin{figure}
     \centering   
     \textbf{Outgoing PDC state QFI for TLS parameter $\omega_0$ for $\Delta\neq0~(\Gamma_{\perp}/\Gamma = 0.5)$}\par\medskip
         \includegraphics[width=0.49\textwidth]{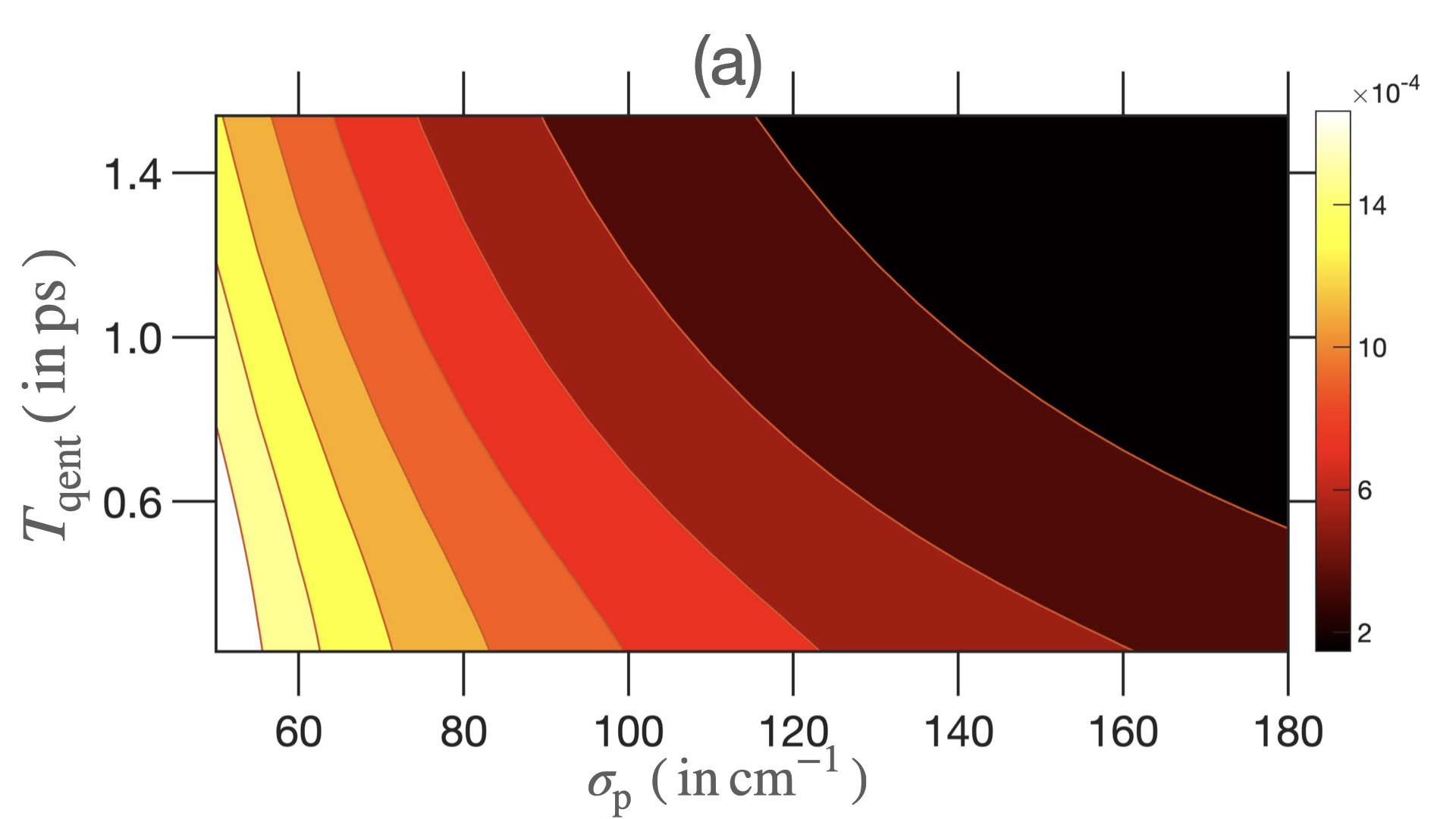}    
         \includegraphics[width=0.49\textwidth]{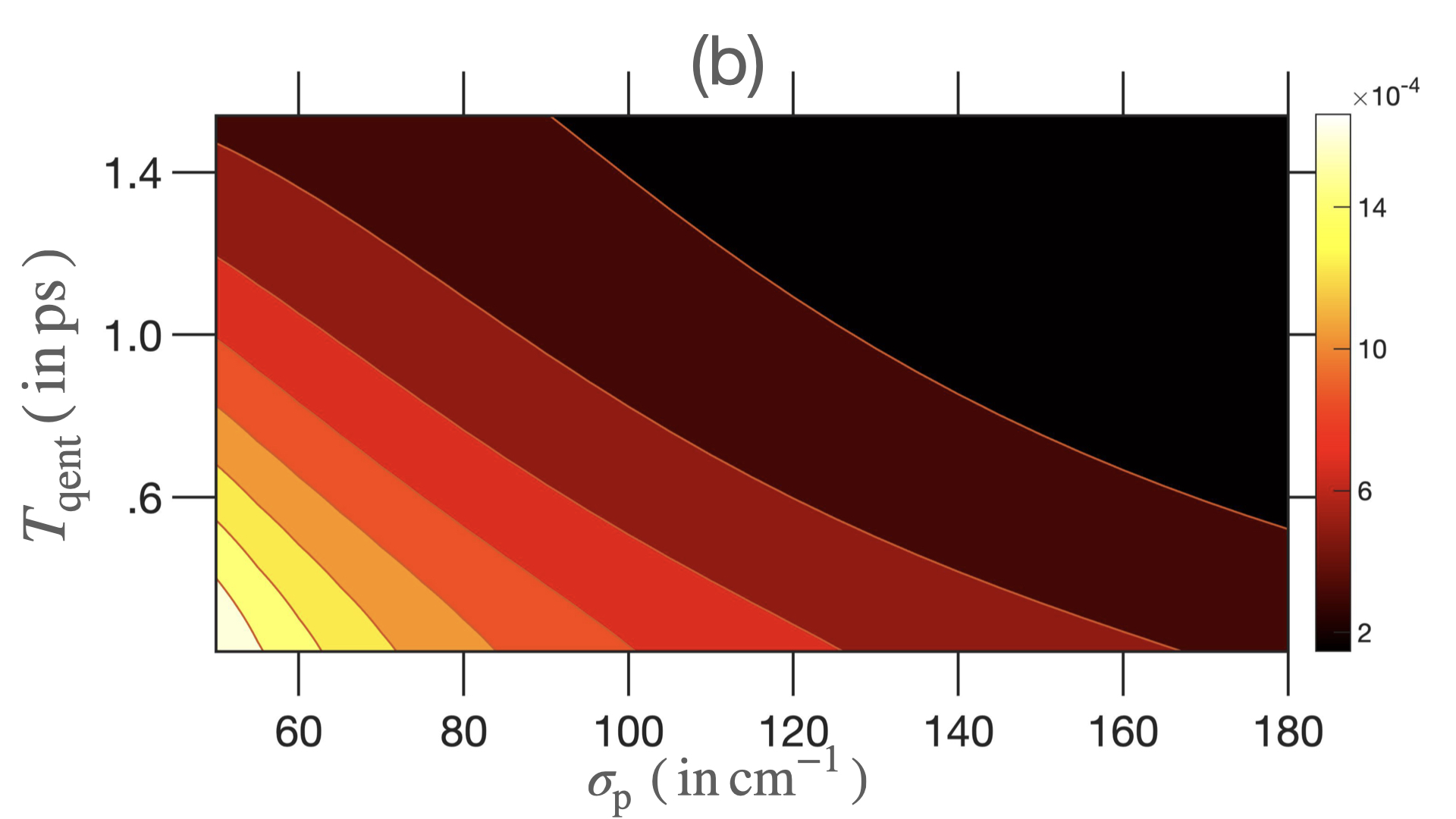}
\caption{Outgoing entangled beam QFI  $\mathcal{Q}(\omega_0;\ket{\Phi_{\mathrm{PDC,out}}})$, calculated numerically using Eq.~(\ref{eq:rhoPouttwinQFI}), for varying entanglement time $T_{\mathrm{qent}}$, and classical pumpwidth $\sigma_{\mathrm{p}}$, for TLS parameter $\omega_0$ and detunings (a) $\Delta = 20\,$THz, and (b) $\Delta = 100\,$THz. ~($\Gamma = 0.15$,\,THz,\,$\Gamma_{\perp}/\Gamma = 0.5.$)}  
\label{fig:PDCheatmap_TLS_omega0Deltaneq0_gammperp5}
\end{figure}

\begin{figure}
     \centering   
     \textbf{Outgoing PDC state QFI for TLS parameter $\omega_0$ for $\Delta\neq0~(\Gamma_{\perp}/\Gamma = 10.0)$}\par\medskip
         \includegraphics[width=0.49\textwidth]{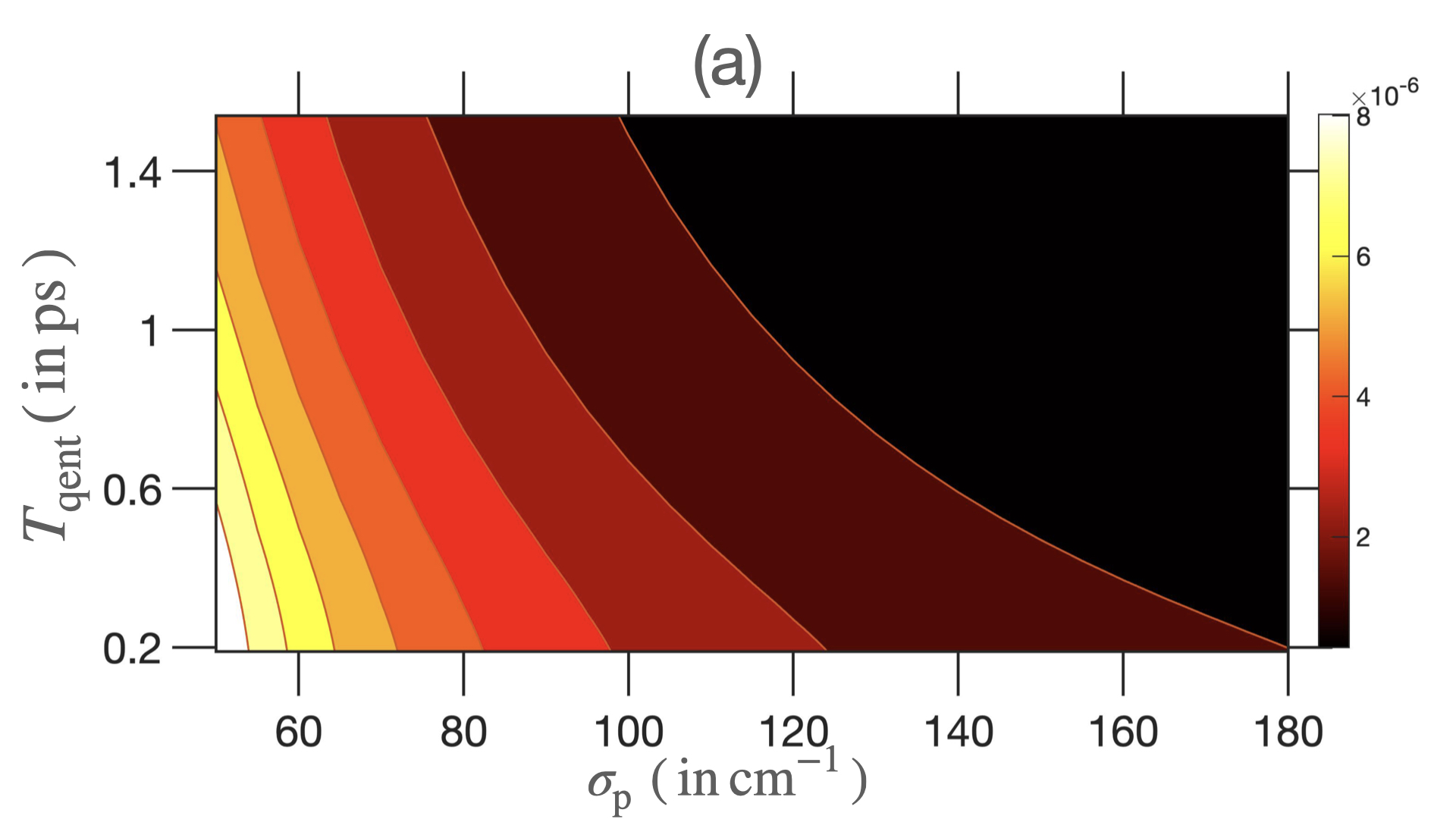}    
         \includegraphics[width=0.49\textwidth]{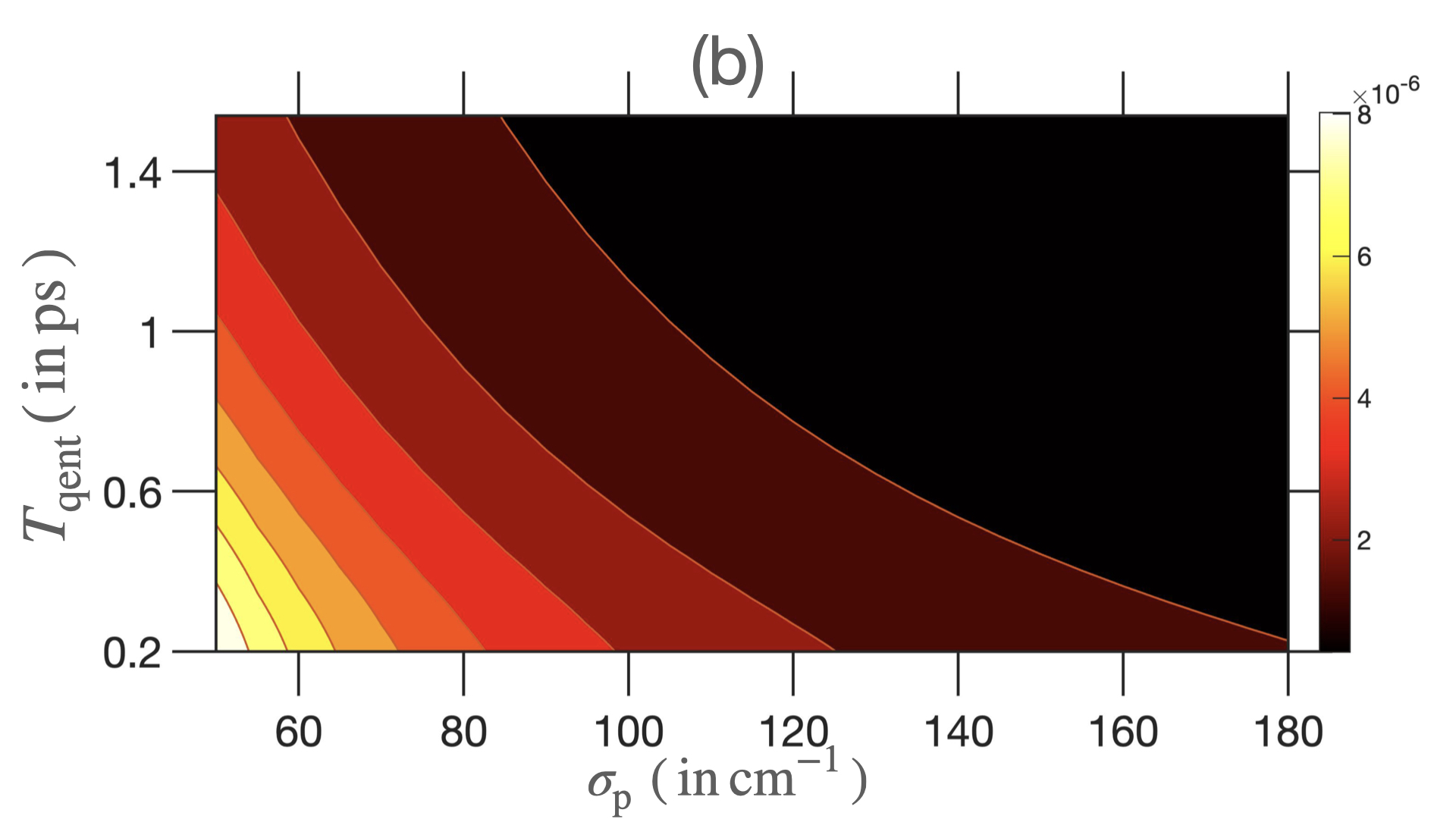}
\caption{Outgoing entangled beam QFI  $\mathcal{Q}(\omega_0;\ket{\Phi_{\mathrm{PDC,out}}})$, calculated numerically using Eq.~(\ref{eq:rhoPouttwinQFI}), for varying entanglement time $T_{\mathrm{qent}}$, and classical pumpwidth $\sigma_{\mathrm{p}}$, for TLS parameter $\omega_0$ and detunings (a) $\Delta = 20\,$THz, and (b) $\Delta = 100\,$THz. ~($\Gamma = 0.15$\,THz,\,$\Gamma_{\perp}/\Gamma = 10.0.$)}  
\label{fig:PDCheatmap_TLS_omega0Deltaneq0_gammperp10}
\end{figure}

We can also study the effect of non-zero detuning on TLS estimation for the free space scenario, characterised by non-zero coupling to E space $\Gamma_{\perp}>0$. In Figures \ref{fig:PDCheatmap_TLS_GammaDeltaneq0_gammperp5} and \ref{fig:PDCheatmap_TLS_GammaDeltaneq0_gammperp10}, we plot $\Gamma$-QFI for the same grid of PDC characteristics $\sigma_p$ and $T_{\mathrm{qent}}$ as Figure \ref{fig:freespaceQFIGam}. Again, the trend for QFI values against PDC chacteristics $\sigma_p$ and $T_{\mathrm{qent}}$ remains unchanged, with moderately diminishing values for the QFI as detuning increases. A similar effect is seen for the TLS parameter $\omega_0$ in Figures \ref{fig:PDCheatmap_TLS_omega0Deltaneq0_gammperp5} and \ref{fig:PDCheatmap_TLS_omega0Deltaneq0_gammperp10} for free space couplings $\Gamma_{\perp}/\Gamma = 0.5$ and $\Gamma_{\perp}/\Gamma = 10.0$ respectively.

\section{Light-CD Interaction In Excitonic Basis}\label{appendix:CDdetails}
The coupled dimer~(CD) system is comprised of two two-level systems, coupled to each other via an attractive Coulomb interaction, so that the bare molecular Hamiltonian is
\begin{equation}\label{eq:cdhamiltonian}
    H^{\mathrm{CD}} = \sum_{j=a,b}\hbar\omega_j\ket{j}\bra{j} +  \hbar(\omega_a+\omega_b)\ket{f}\bra{f} +  J\,(\ket{a}\bra{b} + \ket{b}\bra{a}),
\end{equation}
where $J$ is the coupling strength between the two sites labelled $a$ and $b$, whose excited levels are respectively $\ket{a}$ and $\ket{b}$, and $\ket{f}$ is the doubly excited state, characterised by level energy $\hbar(\omega_a+\omega_b)$, obtained under the assumption of zero binding energy meaning that there is no interaction between the two excitations. 

This Hamiltonian is obtained by setting $P=2$ in the $P$-site Hamiltonian in Eq.~(\ref{eq:matterhamiltonian}), and is a precursor to the far more complex Frenkel-Holstein Hamiltonians~\citep{yang2002influence,ishizaki2009unified,ishizaki2009theoretical,tempelaar2014vibrational} that are used to model complex dynamics~(including excitonic energy transport~(EET) and long-lived quantum beats) in photosynthetic LHCs. In these more involved Hamiltonians, also included are site-dependent reorganisation energy terms, and interaction terms corresponding to phonon baths comprised of harmonic oscillators at each site whose displacement couples linearly with each excitation~\citep{ishizaki2009unified}. This is in addition to modelling each pigment site as a two-level system to account for the lowest transition, as well as interstitial coupling terms that appear in both the $n$-site and coupled dimer Hamiltonians in Eqs.~(\ref{eq:matterhamiltonian}) and (\ref{eq:cdhamiltonian}) respectively.

\begin{figure}[h!]
        \fbox{\includegraphics[width=1.0\textwidth]{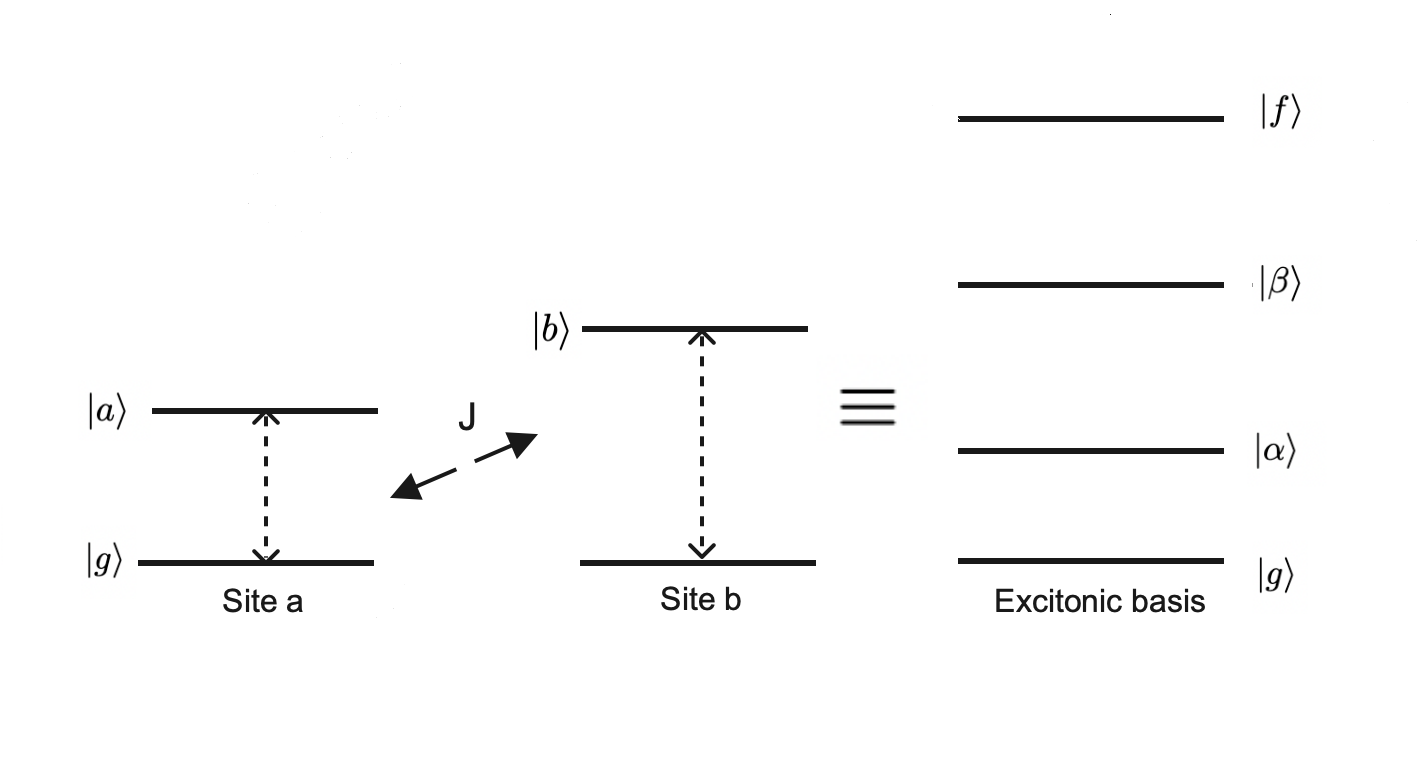}}
        \caption{ Level diagram of the CD system comprised of two two-level systems that have a common ground set to zero energy, and excited levels $\ket{a}$ and $\ket{b}$, coupled to each other via Coulomb interaction of strength $J$. In the diagonal excitonic basis effectively in the SEM space, $\ket{\alpha}$ and $\ket{\beta}$ are delocalised over both sites a and b.}
        \label{fig:CDleveldiagram}
\end{figure}

For an analytically tractable description of CD dynamics, we transform to the diagonal basis for the CD system, called the excitonic basis, through the eigendecomposition of the Hamiltonian in Eq.~(\ref{eq:cdhamiltonian}), 
\begin{equation}
    H^{\mathrm{CD}} = \sum_{j=\alpha,\beta}\hbar\omega_j\ket{j}\bra{j} + \hbar(\omega_a+\omega_b)\ket{f}
\end{equation}
where, effectively, we have diagonalised in the singly-excited manifold~(SEM) space only~($\ket{g}$ and $\ket{f}$ states are already eigenstates of $H^{\mathrm{CD}}$), yielding SEM eigenstates $\ket{\alpha}$ and $\ket{\beta}$. The delocalised excitonic states can be explicitly related to the basis kets $\ket{a}$ and $\ket{b}$ by the following relations~\citep{yuen2014ultrafast},
\begin{equation}\label{eq:excitonicsitebasis}
    \ket{\alpha} = \cos\Theta\ket{a} + \sin\Theta\ket{b},~\ket{\beta} = -\sin\Theta\ket{a} + \cos\Theta\ket{b}, 
\end{equation}
where $\Theta = \frac{1}{2}\arctan\left(\frac{2J}{\delta}\right)$ and  $\delta = \hbar(\omega_a-\omega_b)$, and the corresponding eigenvalues are $\omega_{\alpha} = \bar{\omega} - (\delta/2)\sec2\Theta$, $\omega_{\beta} = \bar{\omega} + (\delta/2)\sec2\Theta$ respectively for the $\ket{\alpha}$ and $\ket{\beta}$ state. For reference, CD site and excitonic bases are visually represented as level diagrams in Figure \ref{fig:CDleveldiagram}.

The transition dipole moment operator $\vec{d}$ of the CD system that couples with the incoming electric field has the following form in the site basis,
\begin{equation}
    \vec{d} = \sum_{i=a,b}\vec{\mu}_{ig}\,\ket{i}\bra{g} + \sum_{i=a,b}\vec{\mu}_{fi}\,\ket{f}\bra{i} + \mathrm{h.c.} 
\end{equation}
which we can transform to the excitonic basis using Eq.~(\ref{eq:excitonicsitebasis}) where it has the analogous form,
\begin{equation}
    \vec{d} = \sum_{i=\alpha,\beta}\vec{\mu}_{ig}\,\ket{i}\bra{g} + \sum_{i=\alpha,\beta}\vec{\mu}_{fi}\,\ket{f}\bra{i} + \mathrm{h.c.} 
\end{equation}
such that the various vector dipole elements transform via the following $\Theta$-rotation matrix,
\begin{align}
    \begin{bmatrix} \vec{\mu}_{\alpha g} \\ \vec{\mu}_{\beta g} \end{bmatrix} &= \begin{bmatrix} ~ \cos\Theta & \sin\Theta \\ -\sin\Theta & \cos\Theta \end{bmatrix}~~ \begin{bmatrix} \vec{\mu}_{ag} \\ \vec{\mu}_{bg}\end{bmatrix}, \nonumber\\
    \begin{bmatrix} \vec{\mu}_{f\alpha} \\ \vec{\mu}_{f\beta} \end{bmatrix} &= \begin{bmatrix} -\sin\Theta & \cos\Theta \\ ~~\cos\Theta & \sin\Theta \end{bmatrix}~~ \begin{bmatrix} \vec{\mu}_{fa} \\ \vec{\mu}_{fb}\end{bmatrix}.
\end{align}
As we will restrict our discussion to the use of pulses carrying single excitation to the CD system, and the CD system itself being in the ground state $\ket{g}$ at $t=0$, the dipole operator elements $\vec{\mu}_{fi}$ that link the SEM and doubly excited level $\ket{f}$ can be dropped altogether. We will then abbreviate the GSM-SEM transition dipoles for brevity as $\vec{\mu}_{ig} \equiv \vec{\mu}_i = |\vec{\mu}_i|\hat{\mu_i},\,i=\alpha,\beta$ for brevity.

In order to describe the light-CD interaction, we transform to the interaction picture with respect to the zeroth-order Hamiltonian, 
\begin{equation}\label{eq:zerothHamiltonianCD}
    H_0^{\mathrm{CD}} = \sum_{i=\alpha,\beta}\hbar\bar{\omega}_{\mathrm{S}}\ket{i}\bra{i} + 2\hbar\bar{\omega}_{\mathrm{S}}\ket{f}\bra{f} +  H^{\mathrm{F}},
\end{equation}
where $\bar{\omega}_{\mathrm{S}}$ is the central frequency of the signal mode, and the free field Hamiltonian is 
\begin{equation}\label{eq:fieldHamiltonian_appendix}
    H^{\mathrm{F}} = \int d\omega\, \hbar\omega\,a_{\mathrm{S}}^{\dag}(\omega)a_{\mathrm{S}}(\omega) + \int d\omega\,\hbar\omega\,b^{\dag}(\omega)b(\omega),
\end{equation}
where we have dropped the free field term corresponding to idler mode because of the explicit assumption that the CD system only interacts with a single mode in this input-output setup, and the operator $b(t)$ corresponds to the environmental modes E. The interaction picture Hamiltonian is then
\begin{align}\label{eq:CDfieldHamiltonian}
    H^{\mathrm{CD}}(t) &= H^{\mathrm{CD}}_{I} -  \vec{d}. \vec{E}(t) \noindent\nonumber\\
    &= \sum_{i=\alpha,\beta}\hbar\Delta_i\ket{i}\bra{i} + \hbar(\Delta_{\alpha}+\Delta_{\beta})\ket{f}\bra{f}   -i\hbar\, [\sqrt{\Gamma}\,\Sigma^{\dag}\,a_{\mathrm{S}}(t)\otimes\mathds{1}^{\mathrm{I}}\otimes\mathds{1}^{\mathrm{E}} + \sqrt{\Gamma_{\perp}}\,\Sigma^{\dag}\,\mathds{1}^{\mathrm{S}}\otimes\mathds{1}^{\mathrm{I}}\otimes b(t) - \mathrm{h.c}]
\end{align}
where
\begin{align}\label{eq:CDdipole}
    \Sigma^{\dag} &= \sqrt{\frac{\omega_c \mu_0^2}{2\epsilon_0 c A_0 \hbar}}~(\lambda_{\alpha}\sigma_+^{\alpha} + \lambda_{\beta}\sigma_+^{\beta})\nonumber\\
    &= \sqrt{2\Gamma}~(\lambda_{\alpha}\sigma_+^{\alpha} + \lambda_{\beta}\sigma_+^{\beta}),\, \mathrm{where}\,\sigma_+^{i} = \ket{i}\bra{g},\lambda_{i} = (\hat{e}.\hat{\mu}_{i})~\left(\frac{|\vec{\mu}_i|}{\mu_0}\right)~( i=\alpha,\beta)
\end{align}
is the collective SEM raising operator, and
\begin{equation}
    H^{\mathrm{CD}}_{I} = \sum_{i=\alpha,\beta}\hbar\Delta_i\ket{i}\bra{i} + \hbar(\Delta_{\alpha}+\Delta_{\beta})\ket{f}\bra{f}  
\end{equation}
where $\Delta_i = \omega_i - \bar{\omega}_{\mathrm{S}}\,(i=\alpha,\beta)$ are the detunings from the central signal pulse frequency of the excitonic levels, and $a(t)$ are white-noise operators previously defined in Eq~(\ref{eq:whitenoisedef}) corresponding to the incoming paraxial mode. In obtaining these, we have re-centered the frequency integrals with the transformation $\omega \rightarrow \omega-\bar{\omega}_{\mathrm{S}}$ so that the field operators $a_{\mathrm{S}}(\omega)$ and $b(\omega)$ are centred around $\omega=0$. The extension of the frequency integrals to all frequencies from $-\infty$ to $+\infty$ is a consequence of the fact that the CD system-quantum pulse interaction is assumed to be peaked around the central frequency of the oncoming pulse, allowing us to invoke the white noise approximation of SVEA.


\section{Characteristic Matter Function For Bare CD Hamiltonian}\label{appendix:characteristicfnCD}

Closed expressions for the characteristic function $f^{\mathrm{CD}}(t_1)$ can be calculated explicitly for the bare CD Hamiltonian using the following formula for matrix exponential of a $2\times2$ matrix $M = \begin{pmatrix} a & b \\ c & d \end{pmatrix} \in \mathbf{C}^{2\times2}$~\citep{bernstein1993some}
\begin{equation}
    e^M = e^{\frac{a+d}{2}}\,\begin{pmatrix} \cosh\upsilon + \frac{a-d}{2}\,\frac{\sinh \upsilon}{\upsilon} & b\,\frac{\sinh\upsilon}{\upsilon} \\  c\,\frac{\sinh\upsilon}{\upsilon} & \cosh\upsilon - \frac{a-d}{2}\,\frac{\sinh \upsilon}{\upsilon}\end{pmatrix}.
\end{equation}
where $\upsilon = (1/2)\sqrt{(a-d)^2 + 4bc}$.

For the bare CD Hamiltonian, the characteristic function can be expressed as the expectation value of the matrix exponential
\begin{equation}
    f_{\mathrm{CD}}(t_1) = \begin{pmatrix} \lambda_{\alpha} & \lambda_{\beta} \end{pmatrix} \,\mathrm{exp}\left[ -\frac{i H^{\mathrm{CD}}_{I}\,t_1}{\hbar} - \frac{1}{2}\Sigma^{\dag}\Sigma\,t_1\right]\,\begin{pmatrix} \lambda_{\alpha} \\ \lambda_{\beta}  \end{pmatrix}  
\end{equation}
which is then 
\begin{equation}\label{eq:characteristicfn}
    f_{\mathrm{CD}}(t_1) = e^{-i\Tilde{\Delta}t_1-\frac{\Gamma t_1}{2}(\lambda_a^2+\lambda_b^2)}\,\left\{ (\lambda_a^2 + \lambda_b^2)\cosh\upsilon + \frac{\sinh\upsilon}{\upsilon}\left[ (\lambda_{\alpha}^2-\lambda_{\beta}^2)\frac{a-d}{2} - 2\lambda_{\alpha}^2\lambda_{\beta}^2\,\Gamma t_1 \right]\right\}
\end{equation}
where $\Tilde{\Delta} = (\Delta_{\alpha}+\Delta_{\beta})/2$ is the averaged detuning, and the various terms~(in terms of the CD parameters) are
\begin{equation}
    \upsilon = \frac{t_1}{2}\,\sqrt{\frac{-\delta^2\sec^2 2\Theta}{\hbar^2} + \Gamma^2(\lambda_a^2+\lambda_b^2)^2 - \frac{2i\delta\Gamma}{\hbar}\sec 2\Theta\,(\lambda_{\alpha}^2-\lambda_{\beta}^2)  }
\end{equation}
and 
\begin{equation}
    \frac{a-d}{2} = \frac{i\delta t_1\sec 2\Theta }{2\hbar} - \frac{\Gamma t_1}{2}(\lambda_{\alpha}^2-\lambda_{\beta}^2). 
\end{equation}
The  QFI  for  the  outgoing  single  photon  state  that  scatters  off  of  the  CD  system is proportional to a convolution of the incoming field envelope, and the parametric derivative of the characteristic function $f_{\mathrm{CD}}(t_1)$ which can also be calculated explicitly for the bare CD Hamiltonian for the $J$ parameter,
\begin{align}\label{eq:Jdercharacteristicfn}
    &\frac{\partial}{\partial J}\,f_{\mathrm{CD}}(t_1) = e^{-i\Tilde{\Delta}t_1-\frac{\Gamma t_1}{2}(\lambda_a^2+\lambda_b^2)}\bigg[~ \frac{\partial\upsilon}{\partial J}\,\bigg\{ (\lambda_a^2+\lambda_b^2)\sinh\upsilon + \left( \frac{\cosh\upsilon}{\upsilon} - \frac{\sinh\upsilon}{\upsilon^2}\right)\left((\lambda_{\alpha}^2-\lambda_{\beta}^2)\frac{a-d}{2} - 2\lambda_{\alpha}^2\lambda_{\beta}^2\,\Gamma t_1 \right) \bigg\}  \nonumber\\
    &+ \frac{\delta}{\delta^2+4J^2} \frac{\sinh\upsilon}{\upsilon}\left\{ 4\lambda_{\alpha}\lambda_{\beta}\frac{a-d}{2} + (\lambda_{\alpha}^2-\lambda_{\beta}^2)\frac{i\delta t_1}{\hbar}\sec 2\Theta\tan 2\Theta + 2\Gamma t_1\,\lambda_{\alpha}\lambda_{\beta}(\lambda_{\alpha}^2-\lambda_{\beta}^2)   \right\} ~   \bigg]
\end{align}
where
\begin{equation}
    \frac{\partial\upsilon}{\partial J} = \frac{\delta}{\delta^2+4J^2}\,\frac{t_1^2}{8\upsilon}\,\left\{ -\frac{4\delta^2}{\hbar^2}\sec^2 2\Theta\tan 2\Theta - \frac{4i\delta\Gamma}{\hbar} \sec 2\Theta\tan 2\Theta (\lambda_{\alpha}^2-\lambda_{\beta}^2) - \frac{8i\delta\Gamma}{\hbar} \sec 2\Theta \lambda_{\alpha}\lambda_{\beta}  \right\}.
\end{equation}

\section{Relation Between QFI of PDC State and Post-Selected Biphoton State}\label{appendix:relationPDCbiphoton}
The vacuum term in the two-photon PDC state in Eq.~(\ref{eq:PDCstate1}) does not contribute to the detected signal, and is often dropped in theoretical analyses by post-selecting for the detected two-photon states only. The post-selected biphoton state, renormalised to ensure a unit norm, can be expressed in terms of the PDC JTA as:
\begin{equation}\label{eq:biphotonPDC}
    \ket{\Phi_{\mathrm{biph}}} = \frac{1}{\sqrt{\Lambda}}\,\int dt_{\mathrm{S}} \int dt_{\mathrm{I}} \,\Phi_{\mathrm{PDC}}(t_{\mathrm{S}},t_{\mathrm{I}})\,a_{\mathrm{S}}^{\dag}(t_{\mathrm{S}})a_{\mathrm{I}}^{\dag}(t_{\mathrm{I}})\ket{0}
\end{equation}
where $\Lambda = \int dt_{\mathrm{S}} \int dt_{\mathrm{I}}\, \Phi_{\mathrm{PDC}}^*(t_{\mathrm{S}},t_{\mathrm{I}})\Phi_{\mathrm{PDC}}(t_{\mathrm{S}},t_{\mathrm{I}})$ is the normalisation factor for the post-selected state. 
We can then work out the outgoing state corresponding to the biphoton state in Eq.~(\ref{eq:biphotonPDC}) for arbitrary matter systems with Hamiltonian $H_I^{\mathrm{M}}$, also in terms of the PDC JTA:
\begin{equation}
    \ket{\Phi_{\mathrm{biph,out}}} = \frac{1}{\sqrt{\Lambda}}\,\int dt_{\mathrm{S}} \int dt_{\mathrm{I}} \,\Phi_{\mathrm{PDC,out}}(t_{\mathrm{S}},t_{\mathrm{I}})\,a_{\mathrm{S}}^{\dag}(t_{\mathrm{S}})a_{\mathrm{I}}^{\dag}(t_{\mathrm{I}})\ket{0}
\end{equation}
where $\Phi_{\mathrm{PDC,out}}(t_{\mathrm{S}},t_{\mathrm{I}}) = \sum_{n}\,r_{n,\mathrm{PDC}}\,\phi_{n,\mathrm{PDC}}^{\mathrm{S}}(t_{\mathrm{S}})\,h_{n}^{\mathrm{I}}(t_{\mathrm{I}})$. This can then be used to calculate the corresponding QFI for parameter $\theta$,
\begin{align}\label{eq:biphotonPDCQFI}
    &\mathcal{Q}(\theta;\ket{\Phi_{\mathrm{biph,out}}}) = \nonumber\noindent\\
    &\frac{4}{\Lambda}\bigg[\int dt_{\mathrm{S}} \int dt_{\mathrm{I}}~ \frac{\partial \Phi_{\mathrm{PDC,out}}(t_{\mathrm{S}},t_{\mathrm{I}})^*}{\partial \theta}~\frac{\partial \Phi_{\mathrm{PDC,out}}(t_{\mathrm{S}},t_{\mathrm{I}})}{\partial \theta} - \frac{1}{\Lambda}\left|\int dt_{\mathrm{S}} \int dt_{\mathrm{I}} \frac{\partial \Phi_{\mathrm{PDC,out}}(t_{\mathrm{S}},t_{\mathrm{I}})^*}{\partial \theta} \Phi_{\mathrm{PDC,out}}(t_{\mathrm{S}},t_{\mathrm{I}}) \right|^2~ \bigg].
\end{align}
The outgoing PDC state QFI and the post-selected state QFI are then related to each other as
\begin{align}\label{eq:PDCBPrelation}
     \mathcal{Q}(\theta;\ket{\Phi_{\mathrm{biph,out}}}) = \frac{N_{\mathrm{PDC}}}{\Lambda} \mathcal{Q}(\theta;\ket{\Phi_{\mathrm{PDC,out}}}) + \frac{4(\Lambda - 1)}{\Lambda^2}\,\left|\int dt_{\mathrm{S}} \int dt_{\mathrm{I}} \frac{\partial \Phi_{\mathrm{PDC,out}}(t_{\mathrm{S}},t_{\mathrm{I}})^*}{\partial \theta} \Phi_{\mathrm{PDC,out}}(t_{\mathrm{S}},t_{\mathrm{I}}) \right|^2.
\end{align}
It is interesting to note the contrasting nontriviality of this relation~(where the transformation between the two QFIs depends on the value of the true value of the parameter $\theta$) vis-\`{a}-vis the transformation of the QFI function when the parameter is rescaled~(in which case the QFI is rescaled by the square of the constant scaling factor $\mathcal{Q}(\theta/c;\ket{\psi}\bra{\psi}) = (1/c^2)\mathcal{Q}(\theta;\ket{\psi}\bra{\psi}$). 

The biphoton normalisation factor $\Lambda$ is proportional to the rate of entangled photon production given by $(\alpha_{\mathrm{pump}}^2/\hbar^2)$, which is typically a very small number given that the PDC process only converts between one in $10^6$ to $10^{11}$ pump photons into entangled daughter photons, depending on the particular nonlinear crystal used, and other experimental variables. In this text, for our choice of $\alpha_{\mathrm{pump}}/\hbar = 0.01$, we can safely conclude that $\Lambda\ll1$, and (by the same token) $N_{\mathrm{PDC}}\approx1$. This yields the simpler relation between the two QFIs,
\begin{align}\label{eq:PDCBPrelation_simplified}
      \mathcal{Q}(\theta;\ket{\Phi_{\mathrm{PDC,out}}}) \approx
     \frac{\Lambda}{N_{\mathrm{PDC}}}  \mathcal{Q}(\theta;\ket{\Phi_{\mathrm{biph,out}}})  +  \frac{4}{\Lambda N_{\mathrm{PDC}}}\,\left|\int dt_{\mathrm{S}} \int dt_{\mathrm{I}} \frac{\partial \Phi_{\mathrm{PDC,out}}(t_{\mathrm{S}},t_{\mathrm{I}})^*}{\partial \theta} \Phi_{\mathrm{PDC,out}}(t_{\mathrm{S}},t_{\mathrm{I}}) \right|^2.
\end{align}

\end{document}